\documentstyle[sprocl,epsf,twoside]{article}

\def\beq{\begin{equation}}   
\def\eeq{\end{equation}}

\def\lsim{\mathrel{\rlap{\lower3pt\hbox{\hskip0pt$\sim$}}
    \raise1pt\hbox{$<$}}}         
\def\gsim{\mathrel{\rlap{\lower4pt\hbox{\hskip1pt$\sim$}}
    \raise1pt\hbox{$>$}}}         

\begin{document}
\sloppy

\textheight=6.98truein   
\thispagestyle{empty}

\begin{flushright}
JLAB-THY-00-33\\
October 19, 2000 
\end{flushright}

\vspace{1cm}
\begin{center}
{\Large \bf GENERALIZED PARTON DISTRIBUTIONS}
\end{center}\vspace{1cm}
\begin{center}
{  A.V. RADYUSHKIN$^{1,2}$ \footnotemark
}  \\ \vspace{1cm}
{\em $^1$Physics Department, Old Dominion University,} \\ 
{\em Norfolk, VA 23529, USA}
 \\ 
{\em $^2$Theory Group, Jefferson Lab,}\\{\em Newport News, VA 23606, USA}
 
\end{center}

 \footnotetext{ To be published in the
Boris Ioffe Festschrift ``At the Frontier of Particle Physics / Handbook
of QCD'', edited by M. Shifman (World Scientific, Singapore, 2001).}

\newpage

\pagestyle{myheadings}  
\markboth{\small \em Handbook of QCD / Volume 2}{\small \em 
 Generalized Parton Distributions}

\title{GENERALIZED PARTON DISTRIBUTIONS}

\author{A.V. RADYUSHKIN}

\address{Physics Department, Old Dominion University,
\\ Norfolk, VA 23529, USA 
 \\  and  \\
 Theory  Group, Jefferson Lab, Newport News,VA 23606, USA}

\maketitle\abstracts{
Applications of perturbative QCD to deeply virtual Compton 
scattering and hard exclusive electroproduction  processes 
require a generalization of the  usual parton distributions for 
the case when long-distance information is accumulated in  
nondiagonal matrix elements  of quark and gluon light-cone  
operators. I describe  two types of nonperturbative functions 
parametrizing such matrix elements: double distributions 
and skewed  parton distributions. I    discuss their general 
properties, relation to the usual parton densities and form 
factors,  evolution equations for both types of generalized 
parton distributions (GPD), models for GPDs and their applications 
in  virtual and real Compton scattering. }

\vspace{1cm}

\tableofcontents
\newpage

\section{Introduction}

The standard feature of  
applications of perturbative QCD to hard
processes  is the introduction of  phenomenological
functions  accumulating information about
nonperturbative long-distance dynamics.    The well-known 
 examples are the 
 parton distribution functions\,\cite{feynman} $f_{p/H}(x)$ 
used in  perturbative QCD approaches to hard
inclusive processes, and 
distribution amplitudes  
 $\varphi_{\pi}(x)$, 
$\varphi_{N}(x_1 , x_2 , x_3) $  which naturally
emerge in the  asymptotic  QCD
analyses\,$^{2-7}$
of hard exclusive processes.
More recently,  it was argued that the  
gluon distribution function $f_g(x)$ used for description
of hard inclusive processes also determines 
the amplitudes  of  hard exclusive $J/ \psi$ (Ref. 8) 
 and  $\rho$-meson \mbox{(Ref. 9)}  
electroproduction. 
Later, it was proposed~\cite{ji,ji2} 
to use  another exclusive process of  
deeply virtual Compton scattering $\gamma^* p \to \gamma p' $
 (DVCS) for measuring   
quark  distribution functions inaccessible 
in  inclusive measurements
(earlier discussions of nonforward Compton-like   
amplitudes $\gamma^* p \to \gamma^* p' $ 
with a  virtual photon or $Z^0$ in the
final state  can be found   in Refs. 12--14). 
The important feature (noticed  long ago~\cite{barloe,glr}) 
 is that  kinematics of hard elastic 
electroproduction processes (DVCS can be also treated as one of them) 
requires 
the presence of the longitudinal (or, more precisely,
light-cone ``plus'') component
in the  momentum transfer 
$r\equiv p - p'$ from the initial hadron to the final:
$r^+ = \zeta p^+$. For DVCS and $\rho$-electroproduction 
in the region       $Q^2 \gg |t|, m_H^2$,
the longitudinal momentum asymmetry (or ``skewedness'') 
parameter $\zeta$
coincides~\cite{afs} with  the
Bjorken variable $x_{Bj} = Q^2/2(pq)$ associated with 
the virtual photon momentum $q$. 
This means that kinematics of the nonperturbative matrix element
$\langle p' | \ldots | p \rangle$ 
is asymmetric (skewed). In particular, 
the gluon distribution which appears in  
hard elastic diffraction  amplitudes
differs  from that studied in inclusive 
processes.\cite{gluon,npd} In the latter case, 
one has  a symmetric situation when the same
momentum $p$ appears in both brackets of 
the hadron matrix element
$\langle p | \ldots | p \rangle$. 
 Studying  the DVCS process,
 one deals with essentially {\it off-forward}~\cite{ji,ji2}
or {\it nonforward}~\cite{npd,compton} 
kinematics for the matrix element $\langle p' | \ldots | p \rangle$. 
Perturbative quantum chromodynamics (PQCD)
provides an appropriate theoretical framwork.
The basics of the    PQCD approaches   
incorporating  the new {\it generalized}   parton distributions (GPDs) 
were formulated in Refs. 10,11,16--19. 
A detailed analysis of PQCD factorization 
for hard meson electroproduction processes
was given in Ref. 20.  

Our goal in the present paper is to review
 the formalism of the generalized 
parton distributions 
based on  the approach outlined in 
our  papers 16--18,21--25. 

Its  main idea  is that  
constructing a consistent PQCD approach  for 
 amplitudes of hard exclusive electroproduction processes
one should treat the initial 
momentum $p$ and 
 the momentum transfer $r$ on 
equal footing by  introducing 
{\it double distributions} (DDs)  
$F(x,y)$, which specify the fractions of  $p^+$ 
and  $r^+$ 
carried by the active parton 
of the parent hadron.
These distributions have hybrid properties:
they look like distribution functions 
with respect to $x$ and like  distribution amplitudes
with respect to $y$.

The other possibility is to treat
 the proportionality  coefficient 
$\zeta\equiv r^+/p^+$ as an 
independent parameter  and    introduce 
an alternative description in terms
of   {\it nonforward parton distributions}\,\cite{npd,cfs}  (NFPDs) 
${\cal F}_{\zeta}(X;t)$ 
with $X=x+y \zeta$ being the total 
fraction of the initial hadron momentum  $p^+$ 
taken  by the initial  parton.
The shape of  NFPDs  explicitly
depends on the parameter $\zeta$ characterizing the  skewedness
of the relevant nonforward matrix element.
This parametrization 
is similar to that proposed  by 
X. Ji,\cite{ji,ji2,jirev} who introduced 
 off-forward parton distributions (OFPDs) $H(\tilde x,\xi;t)$
in which the parton momenta and  the skewedness
parameter $\xi\equiv r^+ / 2 P^+$ 
are measured in units of the average 
hadron momentum $P=(p+p')/2$. 
  OFPDs and NFPDs 
can be   treated 
as particular forms  of {\it skewed } parton 
distributions (SPDs).
One can also introduce the version of  DDs (``$\alpha$-DDs'',
see Ref. 22) 
in which the active parton momentum is  written in terms of symmetric 
variables:  
$k^+= xP^+ + (1+\alpha) r^+/2$.    

The paper is organized as follows.
In Sec. 2,   I recall the basic properties of 
``old''  parton distributions, i.e., I discuss the  usual parton densities 
$f(x)$ 
and the  meson distributions amplitudes $\varphi (\alpha)$. 
In Sec. 3, I consider  deeply virtual Compton scattering
 as a characteristic process involving nonforward matrix 
 elements of light--cone operators. I introduce double distributions
 $f(x,\alpha;t)$ and discuss their general 
 properties. The alternative description
 in terms of skewed parton distributions
 $H(\tilde x,\xi;t)$ is described in Sec. 4.
Models for double and skewed distributions
based on relations between GPDs and the usual 
parton densities are constructed in Sec. 5.
The evolution of GPDs at the leading logarithm 
level is studied in Sec. 6. 
In Sec. 7, I discuss recent studies 
of  DVCS amplitude at twist-2 and 
twist-3 level.  In Sec. 8, the 
 GPD formalism is applied to real Compton scattering
at large momentum transfer. 
In the concluding section (Sec.9),
I briefly outline other  developments in the theory
of generalized parton distributions
and their applications.  
 
 \section{``Old'' Parton Distributions }

\subsection{Parton Distribution Functions}

The parton distribution function $f_{a/H}(x)$
gives the probability that  a fast-moving 
hadron $H$, having  the momentum $p$,   contains  a parton $a$
carrying the momentum $xp$ and any other 
partons $X$ (spectators) carrying together the remaining 
momentum $(1-x) p$. Schematically,    
  $$f_{a/H}(x) \sim  \sum_{``X"} |\Psi \{H (p)  \to a(xp) +  ``X"\}|^2 \ , $$ 
where the summation is over all possible 
sets of spectators and $\psi \{H\to a X\}$ is the probability 
amplitude for the splitting process
$H \to aX$. 
\begin{figure}[h] 
 \begin{center} 
\mbox{ 
\epsfxsize=6cm
 \epsfysize=4.5cm
  \epsffile{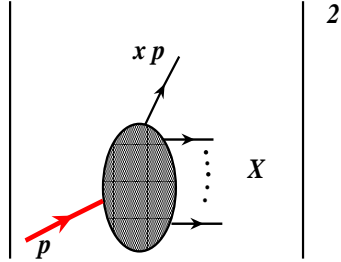}}
\end{center}
\caption{\label{pionda}  Parton distribution function. }  
\end{figure} 
The summation over $X$ 
reflects the {\em inclusive} nature 
of the description of the hadron structure by 
the  parton distribution functions $f_{a/H}(x)$. 
The parton distribution functions $f_{a/H}(x) $
have been intensively studied in experiments
on hard inclusive processes
for the last 30 years. 
The classic process in this respect  is the 
deep inelastic scattering  (DIS) 
$eN \to e' X$ whose structure functions
are directly expressed in terms of $f_{a/H}(x) $.

The standard approach is to write 
the DIS structure function as the imaginary part
 of the virtual forward Compton amplitude $T_{\mu\nu}(q,p)$.
 In the most general nonforward case, the
 virtual Compton scattering amplitude is derived from the correlation 
function of two electromagnetic  currents $J^\mu (x), J_\nu (y)$ 
\begin{eqnarray}
T_{\mu\nu}  &=& 
i \int d^4 x \; \int d^4 y \; e^{-i (q x) + i(q'  y)}
\langle p' | {\rm T} \left\{ J_\mu (x) J_\nu (y) \right\}
| p \rangle \ .
\label{corr}
\end{eqnarray}
In the  forward limit,
the ``final'' photon  has the momentum $q$ coinciding with
that of the initial one. The momenta $p, p'$ of the initial
and ``final'' hadrons also coincide. 
The light-cone dominance of the 
virtual  forward Compton amplitude is secured by high
virtuality of the photons $-q^2 \equiv Q^2 > 1 $ GeV$^2$ and large 
 total center-of-mass (cms)  energy of the photon-hadron system
$s= (p+q)^2$. The latter   should be above resonance region,
with the Bjorken ratio $x_{Bj} \equiv  Q^2/2(pq)$  fixed.

 An  efficient  way  to study 
the behavior of Compton amplitudes
in the Bjorken limit is to use the light--cone expansion 
for the 
product $$
\Pi_{\mu\nu} (x, y)  \equiv  i {\rm T} J_{\mu}(x) J_{\nu} (y)  $$ 
of two vector currents in the coordinate representation.    
The leading order contribution is given  
by two ``handbag'' diagrams,  
\begin{eqnarray} 
 \Pi_{\mu\nu} (z\,|\, X) 
 =   \frac{ z_\rho}{\pi^2 z^4} \biggl\{ s_{\mu\rho\nu\sigma} 
 {\cal O}_{\sigma} (z\,|\,X) - 
 i \epsilon_{\mu\rho\nu\sigma}  
{\cal O}_{5\sigma} (z\,|\,X)
\biggl \} \  , 
\label{handbag}
\end{eqnarray}
where
$s_{\mu\rho\nu\sigma} = g_{\mu\rho} g_{\nu\sigma} - g_{\mu\nu} g_{\rho\sigma}
+ g_{\mu\sigma} g_{\nu\rho} $, \,  $X  = (x + y)/2$,  
$z  =  y - x $,  and 
\begin{eqnarray}
{\cal O}_{\sigma} (z\,|\,X) &=& \frac1{2} \,
\left[ \bar\psi (X-z/2) \gamma_\sigma 
\psi (X+z/2) \; - \; (z \rightarrow -z) \right] \  ,
\nonumber \\
{\cal O}_{5\sigma} (z\,|\,X) &=& \frac1{2} \,
\left[ \bar\psi (X-z/2) \gamma_\sigma \gamma_5
\psi (X+z/2) \; + \; (z \rightarrow -z) \right] \  .
\end{eqnarray}

 Formally,  the parton distribution functions
 provide  parametrization of the forward 
matrix elements of quark and gluon 
operators on the light cone. 
For example,  in the parton helicity averaged case
(corresponding to the  vector operator ${\cal O}_{\sigma}$) 
the $a$-quark/antiquark distributions are  defined by 
\begin{eqnarray} 
&& \langle p\, \,  |     \,  \, \bar \psi_a(-z/2) \hat z 
E(0,z;A)  \psi_a(z/2) \, \,  |     \,  \, p \rangle \,   
\nonumber \\ && 
 =  \bar u(p)  \hat z u(p)  
   \int_0^1  \,  
 \left ( e^{-ix(pz)}f_a(x) 
  -   e^{ix(pz)}f_{\bar a}(x)
\right ) \, dx \, + O(z^2) , 
\label{33} \end{eqnarray}  
 where $E(0,z;A)$ is the standard path-ordered exponential
 (Wilson line) of the gluon field $A$ 
 which secures gauge invariance 
 of the nonlocal operator. 
 In what follows, we will not write it explicitly.  
  Throughout, we use  the ``Russian hat''
 notation $\hat z \equiv \gamma_\mu z^\mu$.
  
 The non-leading (or higher-twist) $O(z^2)$ terms 
 in the above representation
 soften the light--cone singularity 
 of the Compton amplitude,  which    results in  
 suppression by  powers of $1/Q^2$
 (see Sec. 7 for a  discussion
 of twist decomposition and higher--twist corrections).

  The exponential factors $\exp[\mp ix(pz)]$ accompanying
the quark and antiquark distributions 
reflect the  fact that the field $\psi (z/2) $
appearing in the  operator
$\bar \psi (-z/2)  \ldots \psi (z/2) $  
consists of  the quark annihilation 
operator (a quark with momentum $xp$
comes into  this point) and the antiquark
creation operator (i.e., an antiquark with momentum
$xp$ goes out of this point).
To get the relative signs with which  quark 
and antiquark distributions appear in these definitions,
we should take into account 
that antiquark creation and annihilation
operators appear in $\bar \psi (-z/2)  \ldots \psi (z/2) $ 
in   opposite  order.

In a similar way, one can introduce 
 polarized quark densities $\Delta f_{a,\bar a} (x)$  which parametrize
the forward matrix element of the axial operator
$\bar \psi_a(-z/2) \hat z \gamma_5  
E(0,z;A)  \psi_a(z/2)$.

Combining  the parametrization (\ref{33}) 
with the Compton amplitude (\ref{handbag}) 
   one obtains  the 
QCD parton representation
\begin{equation}
 T^{\mu \nu} (p,q) = \sum_a \int_0^1  f_a(x) \, t^{\mu \nu}_a(xp,q) \, 
 dx \biggl\{1+O(1/Q^2)\biggr\}
 \label{DISfactorize} 
 \end{equation} 
 for the hadron amplitude in terms of the 
 perturbatively calculable hard parton amplitude $t^{\mu \nu}_a(xp,q)$ 
 convoluted with the parton distribution
functions $f_a(x)$ ($a=u,d,s,g,  \ldots$) 
which describe/parametrize nonperturbative
information about hadronic structure.
The short-distance part of the handbag 
contribution is given by the  hard quark
propagator proportional to $1/[(xp+q)^2+i \epsilon]$.
Its imaginary part  contains the  $\delta (-Q^2 +2x (qp))$
factor (terms $O(p^2)$  are neglected) which 
selects just the $x=x_{Bj}$  value from the $x$-integral. 
As a result, the DIS cross section is directly expressed in terms of
$f_{a/H}(x_{Bj})$.
\begin{figure}
   \epsfxsize=12cm
 \epsfysize=4.5cm
 \centerline{\epsfbox{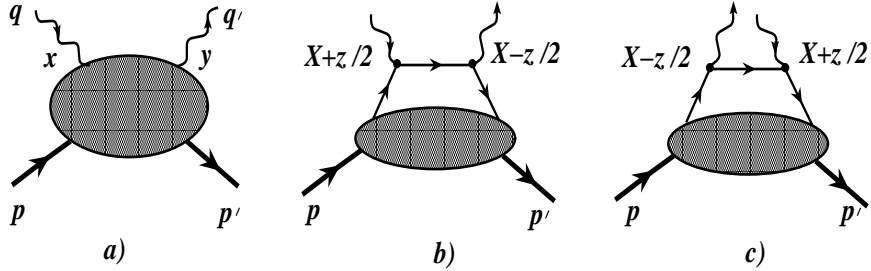}  }
{\caption{\label{fig:cohan} $a)$ General Compton amplitude;
$b)$ $s$-channel handbag diagram;  $c)$ $u$-channel handbag diagram.
   }}
\end{figure} 
 Note, however, that the factorized representation  
is valid  for the full Compton amplitude 
in which the parton densities are integrated 
over $x$.  In other words, the variable $x$ 
in Eq.(\ref{DISfactorize})  has the meaning of the momentum
fraction carried by the parton, but it is not equal
to the Bjorken parameter $x_{Bj}$. 

 The basic parametrization (\ref{33}) 
can be also written as an integral from $-1$ to 1 
with a common exponential  
 \begin{eqnarray} 
\langle p\, \,  |     \,  \, \bar \psi_a(-z/2) \hat z 
  \psi_a(z/2) \, \,  |     \,  \, p \rangle \,  |     \,  
 =  \bar u(p)  \hat z u(p)  
   \int_{-1}^1  \,  
  e^{-ix(pz)}\tilde f_a(x) 
 \, dx \, + O(z^2). 
\label{33c} \end{eqnarray} 
 For $x>0$, the distribution function 
$\tilde f_a(x)$ coincides with
the quark distribution  $f_a(x)$,
while for $x<0$ it is given by (minus)
the antiquark distribution    $-f_{\bar a}(-x) $.

\subsection{Distribution amplitudes }

To give an example of another important type
of nonperturbative functions 
describing hadronic structure,
namely, the {\it distribution amplitudes}, 
let us consider the 
$\gamma^* \gamma \pi^0$ form factor.
 It is usually measured in $e^+ e^-$
 collisions,  but for our purposes 
 it is more convenient to 
 represent it through the process
 in which electron is scattered
 with large momentum transfer
 $q$ off the pion target  
 producing a real photon in the final state.
 The pion, in particular,  can belong to 
 the cloud surrounding a nucleon. 
 In this case, the $\gamma^*  \pi^0 \to \gamma$  subprocess 
 is a part of deeply virtual Compton scattering
 (DVCS)  which will be considered in more detail later on. 

  The leading 
 order contribution for large $Q^2$ is given by two handbag 
 diagrams, and in the coordinate 
 representation one deals with the 
 same Compton amplitude (\ref{handbag}).
 The only difference is that 
 the nonlocal operators should
 be sandwiched between the one-pion state
 $|\, \pi (r) \, \rangle$ 
 ($r$ is the pion momentum) and the vacuum
 $\langle \, 0 \, |$.  
  Since the pion is a pseudoscalar 
 particle, only the  axial nonlocal operator 
${\cal O}_{5 \sigma}$ contributes, and 
 the pion distribution amplitude 
 $\varphi_{\pi}  (\alpha)$ 
 is the function parametrizing 
 its  matrix element 
  \begin{eqnarray} 
\langle 0\, \,  |     \,  \, \bar \psi (-z/2) \hat z \gamma_5
 \psi  (z/2) \, \,  |     \, \pi (r) \,  \rangle \,       \,  
 =  (rz)   
   \int_{-1}^1  \,  
  e^{-i\alpha (rz)} \varphi_{\pi}  (\alpha)
 d \alpha \, +O(z^2).  
\label{fpi}
 \end{eqnarray} 
One can interpret $\varphi_{\pi}  (\alpha)$ 
as  the probability amplitude 
$$\Psi \{\pi (r) \to q(yr)+ \bar q((1-y)r) \}$$
to find the pion in a quark-antiquark state
with the pion momentum $r$ shared 
in fractions $y\equiv (1+\alpha)/2$ and 
 $1-y = (1-\alpha)/2$. 
 Since the function $\varphi_{\pi}  (\alpha)$ is even in $\alpha$, 
 the  use of the  relative fraction   $\alpha$ has some  advantages
 when the symmetry properties are concerned.

\begin{figure}
   \epsfxsize=10cm
 \epsfysize=4.5cm
 \centerline{\epsfbox{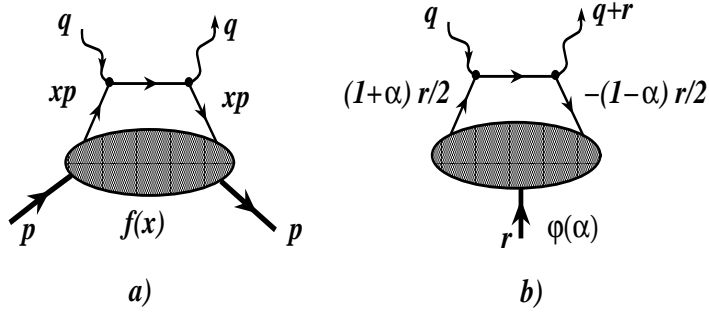}  }
{\caption{\label{fig:dist} Handbag diagrams and parton picture.  $a)$ 
Virtual forward Compton amplitude expressed through 
 the usual parton densities $f(x)$.
$b)$ Form factor  $\gamma^* \gamma \pi^0$ written 
in terms of the distribution amplitude 
$\varphi (\alpha)$. 
   }}
\end{figure}

\section{Double distributions}

\subsection{DVCS and DDs} 

Now, let us consider deeply virtual Compton
scattering (DVCS), an  exclusive process 
$\gamma^* (q) N (p)  \to \gamma (q') N(p')$ 
in which a highly virtual initial photon 
 produces a real photon in the final state.
 The initial state of this reaction 
 is in the  Bjorken kinematics:  $Q^2 \equiv -q^2$
and $(pq)$ are large, while the  ratio 
$x_{Bj} \equiv Q^2/2(pq)$  is fixed,
just like in DIS.   
 An extra variable is the momentum transfer
 $r \equiv p-p'$.  The simplest case is when   
  $t \equiv r^2$ is small. 
 This does not mean, of course, that 
 the components of $r$ must be small:
 to convert a highly virtual photon
 into a real one, one needs 
 $r$ with a large projection on $q$.
 Indeed, from $q'^2=0$ it follows that
 \beq (q+r)^2=-Q^2 +2(qr)+t =0,\eeq
  i.e. $(qr)\approx -Q^2/2$ for small $t$. 
  In the $t\to 0$ limit, we can  
  write $(qr) = x_{Bj} (qp)$.
  Taking the initial momentum $p$
  in the light cone ``plus'' direction
  and the momentum $q'$ of the final photon
  in the light cone ``minus'' direction,
  we conclude that the momentum transfer 
  $r$ in DVCS kinematics must have a
  non-zero plus component: $r^+ = \zeta p^+$,
  with $\zeta = x_{Bj}$. 

For large $Q^2$, the leading order contribution is given 
again by   handbag diagrams.
A new feature is that the nonperturbative part is described 
by nonforward matrix elements $\langle \, p-r \, | \ldots |\, p \, \rangle$ 
of  ${\cal O}_{\sigma}$ and ${\cal O}_{5 \sigma}$ operators. 
These matrix elements are parametrized by generalized
parton distributions (GPDs). 
It is instructive to 
treat GPDs as  kinematic ``hybrids''
of the usual parton densities $f(x)$ 
and distribution amplitudes $\varphi (\alpha)$. 
Indeed, $f(x)$ corresponds to the  forward limit 
$r=0$,  when  the momentum $p$ flows only in the $s$-channel
and the outgoing parton carries the momentum $xp^+$. 
  On the other hand, if we take $p=0$,
  the momentum $r$ flows in the  $t$-channel only
  and is shared in fractions $(1+\alpha)r^+/2$ and $(1-\alpha)r^+/2$.
  In   general case, we deal with superposition
  of two momentum fluxes: the plus component 
  $k^+$ of the parton momentum $k$  
  can be written as $ x p^+ +(1+\alpha)r^+/2$.
  To fully incorporate the symmetry
  properties of  nonforward matrix
  elements,  it is  convenient to introduce 
  the symmetric momentum variable
  $P=(p+p')/2$ and write the parton momentum 
  as \beq k^+ =  x P^+ + (1+\alpha)r^+/2. \eeq  
This decomposition  corresponds to the 
following parametrization    
\begin{eqnarray} && 
\langle P-r/2 |  \bar \psi_a (-z/2) \hat z
\psi_a (z/2) |P+r/2 \rangle \, \nonumber \\&&  = 
\bar u(p') \hat z u(p) \, \int_{-1}^1 \, dx \int_{-1+|x|}^{1-|x|} 
 e^{-ix(Pz)-i\alpha (rz)/2}  \,  f_a(x,\alpha;t) 
 \,
d\alpha  \,  \nonumber \\ && 
 + O(r) \,\, {\rm terms}\ +O(z^2), 
 \label{17} \end{eqnarray}
where $f_a(x,\alpha;t)$ is  the 
{\it double distribution} (DD).\cite{ddee,sssdd} 
For the moment, we  do not write explicitly  
 ``$O(r)$''  terms
corresponding to $\bar u(p') 
\sigma_{\alpha \beta} r^{\alpha} z^{\beta}  u(p)$
and  $(rz) \bar u(p')   u(p)$ structures. 

\begin{figure}
   \epsfxsize=12cm
 \epsfysize=4.5cm
 \centerline{\epsfbox{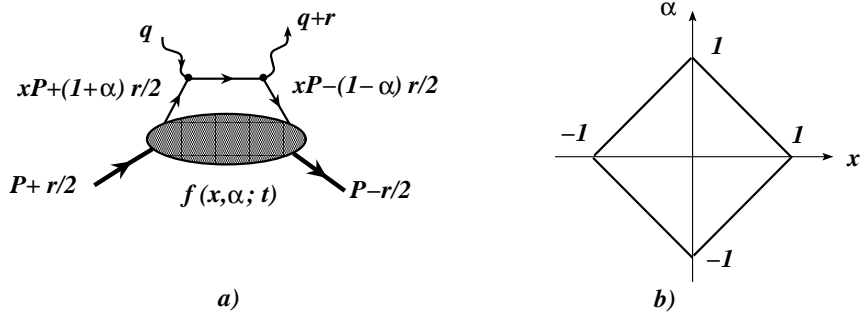}  }
{\caption{\label{fig:dd} 
$a)$ {Parton description of deeply virtual Compton scattering 
in terms of  double distributions
 $f(x,\alpha;t)$. $b)$ Support region for $f(x,\alpha;t)$.
  }}}
\end{figure}

\subsection{General properties of DDs}

The support area for $f_a(x,\alpha;t)$ is shown 
on Fig.~\ref{fig:dd}b. For any Feynman 
diagram, the spectral constraint 
$|x| + |\alpha| \leq 1$   can be   proved 
in the $\alpha$-representation\,\cite{npd} using the approach of 
Ref. 26. 
 Comparing Eq. (\ref{33}) 
with the $r =0$ limit of the DD definition (\ref{17})
gives  the   
``reduction formulas'' 
relating   the double distribution $f_{a}(x,\alpha;t=0)$  
to the quark and antiquark parton  densities
\begin{eqnarray} 
\int_{-1+x}^{1-x} \, f_{a}(x,\alpha;t=0)|_{x>0} \, d \alpha = 
 f_{a}(x)  \\ 
\int_{-1+|x|}^{1-|x|} \, f_{a}(x,\alpha;t=0)|_{x< 0}\, d \alpha = 
-  f_{\bar a}(-x) \,  \label{eq:redfsym} \, .
 \end{eqnarray}
Hence, the  positive-$x$
and negative-$x$  components of the double 
distribution  $f_{a}(x,\alpha;t=0)$ 
can be treated as nonforward generalizations of 
quark and antiquark densities, respectively.
The usual ``forward''  densities 
$f_a(x)$ and $f_{\bar a} (x)$ 
are thus given by integrating 
$f_a(x,\alpha\,;\,t=0)$  over vertical lines
$x = {\rm const} $ for $x>0$ and $x<0$, respectively.
In principle, we cannot exclude 
the  third possibility that  the functions 
$f_{a}(x,\alpha;t) $    have  singular terms at $x=0$ 
proportional 
to $\delta (x)$ or its derivative(s).
Such terms  
have no projection  onto the usual parton densities.
We will denote them by $f_M (x,\alpha;t)$ $-$  they may 
be interpreted as coming from the $t$-channel 
meson-exchange type contributions.  
In this case, the partons  just share 
the plus component of the momentum transfer $r$:
information about 
the magnitude of the  initial hadron momentum
is lost if the exchanged particle can be described
by a pole propagator \mbox{$ 1/(t-m_M^2)$.}
Hence, the meson-exchange  contributions to a double distribution
may look like
\begin{equation}
 f_M^+(x,y;t) \sim  \delta (x) \,
 \frac{\varphi_{M}^+ (\alpha)}{m_M^2 -t } \ \  \ 
{\rm or }  \ \  \    f_M^-(x,\alpha;t) 
\sim  \delta ' (x) \, \frac{\varphi^-_{M
}(\alpha)}{m_M^2 -t } \ \  , \  \  {\rm etc.} \, , 
\end{equation} 
where $\varphi_M^{\pm}  (\alpha)$ are  the functions related to the 
distribution amplitudes of the relevant mesons $M^{\pm}$.
The  two examples above correspond to $x$-even and $x$-odd
parts of the double distribution $f  (x,\alpha;t)$. 

Due  to hermiticity
and time-reversal invariance properties of
nonforward matrix elements,  the  DDs
are even functions of $\alpha$,
$$
\tilde f_a(x,\alpha;t) = \tilde f_a(x, - \alpha;t) \, . 
$$
In
particular, the  functions  $\varphi_M^{\pm}(\alpha)$ 
for singular contributions
$f_M^{\pm}(x,\alpha;t)$ are  even functions  
$\varphi_M^{\pm}(\alpha) = \varphi_M^{\pm}(-\alpha) $ of $ \alpha$
both for  $x$-even and $x$-odd parts.

Note that the $\mu \leftrightarrow \nu$ 
symmetric part of the DVCS amplitude 
contains only the $C$-even operators 
${\cal O}^{\sigma}_a$.  
Their matrix elements 
are parametrized 
\begin{eqnarray} 
&& \langle \,  p',s' \,  |     \,  \,z^{\sigma}  
{\cal O}_{ a} (-z/2,z/2)\, 
\,  |     \,  \, p,s \rangle \,  |     \, _{z^2=0} \nonumber \\  &&  
= \bar u(p',s')  \hat z 
 u(p,s) \,  \int_{-1}^1  dx  \int_{-1+|x|}^{1-|x|}  \, 
    e^{-ix(Pz)-i\alpha (r z)/2} 
\,   f_a^S(x,\alpha \, ; \, t)  
 \, d\alpha \nonumber \\ &&  + O(r)\, \, {\rm terms}
\label{311Q} 
\end{eqnarray} 
by the DDs 
$$f_a^S(x,\alpha \, ; \, t) \! =\! {\rm sign} 
(x) [f_a(|x|,\alpha \, ; \, t) +  
f_{\bar a} (|x|,\alpha \, ; \, t)]$$ 
which are   odd functions of $x$.
In applications to the hard meson electroproduction
one also needs valence-type DDs
$$f_a^V(x,\alpha \, ; \, t)\! =  \!
 [f_a(|x|,\alpha \, ; \, t) - 
f_{\bar a} (|x|,\alpha \, ; \, t)]$$
parametrizing matrix elements 
of $C$-odd operators 
$\bar \psi_a (-z/2) \hat z  \psi_a (z/2) + 
\bar \psi_a (z/2) \hat z  \psi_a (-z/2)$.  

 \section{Skewed parton  distributions}

 \subsection{General definition} 
 
An important parameter for nonforward matrix 
elements is the  coefficient of proportionality 
$\zeta = r^+/p^+$  (or $\xi = r^+/P^+$)    between 
the plus components of the momentum transfer 
and the initial (or  average) hadron  momentum. 
It  specifies 
the {\it skewedness}  of the matrix elements.
The two skewedness parameters are related by
\begin{equation} 
\xi = \frac{\zeta}{2-\zeta} \  . 
\end{equation}

The characteristic   feature  implied by the definition of 
the double distribution 
 (\ref{17})
 is the absence 
of the $\xi$-dependence in $f(x,\alpha;t)$. 
An alternative way to parametrize 
nonforward matrix elements of light-cone operators
is to use   
  $\xi$  and the {\it total } 
 momentum fraction  $ \tilde x \equiv  x +\xi \alpha$  
  as  independent 
variables and introduce 
{\it skewed} parton distributions (SPDs).
The shape of  SPDs  explicitly
depends on the skewedness 
of the relevant nonforward matrix element.

There are
two types of SPDs: off-forward parton distributions\,\cite{ji,ji2} (OFPDs) 
$H(\tilde x,\xi;t)$ and nonforward 
parton distributions\,\cite{gluon,npd} (NFPDs) ${\cal F}_{\zeta} (X;t)$. 
The basic difference is that 
 the skewedness
parameter $\xi$ ($\zeta$) and the parton momentum $k^+$ 
in  the OFPD (NFPD) formalism is
measured in units of the average (initial) momentum $P^+$ ($p^+$).
Hence, there are one-to-one  relations between  OFPDs and NFPDs.\cite{npd}
 We  start with  OFPDs because 
 they  have  simpler symmetry properties. 
 
The relation 
between OFPDs 
$ H(\tilde x,\xi;t)$ 
and DDs  $f(x,\alpha;t)$  is  given just by
the change of variables from $\{x,\alpha \}$ to
$\{ \tilde x , \xi \}$, 
\begin{equation} 
  H (\tilde x,\xi; t)=  \int_{-1}^1 dx \int_{-1+|x|}^{1-|x|}  
\, \delta (x+ \xi \alpha - \tilde x) \, 
 f (x,\alpha;t) \, d\alpha  \, . \label{offfor} 
 \end{equation}
 
If we require that 
the light-cone plus components  
of both the momentum transfer
$r$ and the final hadron momentum  $p - r$ are 
positive (which is the case for DVCS), 
then   $0 \leq \zeta \leq 1$ and $0 \leq \xi \leq 1$.
Using the spectral property   $0 \leq |x|+ |\alpha| \leq 1$
of  double distributions  we obtain that 
the OFPD variable $\tilde x $ satisfies the    
constraint $0\leq |\tilde x|  \leq 1$.
Note also   that Eq. (\ref{offfor}) formally allows to construct 
$\tilde H (\tilde x,\xi; t)$
both for positive and negative values of $\xi$.
Since the DDs $\tilde f (x,\alpha;t)$ are even functions 
of $\alpha$, the OFPDs  
$ H (\tilde x,\xi; t)$ are even functions 
of $\xi$: 
$$
\tilde H(\tilde x,\xi;t) = \tilde H(\tilde x,- \xi;t )  \, .
$$
This result was originally obtained by X.~Ji\,\cite{jirev} 
who   used hermiticity and time reversal
invariance  properties in the direct  definition of OFPDs
\begin{eqnarray} && 
\langle p' |  \bar \psi_a (-z/2) \hat z
\psi_a (z/2) |p \rangle \, \nonumber \\ 
&& = 
\bar u(p') \hat z u(p)  \int_{-1}^1 \,  
 e^{-i\tilde x(Pz)}  \, H_a (\tilde x,\xi; t) \,  d  \tilde  x 
 + O(r)  \,\,   {\rm term}\ +O(z^2)  .  
 \label{ofpd} 
\end{eqnarray}
 
 \begin{figure}[htb] 
 \begin{center} 
\mbox{ 
\epsfxsize=12cm
 \epsfysize=4.5cm
   \epsffile{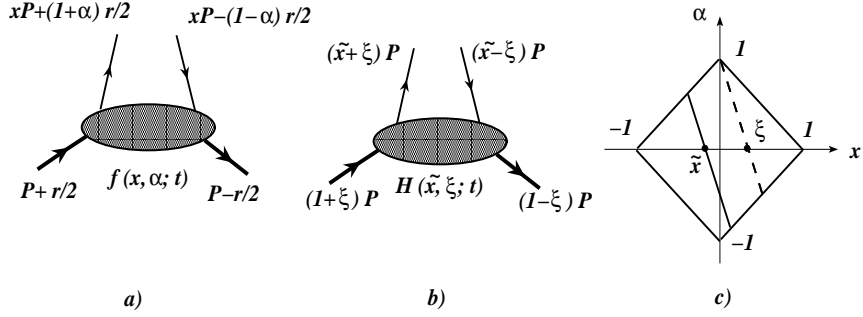}}
\end{center}
\caption{\label{fig:spd} Description 
of the nonforward matrix element
in terms of $a)$  double distribution 
$f(x,\alpha;t)$ and  $b)$ off-forward parton
distribution $H(\tilde x,\xi;t)$. $c)$ Integration 
lines for integrals relating OFPDs and DDs.}  
\end{figure}

 \subsection{Structure of SPDs} 
 
The parton interpretation of 
 the OFPD definition   is the following: the  quark going
 out of the parent hadron carries the 
 fraction $\tilde x + \xi$ of the 
 average hadron momentum $P^+$, while 
 the momentum of the ``returning'' quark  
 is $(\tilde x - \xi)P^+$. For definiteness,  
 we shall assume below
that 
$\xi$ is positive.   
 Treating a quark
 with a negative momentum as an
 antiquark, we   can   distinguish 3 components
 of $H (\tilde x,\xi; t)$.  
 For $\tilde x > \xi$ 
 both the initial fraction $\tilde x + \xi$ and the final 
 one $\tilde x - \xi$ are positive. 
 Hence, $H (\tilde x,\xi; t)$ in this region 
 corresponds to a modified quark distribution.
 Similarly, for $\tilde x < - \xi$
 both fractions are negative,  and 
 $H (\tilde x,\xi; t)$ can be treated as 
an  antiquark distribution. 
 In the third (middle) region, $-\xi < \tilde x < \xi$,
 the two fractions have opposite signs,
 and $H (\tilde x,\xi; t)$  describes 
 splitting of a quark-antiquark pair 
 from the initial hadron. 
 The total momentum carried by the $\bar qq$ pair
 is $2 \xi P^+ = r^+$, and it is shared
 in fractions $(\tilde x + \xi)P^+ = (1+\beta)r^+/2$
 and $(\xi - \tilde x )P^+ = (1-\beta)r^+/2$,
 where $\beta \equiv  \tilde x /\xi$. 
 Note that $|\beta| <1 $ in the 
 $|\tilde x | <\xi $ region. 
  Hence,
the third component  can be interpreted as the
probability amplitude for the initial hadron with momentum 
$(1+\xi)P^+$ to split into the final hadron with momentum $(1-\xi)P^+$
and  a two-parton state with total momentum $r^+$
shared by the partons of the pair. 
Thus, we may expect that $H (\tilde x,\xi; t)$  
in the middle region looks more like a distribution amplitude.

The relation between DDs and SPDs can be 
illustrated on the DD support  rhombus 
$|x|+|\alpha| \leq 1$ (see Fig. \ref{fig:spd}c).
The  delta-function in Eq. (\ref{offfor}) specifies 
the line of  integration in the
$\{ x, \alpha \}$ plane. 
 To get $H (\tilde x,\xi; t)$,
 one should integrate $f(x,\alpha)$ over $\alpha$ along a straight
line  $x=\tilde x - \xi \alpha$. 
Fixing some value of  $\xi$,
one deals with  a set of parallel lines  intersecting  the $x$-axis
at $x=\tilde x $. The upper limit of the $\alpha$-integration
is determined by intersection of this line
either with the line $x+\alpha=1$ (this happens if 
$\tilde x  > \xi$)  or with the line 
$-x+\alpha=1$  (if $\tilde x  < \xi$).
Similarly, the  lower limit of the $\alpha$-integration
is set by the line $x-\alpha=1$ for 
$\tilde x  > - \xi$   or by the line 
$x+ \alpha= - 1$  for $\tilde x  < - \xi$.
The lines corresponding to $\tilde x = \pm \xi $
separate the rhombus  into three parts
generating the  three components of $H (\tilde x,\xi; t)$: 
  \begin{eqnarray}
H_a (\tilde x,\xi;t) = 
\theta(\xi \leq \tilde x \leq 1) 
 \int_{-\frac{1- \tilde x}{1+\xi} }^{\frac{1- \tilde x}{1-\xi} }
f_{a} (\tilde x - \xi \alpha,  \alpha ) \, d \alpha \nonumber \\
+ \theta(-\xi \leq \tilde x \leq \xi) 
 \int_{-\frac{1- \tilde x}{1+\xi} }^{ \frac{1+ \tilde x}{1+\xi}} 
f_{a } (\tilde x - \xi \alpha,  \alpha ) \, d \alpha \, \nonumber \\
+ \theta(-1 \leq\tilde x \leq  - \xi ) 
 \int_{-\frac{1+ \tilde x}{1-\xi} }^{\frac{1+\tilde x}{1+\xi} }
f_{a} (\tilde x - \xi \alpha,  \alpha ) \, d \alpha
.
\label{710}  
 \end{eqnarray}

Recall that integrating the  DD $f(x,\alpha;t=0)$
over a vertical line gives the usual parton density 
$f(x)$.  To get    
 the $t=0$   SPDs  
one should scan    the same DD 
along  the lines having  a  $\xi$-dependent slope. 
Thus, one can try to build 
models for SPDs using information about 
usual parton densities.  
Note, however, that the usual parton densities
are insensitive to the meson-exchange type contributions  
 $H_M (\tilde x,\xi; t)$
coming from the singular $x=0$ parts of DDs.
Thus, information contained in SPDs originates from two  
physically different sources: meson-exchange type contributions  
 $H_M (\tilde x,\xi; t)$
and  the functions 
$H_M (\tilde x,\xi; t)$
 obtained by scanning the $x \neq 0$  parts of  
DDs $f(x,\alpha;t)$. 
The support of exchange contributions is restricted 
to $|\tilde x|  \leq \xi$. Up to rescaling, the function
$H_M (\tilde x,\xi; t)$   
has the same shape for all $\xi$,
 e.g., $\varphi_M(\tilde x / \xi;t)/|\xi|$.
For any nonvanishing $\xi$, these exchange terms  become  invisible 
in the forward limit $\xi\to 0$.  
On the other hand, 
interplay between $\tilde x$ and $\xi$ 
dependences  of the component resulting from  
integrating the   $x\neq 0$ part of DDs
is quite nontrivial.  Its 
support in general covers the whole   $-1 \leq \tilde x \leq 1$ region
for all $\xi$ including the forward limit 
$\xi$ in which they convert into 
 the usual (forward) densities
$f^a(x)$, $f^{\bar a}(x)$. The latter 
are   
rather well known  from inclusive measurements. 
at small $t$.

\subsection{Polynomiality and analyticity}

In our derivation, DDs are the starting point, 
while SPDs are derived from them by integration.
However, even if one starts directly 
with SPDs,  the latter possess a property
which  can be incorporated only  within  the   
formalism   of double distributions. 
Namely,  the  
$\tilde x^N$ moment of $ H(\tilde x,\xi;t)$
{\it must be} an  $N$th order   polynomial of $\xi$. 
This restriction on the interplay between
$\tilde x$ and $\xi$ dependences of $ H(\tilde x,\xi;t)$ 
follows\,\cite{jirev} from  the simple fact that  the Lorentz  indices
$\mu_1 \ldots \mu_N$ of the  
 nonforward 
matrix element of a local operator $O^{ \mu_1 \ldots \mu_N}$
can be carried either by $P^{\mu_i}$ or by $r^{\mu_i}$.
As a result, 
\begin{eqnarray}  
 && \langle P-r/2 | \phi (0) (\stackrel{
\leftrightarrow}{
\partial^+} )^N
\phi(0)|P+r/2\rangle =
\sum_{k=0}^{N}{N \choose k} (P^+)^{N-k} (r^+)^{k}  A_{Nk} \,  
\nonumber \\[0.1cm]
 && =
(P^+)^{N}\sum_{k=0}^{N} {N \choose k}\xi^k    A_{Nk} \, ,
\label{poly} 
\end{eqnarray}
where ${N \choose k} \equiv N!/(N-k)!k!$ 
is the combinatorial coefficient.
Our  derivation (\ref{offfor}) of 
OFPDs from  DDs
automatically satisfies the 
polynomiality condition (\ref{poly}), since 
\begin{eqnarray}
\int_{-1}^1 H(\tilde x,\xi;t)\, \tilde x^N \, d \tilde x=
\sum_{k=0}^{N} \, \xi^k \, {N \choose k} \, 
\int_{-1}^1 dx \int_{-1 +|x|}
^{1-|x|} \tilde f(x,\alpha) \,  x^{N-k} \alpha^k
\, d\alpha \  .  \end{eqnarray}
Hence, the coefficients $A_{Nk}$ in (\ref{poly}) 
are given by  double moments
of DDs. This  means that modeling 
SPDs one cannot choose the coefficients $A_{Nk}$
arbitrarily: symmetry and support properties of DDs dictate  
a nontrivial interplay between 
$N$ and $k$ dependences of $A_{Nk}$'s.  
 After this observation,   
the use of DDs is a necessary  step
in  building consistent parametrizations of SPDs.

The formalism of  DDs  also allows one to easily establish 
some important  properties of skewed distributions.
Notice  that due to the cusp at the  upper corner
of the DD support rhombus, 
the length of the integration 
line nonanalytically depends on $\tilde x$ for $\tilde x = \pm \xi$. 
Hence, unless a double distribution 
identically  vanishes  in a finite region 
around  the upper corner  of the DD support rhombus,
the $\tilde x$-dependence of the relevant nonforward distribution 
{\it must be nonanalytic}  at the border points $\tilde x = \pm \xi$. 
Still,   the length of the 
integration line is a continuous function of $\tilde x$. 
As a result, 
if the  double distribution $f( x,\alpha;t) $ is not too
singular for small $x$, the skewed 
distribution $H(\tilde x,\xi;t)$
is continuous at the nonanalyticity  
points $\tilde x = \pm \xi$.   Because 
of the  $1/(\tilde x \pm \xi)$ 
factors 
contained in   hard amplitudes, this property  is 
crucial for  PQCD factorization 
in  DVCS and other
hard electroproduction processes.  

Note,  that there may be  also the exchange contributions
 $H_M(\tilde x,\xi;t)$. 
If  it comes from a $\delta (x) \varphi (\alpha)$ type term and 
$\varphi (\alpha)$ 
 vanishes at the end-points $\alpha =\pm 1$,  the   
$H_M(\tilde x,\xi;t)$ part of
SPD  vanishes at $\tilde x = \pm \xi$.
The total function 
$H(\tilde x,\xi;t)$ 
is  then continuous at the nonanalyticity points
$x = \pm \, \xi$.
In the $C$-even  case, the DDs should be odd in $x$,
hence the singular term involves $\delta^{\prime} (x) \varphi(\alpha)$
(or even higher odd derivatives of $\delta (x)$). 
One can  get a continuous SPD in this case only 
if  $\varphi ^ {\prime} (\alpha)$  vanishes at the end points.

\subsection{Nonforward parton distribution functions}

\begin{figure}[ht] 
 \begin{center} 
\mbox{ 
\epsfxsize=12cm
 \epsfysize=3.5cm
   \epsffile{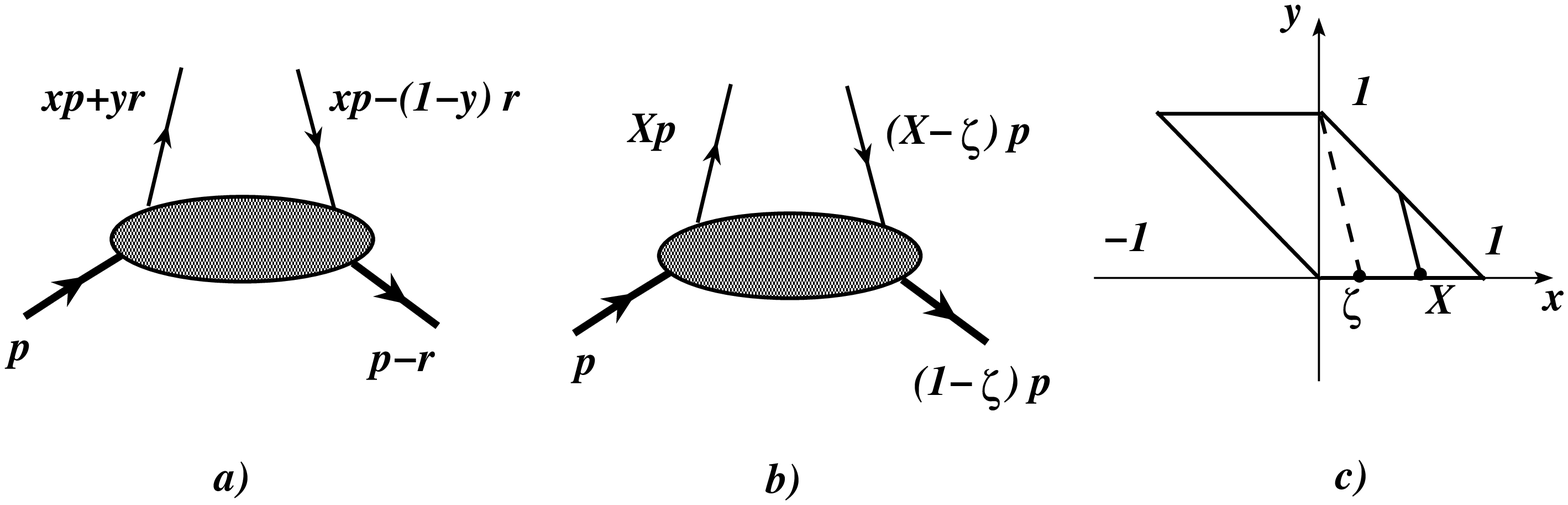}}
\end{center}
\caption{\label{fig:nfpd} Description 
of  nonforward matrix elements 
in terms of $a)$  double distributions 
$F(x,y;t)$ and  $b)$ nonforward parton
distributions ${\cal F}_{\zeta}^{i} (X;t)$. $c)$ Integration 
lines for the integrals relating NFPDs and DDs.}  
\end{figure}

In the 
NFPD formalism,\cite{npd} the skewedness parameter 
$\zeta\equiv r^+/p^+$ and the parton momentum $k^+$ 
are measured in units of the initial momentum 
$p^+$. 
Again, one can start with   double distributions
$F(x,y;t)$ writing the outgoing parton 
momentum as $k^+ =xp^+ + y r^+$
and that of the returning one as $xp^+ - (1-y)  r^+$
(see Fig. \ref{fig:nfpd}a).
The support area for $F(x,y;t)$ is specified by 
  $-1 \leq x \leq 1$, $0 \leq y \leq 1$, $0 \leq x+y
\leq 1$ (see Fig. \ref{fig:nfpd}c).  
   The relation between $ F_{a}(x,y;t=0)$
   and the usual quark and antiquark   densities
   is given by the ``reduction formulas'' 
\begin{equation} 
\int_0^{1-x} \,  F_{a}(x,y;t=0)|_{x>0} \, dy= 
 f_{a}(x) \  ;  \  
\int_{-x}^1 \, F_{ a}(x,y;t=0)|_{x< 0}\, dy= 
-  f_{\bar a}(-x) \,  \label{eq:redf} \, .
 \end{equation}
If we define the ``tilded'' DDs by 
$$ 
 \tilde F_{a}(x,y;t) = F_{a}(x,y;t)_{x>0}  \ 
 ;  \ 
\tilde F_{\bar a}(x,y;t) = -  F_{a}(-x,1-y;t)_{x<0}   \, ,
$$
then $x$ is always positive and 
 the reduction formulas 
\begin{equation} 
\int_0^{1-x} \, \tilde F_{a,\bar a}(x,y;t=0)|_{x \neq 0} \, dy= 
 f_{a,\bar a}(x) 
 \label{34} \end{equation}
have  the same form in both cases. The new antiquark
DDs $\tilde F_{\bar a}(x,y;t)$   
also ``live'' on the triangle
$ 0 \leq x,y, x +y \leq 1$.

The $\alpha \to - \alpha$ symmetry 
of the OFPD-oriented DDs $f(x,\alpha;t)$ 
corresponds to the symmetry of $\tilde F_{a,\bar a}(x,y;t)$ 
with respect to  the interchange 
$y \leftrightarrow 1-x-y$ (``Munich symmetry'') 
established in Ref. 27. 

Using $r^+ = \zeta p^+$ and introducing the total 
momentum fraction $X \equiv x+ \zeta y$,
we define the nonforward parton distributions
(see Fig. \ref{fig:nfpd}b)
 \begin{eqnarray} 
{\cal F}_{\zeta}^{i} (X;t) = \int_0^1 dx \int_0^{1-x} 
\, \delta (x+\zeta y -X) \, \tilde F_i(x,y;t) \, dy \  . 
\label{71}  \end{eqnarray}

The relation between NFPDs
and DDs $\tilde F_i(x,y)$ can be     illustrated
on the DD support  triangle  
 $0 \leq x,y,x+y \leq 1$ (see Fig. \ref{fig:nfpd}c).
To get ${\cal F}_{\zeta} (X;t)$,  one should 
integrate $\tilde F(x,y;t)$ over $y$ along a straight
line  $x=X- \zeta y$. The upper limit of the $y$-integration
is determined by intersection of this line
either with the line $x+y=1$ (this happens if 
$X > \zeta$)  or with the $y$-axis (if $X < \zeta$): 
 \begin{eqnarray} 
{\cal F}_{\zeta}^{i} (X;t) &=&  \theta (X>\zeta) 
\int_0^{(1-X)/(1-\zeta)} 
\,  \, \tilde F_i(X- \zeta y,y;t) \, dy \ \nonumber \\  
&+&   \theta (X<\zeta) 
\int_0^{X/ \zeta } 
\,  \, \tilde F_i(X- \zeta y,y;t) \, dy .  
\label{71m}  \end{eqnarray}

The returning parton  carries the fraction 
$X-\zeta$ of the initial hadron momentum 
$p^+$, which is positive in the
$X > \zeta$ region (where 
NFPDs can be treated as  modified 
parton densities) and negative in the $X < \zeta$
region (where  NFPDs resemble  distribution amplitudes). 

For reference purposes, we present 
the  relations between the NFPD and OFPD variables
\begin{eqnarray} 
\zeta = \frac{2\xi}{1+\xi} ,  \  \  \  
X = \frac{\tilde x + \xi}{1+\xi},  \  \  \ 
X' \equiv X-\zeta = \frac{\tilde x - \xi}{1+\xi} ,  \  \  \ 
\tilde x = \frac{X -\zeta/2}{1-\zeta/2} \  .
\end{eqnarray}

\subsection{Relation to form factors} 

 GPDs depend on the invariant momentum transfer
 $t$, hence, they may    also be treated 
 as generalizations of hadronic form factors.
  In particular, $H_a(\tilde x,\xi;t)$ 
 is related to the Dirac form factor of the proton,\cite{ji} 
\begin{equation} \sum_a e_a \int_{-1}^1   
H_a (x,\xi,t)
 \,  dx  =F_1 (t) \, , \label{eq:f1}
  \end{equation}
where $e_a$ is the electric charge of the
relevant quark.   Just like for form factors,
there are extra generalized parton distributions\,\cite{ji}
$ E_a (x,\xi,t) $ 
corresponding to helicity flip in the 
nonforward matrix element. 
They are related to $F_2(t)$ form factor,
\begin{equation}
\sum_a e_a \int_{-1}^1   
 E_a (x,\xi,t)
 \,  dx =F_2(t) \, .\label{eq:f2} 
\end{equation}
Note that though the shape of GPDs
changes  when $\xi$ is varied,
the integrals (\ref{eq:f1}), (\ref{eq:f2})  
do not depend on $\xi$. 
Since  the $E$ functions are accompanied 
by the $r_{\mu} $ factor in the  parametrization
of the nonforward matrix element
$$\langle p' | \bar \psi \gamma_{\mu} \psi | p \rangle 
\sim \bar u (p') \gamma_{\mu} u (p) ``H" +
\frac1{2M} \bar u (p') \sigma_{\mu \nu } r^{\nu} u (p) ``E" \  , 
$$ 
they are invisible
in deep inelastic scattering  
described 
by  exactly  forward $r=0$ Compton
amplitude.
However, the $t=0, \xi =0 $ limit of the $E$ distributions exists,
$$E_a (x, \xi =0 ; t=0) \equiv e_a(x)\  .$$
In particular, for $t=0$ the integral (\ref{eq:f2}) 
gives  the  anomalous magnetic moment. 
Moreover, the recent interest to DVCS and generalized parton distributions 
is largely due to observation made by X. Ji~\cite{ji} that the integral
\begin{equation}
J_q = \frac12  \sum_a  \int_{-1}^1   
 \left [ f_a (x) +  e_{ a}(x) \right ]  
 \,  x dx \, 
\end{equation}
is related to the total spin and  orbital momentum contribution 
of the quarks into 
the proton spin.  Ji proposed to use deeply virtual Compton scattering
to get access to the $e_{ a}(x)$ functions. 
These function can be also accessed  in hard 
meson electroproduction processes.\cite{cfs} 

The DVCS amplitude contains 
two  other  generalized parton distributions 
$\tilde  H (x,\xi;t)$ and $\tilde  E (x,\xi ; t)$.
They parametrize the nonforward matrix 
element of the axial operator\,\cite{ji}
\beq \langle p' | \bar \psi \gamma_{\mu} \gamma_5 \psi | p \rangle 
\sim \bar u (p') \gamma_{\mu} u (p) ``\tilde H" +
\frac{r_\mu}{2M} \bar u (p')  u (p) ``\tilde E" \  . 
\eeq  
In the forward limit,  the $\tilde  H$-distributions reduce to
the polarized parton densities
$ \tilde  H_a (x,\xi=0 ; t=0)= \Delta f_a (x)$.
After the $x$-integration, the \mbox{$\tilde  H_a (x,\xi=0 ; t=0)$}
distributions   
produce   the flavor components of the axial form 
factor $G_A(t)$.
Similarly, the functions $\tilde  E (x,\xi ; t)$
are related to the pseudoscalar form factor $G_P(t)$.
At small $t$, they are 
dominated by the pion pole term $1/(t-m_{\pi}^2)$. 

 \subsection{Gluon distributions}

In a similar way, we can write  parametrizations
defining double and skewed distributions for  gluonic operators 
 \begin{eqnarray} 
&& \langle P-r/2  \,    |     \,     
z_{\mu}  z_{\nu} G_{\mu \alpha}^a (-z/2) E_{ab}(-z/2,z/2;A) 
G_{ \alpha \nu}^b (z/2)\,   |     \, P+r/2  \rangle \,  
  \nonumber  \\  &&  
= (  P z)\bar u(p')  \hat z 
 u(p) \,   \int_{-1}^1 \, dx \int_{-1+|x|}^{1-|x|} 
 e^{-ix(Pz)-i\alpha (rz)/2}  \,  x f_g (x,\alpha;t) 
 \,
d\alpha  \, + \ldots   \nonumber \\ && 
= (  P z)\bar u(p')  \hat z 
 u(p) \,   \int_{-1}^1   
 e^{-i\tilde x(Pz)}  \,  H_g (\tilde x,\xi;t) \, d\tilde x
 +  \ldots  \ .
\label{gluon} 
 \end{eqnarray} 
 Note, that the gluon SPD $H_g(\tilde x,\xi;t)$ is constructed
  from $  x f_g (x,\alpha;t)$.
 Just like the singlet quark
 distribution, the gluon double distribution   
 $ f_g (x,\alpha;t)$ is an odd function of $x$. 

\section{Modeling GPDs} 

There are two approaches used to model GPDs.
One is based  on a direct calculation of parton distributions
in  specific  dynamical models, 
such as bag model,\cite{jims} chiral soliton model,\cite{bochum}
light-cone formalism,\cite{dikr} etc. 
Another  approach\,\cite{ddee,lech,sinevol} 
is a phenomenological construction
based on reduction formulas
relating GPDs to usual parton densities $f(x), \Delta f (x)$
and form factors $F_1(t), F_2(t), G_A(t), G_P(t)$.
The most convenient way to construct such models is to 
start with double distributions  
$f( x,\alpha;t)$.

\subsection{Modeling DDs} 

Let us consider  the limiting case $t=0$. 
Our   interpretation of
the  $x$-variable  as the  fraction of the 
  $P$  momentum and the reduction formula
  stating that the integral
  of $f_a(x,\alpha)$ over $\alpha$ gives 
  the usual parton density $f_a(x)$  
suggest that the   profile of  $f_a(x,\alpha)$   
in  the $x$-direction follows  the shape 
of $f_a(x)$. 
Thus, it make sense to write 
\begin{equation}
 f(x,\alpha) =  h(x,\alpha) \,   f(x)  \, ,  \label{65n}
 \end{equation}
 where the function $h(x,\alpha)$  
 normalized by 
\begin{equation}
 \int_{-1+|x|}^{1-|x|} h(x,\alpha) \, d\alpha \, =1 
 \end{equation}
characterizes the profile of $f(x,\alpha)$ in the $\alpha$-direction.
 The profile  function should be symmetric with respect to
$\alpha \to -\alpha$ because 
DDs $\tilde f(x,\alpha)$ 
are even in $\alpha$.
For a fixed $x$, the function  $  h(x,\alpha) $ 
describes how   the longitudinal momentum transfer $r^+$
is shared between the two partons. Hence,   
 the shape of $h(x,\alpha)$   
should  look like a symmetric meson 
distribution amplitude $\varphi (\alpha)$.  
Recalling that DDs have the support restricted by  
 $|\alpha| \leq 1- |x| $, to get a 
more complete  analogy
with DAs, 
it makes sense to rescale $\alpha$ as $\alpha = (1-|x|)  \beta$
introducing  the  variable $\beta$ with $x$-independent limits:
$-1 \leq \beta \leq 1$. 
 The simplest model is to assume 
that the $\beta$--profile  is  
 a  universal function  $g(\beta)$ for all $x$. 
Possible simple choices for  $g(\beta)$ may be  $\delta(\beta)$
(no spread in $\beta$-direction),  $\frac34(1-\beta^2)$
(characteristic shape for asymptotic limit 
of nonsinglet quark distribution amplitudes), 
 $\frac{15}{16}(1-\beta^2)^2$
(asymptotic shape of gluon distribution amplitudes), etc.
In the variables $x,\alpha $, this gives   
\begin{eqnarray} && 
h^{(\infty)} (x,\alpha) =  \delta(\alpha) \,   \ , \
h^{(1)}(x,\alpha) = \frac{3}{4} 
\frac{ (1- |x|)^2 - \alpha^2}{(1-|x|)^3}\,, \nonumber \\  
&& h^{(2)}(x,\alpha) = \frac{15}{16} 
\frac{[(1- |x|)^2 - \alpha^2]^2}{(1-|x|)^5} \,   \  . \label{mod123} 
 \end{eqnarray}
 These models can be treated as specific
 cases of the general profile function 
 \begin{equation}
 h^{(b)}(x,\alpha) = \frac{\Gamma (2b+2)}{2^{2b+1} \Gamma^2 (b+1)}
\frac{[(1- |x|)^2 - \alpha^2]^b}{(1-|x|)^{2b+1}} \,  , \label{modn} 
 \end{equation}
whose width is governed by the parameter $b$.

\subsection{Modeling SPDs}

Let us analyze the structure of SPDs 
obtained from these  simple models.
In particular,  taking 
$f^{(\infty )}(x,\alpha) = \delta (\alpha)
f(x)$  
gives the simplest  model   
  $H^{(\infty)}(\tilde x,\xi; t=0) = f(x)$ in which  OFPDs at $t=0$
   have no $\xi$-dependence. 
 For  DDs producing nonforward 
parton distributions ${\cal F}_{\zeta}(X;t=0)$,    this 
 is  equivalent to the $F^{(\infty)}(x,y) = \delta (y -\bar x/2)
f(x)$ model, which gives  
\begin{equation}
{\cal F}_{\zeta}^{(\infty)} (X) = \frac{\theta(X \geq \zeta/2)}{1-\zeta/2}
 f \left (\frac{X-\zeta/2}{1-\zeta/2} \right ) \, ,
 \label{model}
  \end{equation}
i.e.,  NFPDs for non-zero  $\zeta$ are obtained from 
the forward distribution $f(X)\equiv {\cal F}_{\zeta=0} (X)$  
 by   shift and rescaling.

\begin{figure}[htb]
\mbox{
   \epsfxsize=5.5cm
 \epsfysize=4cm
 \hspace{0cm}  
  \epsffile{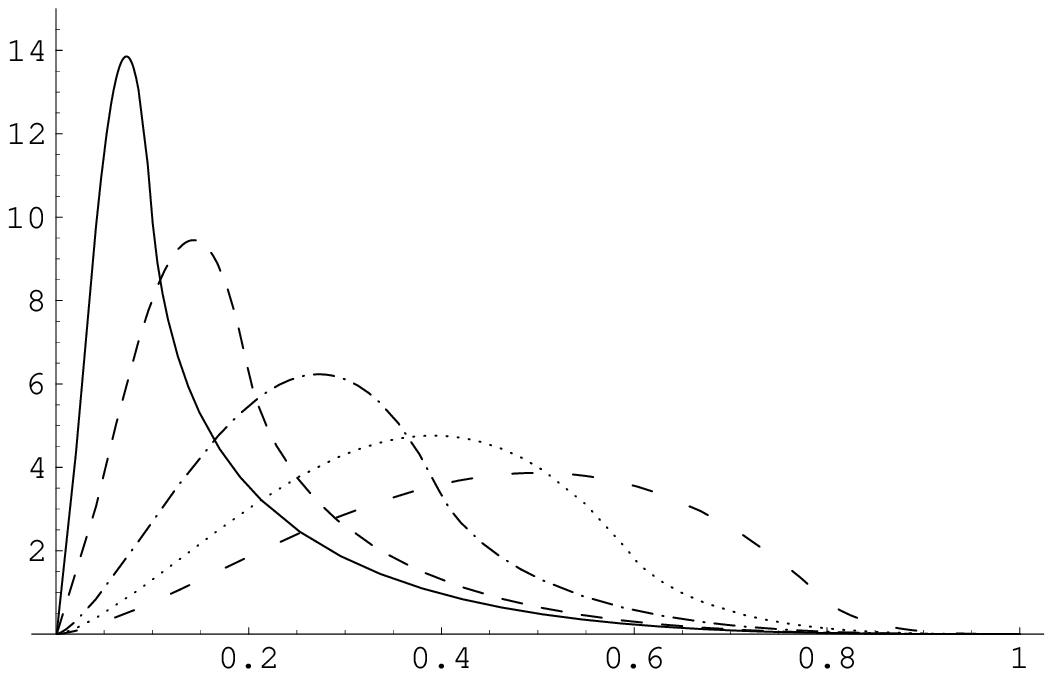}  
   \epsfxsize=5.5cm
 \epsfysize=4cm
 \hspace{0cm}  
  \epsffile{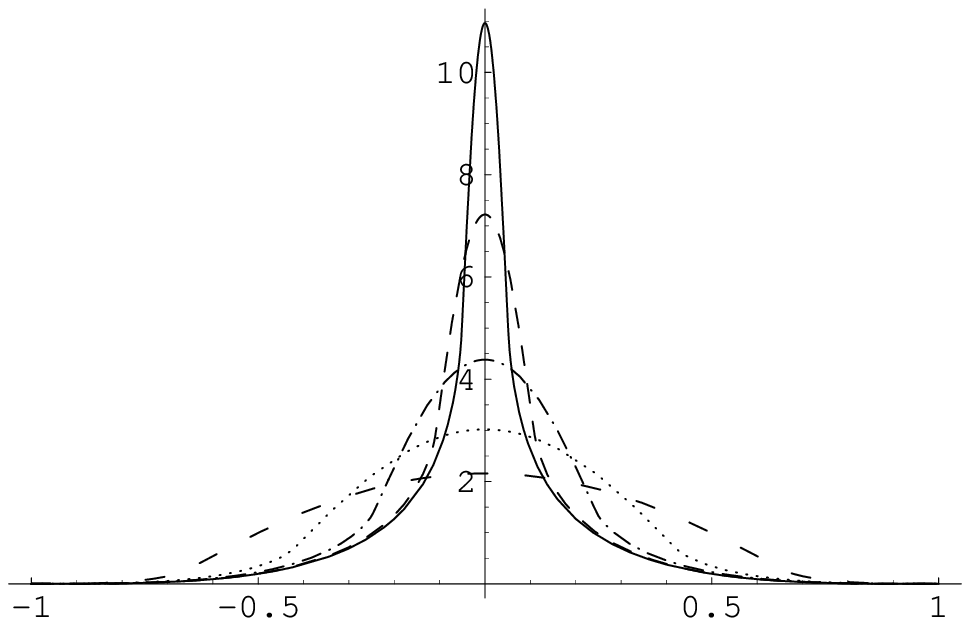}  }
{\caption{\label{q-at-diff-z}
Valence quark  distributions:  NFPDs $F^{q}_\zeta(X)$ (left)
and OFPDs $\tilde H ^1_V(\tilde x,\xi)$ (right) with $a= 0.5$ 
for several values
of $\zeta$, namely,  0.1, 0.2, 0.4, 0.6, 0.8 and corresponding values
of $\xi=\zeta / (2-\zeta)$. Lower curves
correspond to larger values of $\zeta$.}}
\end{figure}

In case of  the $b=1$ and $b=2$  models, simple  analytic results 
can be obtained only for some   explicit forms of  $f(x)$.
For the ``valence quark''-oriented ansatz $\tilde f^{(1)}(x,\alpha)$,
the following choice of a normalized distribution
\begin{equation}  f^{(1)}(x) = 
 \frac{\Gamma(5-a)}{6 \,  \Gamma(1-a)} \, 
  x^{-a} (1-x)^3  \label{74} \end{equation}
is   both  close 
to phenomenological   quark distributions
and   produces a simple expression
for the double distribution since the denominator
$(1-x)^3$ factor in Eq. (\ref{mod123}) is canceled.
As a result, the integral in Eq. (\ref{710})
is easily performed and   we get
\begin{eqnarray}
\tilde H^{1 }_V(\tilde x, \xi)|_{|\tilde x| \geq \xi}  &=& \frac1{\xi^3} 
 \left ( 1- \frac{a}{4} \right ) 
\left \{ \left[ (2-a) \xi (1- \tilde x) (x_+^{2-a} + x_-^{2-a}) \right. \right.
\nonumber \\  
&+& \left. \left.
(\xi^2 -\tilde x)(x_+^{2-a} - x_-^{2-a}) \right ] \, \theta (\tilde x)
+ ( \tilde x \to -\tilde x) \right \}  \label{outs} 
\end{eqnarray}
for  $|\tilde x |\geq \xi$ and 
\begin{equation}
\tilde H^{1 }_V (\tilde x, \xi)|_{|\tilde x| \leq \xi}  = \frac1{\xi^3} 
\left ( 1- \frac{a}{4} \right ) \left \{ x_+^{2-a}[(2-a) \xi (1- \tilde x) +
(\xi^2 -x)] + ( \tilde x \to -\tilde x) \right \} \label{middles}
\end{equation}
in the middle $ -\xi \leq \tilde x \leq \xi$ region.
We use here  the notation  $x_+=(\tilde x + \xi)/(1+\xi)$
  and 
$x_-=(\tilde x - \xi)/(1-\xi)$.
To extend these expressions onto negative values of 
$\xi$, one should  substitute $\xi$ by $|\xi|$.
One can check, however, that no odd powers of $|\xi|$ 
would appear in the $\tilde x^N$ 
moments of $\tilde H^{1V }(\tilde x, \xi)$.
Furthermore, these expressions are explicitly non-analytic 
for $x = \pm \xi$. 
This  is true even if $a$ is integer.
Discontinuity at $x = \pm \xi$, however, appears only 
in the second derivative
of $\tilde H^{1V }(\tilde x, \xi)$, 
i.e., the model curves for $\tilde H^{1V }(\tilde x, \xi)$
look very smooth (see Fig.~\ref{q-at-diff-z}, where
the  
 curves for NFPDs
are also shown).

For   $a=0$, the 
$x>\xi$ part of OFPD has the same $x$-dependence 
as its forward limit, differing from it by an overall $\xi$-dependent 
factor only,
\begin{equation}
\tilde H^{1V }(\tilde x, \xi)|_{a=0} = 
4 \, \frac{(1-|\tilde x|)^3}{(1-\xi^2)^2} 
\, \theta (|\tilde x| \geq \xi) \, 
+ 2\, \frac{\xi +2 -3 \tilde x^2/\xi}{(1+\xi)^2} \, 
 \theta (|\tilde x| \leq \xi)
\, .    \label{(1-x)^3}
\end{equation} 
The  $(1-|\tilde x|)^3$ behavior can be 
trivially continued into the $|\tilde x| < \xi$ 
region. However, the actual behavior
of $\tilde H^{1V }(\tilde x, \xi)|_{a=0}$ in this region 
is given by a different function.
In other words, $ \tilde H^{1V }(\tilde x, \xi)|_{a=0}$
can be represented as a sum of a function analytic at 
border points and a contribution  whose support 
is restricted by $|\tilde x| < \xi$.  
It should be emphasized that despite its DA-like 
appearance, this contribution 
should not be treated as an exchange-type term. 
It is generated by the regular $x \neq 0$ part of the DD,
and, unlike the $\varphi (\tilde x / \xi)/\xi$ functions
its  shape changes with $\xi$, 
the function  becoming very small  for small $\xi$.

\begin{figure}[htb]
\begin{center}\mbox{
   \epsfxsize=8cm
 \epsfysize=5cm
 \hspace{0cm}  
  \epsffile{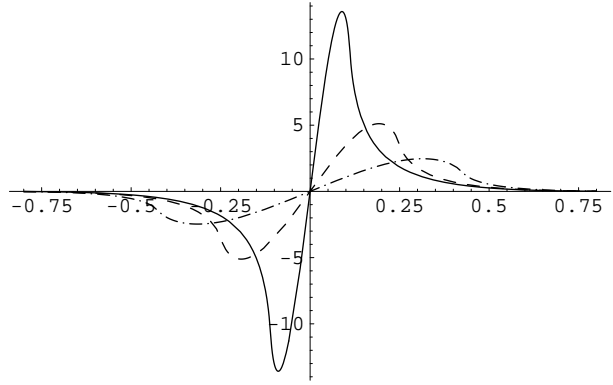}  }
  \end{center}
{\caption{\label{qs-at-diff-z}
Model for singlet quark distributions:  
OFPDs $ H ^1_S(\tilde x,\xi)$  for 
 values of $\xi$ corresponding to $\zeta$ equal to 
 0.2, 0.4, 0.6. Lower curves
correspond to larger values of $\zeta$.}}
\end{figure}

For  the singlet quark distribution, the  DDs
$\tilde f^S( x, \alpha)$ should be odd functions 
of $x$. Still, we can use  the model like (\ref{74}) for 
the $x>0$ part, but 
take $\tilde f^S( x, \alpha)|_{x \neq 0} 
= A \, f^{(1)}( |x|, \alpha)\, {\rm sign} (x)$. 
Note, that the integral (\ref{710})  producing $\tilde H^S(\tilde x, \xi)$ 
in the $|\tilde x| \leq \xi$ region
would diverge  for  $\alpha \to \tilde x /\xi$  
 if $a \geq  1$, which is the usual case 
 for standard parametrizations of singlet quark 
 distributions for sufficiently large $Q^2$. 
However, due to the antisymmetry of  $\tilde f^S( x, \alpha)$
with respect to   $x \to -x$ and its symmetry 
with respect to  $\alpha \to -\alpha$,
the singularity at 
$\alpha = \tilde x /\xi$ can be  integrated  
using the principal value 
prescription   which in this case 
produces the $x\to -x$ antisymmetric version 
of Eqs.~(\ref{outs}) and (\ref{middles}). For $a=0$, 
its middle part  reduces to 
\begin{equation}
\tilde H^{1S}(\tilde x, \xi)|_{|\tilde x| \leq \xi, a=0} = 
 2x\, \frac{3 \xi^2 -2 x^2 \xi - x^2}{\xi^3 (1+\xi)^2} 
  \,  .
\end{equation}
 The shape of singlet SPDs in this model is shown in Fig.
\ref{qs-at-diff-z}

 \subsection{Polyakov-Weiss terms}
 
 Note, that  the operator 
 $\bar \psi (-z/2) \hat z \psi (z/2)$  
  is proportional to $z$.
  Hence, 
  parametrizing its  matrix elements
  in terms of parton 
  distributions, it makes sense to use 
   the structures which are also linear
   in $z$, like \mbox{$\bar u (p') \hat z u(p)$,} 
   \mbox{$\bar u (p') (\hat z \hat r - \hat r \hat z) u(p)$} 
   for the  nucleon target, $(Pz)$ 
   for the pion target,
   etc. 
  However, there is a subtlety emphasized by Polyakov 
  and Weiss.\cite{PW} Namely, 
  using parametrization by DDs, 
 one treats $(Pz)$ and $(rz)$ as independent 
 variables. This means that in the pion case, e.g., 
 one should deal both with $(Pz)$ and $(rz)$ structures:
  \begin{eqnarray}
  \langle P-r/2 \, | \, \bar \psi (-z/2) \hat z \psi (z/2) 
 \, | \, P+r/2 \rangle \nonumber \\ = 
 2 (Pz) 
 \int_{-1}^1 \, dx \int_{-1+|x|}^{1-|x|} 
 e^{-ix(Pz)-i\alpha (rz)/2}  \,  f (x,\alpha;t) 
 \,
d\alpha  \,\nonumber \\ + (rz) \int_{-1}^1 e^{-i\alpha (rz)/2}
D(\alpha;t) \, d \alpha  \ ,
 \end{eqnarray} 
 where the Polyakov-Weiss (PW)\,\cite{PW}  term $D(\alpha;t)$
 accumulates the $r^{\mu_1} \ldots r^{\mu_n}$ 
 parts of the expansion for the matrix elements of the local
 operators 
 $\bar \psi (0) \gamma^{\mu_1} D^{\mu_2} 
 \ldots D^{\mu_n} \psi (0)$.
 There is no   sensitivity in the $D$-term contribution 
 to the value of the average momentum $P$ term: 
 the parton momenta 
 depend  only on $r$.
 Hence, the $D$-term is a particular 
 case of the exchange contributions.
 
 Switching to skewed parton distributions, one 
 deals with just one structure $(Pz)$, and one can 
 incorporate the PW term contribution\,\cite{PW} 
 $D(\tilde x/\xi;t) \theta (|\tilde x | <\xi)$
 into $H(\tilde x, \xi ;t)$.
 
  In the nucleon case,\cite{PW} the additional 
 structure is $(rz) \bar u (p')  u(p)$.
 As a result, the skewed distributions $H,E$ 
  have two components,
  one is obtained from the relevant DD  and another 
  comes from the $D$-term 
 \begin{eqnarray} H(\tilde x, \xi ;t) = H_{DD}(\tilde x, \xi ;t)
 +  D(\tilde x/\xi;t) \theta (|\tilde x | <\xi) \,,\\[0.1cm]
 E(\tilde x, \xi ;t) = E_{DD}(\tilde x, \xi ;t)
 -  D(\tilde x/\xi;t) \theta (|\tilde x | <\xi)\ .
 \end{eqnarray}
 Note, that the $D$-term 
 drops from the Ji's sum rule,\cite{ji,jirev} since  
 $H+E = H_{DD} + E_{DD}$.
 
 It should be noted that
 explicit calculations of skewed parton distributions
 performed within the  chiral soliton model\,\cite{bochum}  
  show that  the middle region behavior of SPDs
  strongly resembles that of  distribution amplitudes.
 
  \subsection{Inequalities}
 
In the case  when $\tilde x > \xi $, 
the  integration line producing  
$H (\tilde x,\xi; t)$ 
 (see Fig.~\ref{fig:spd}c)   is 
inside the space  between two vertical  lines 
giving   the usual parton densities  
 $f(x_-)$ and $f(x_+)$, with  $x_+=(\tilde x +\xi)/(1+\xi)$
and  $x_-=(\tilde x - \xi)/(1- \xi)$: 
 \begin{eqnarray}
H_a (\tilde x,\xi;t) = 
\theta(\xi \leq \tilde x \leq 1) 
 \int_{x_- }^{x_+ }
f_{a} (x, (\tilde x -x)/ \xi  ) \, \frac{dx}{\xi} d x + \ldots  \ \  .
\end{eqnarray} 
The combinations $x_-,x_+$ have a very simple 
interpretation: they measure 
the momentum of the initial or final parton 
in units of the momentum of the relevant 
hadron. 
Assuming a monotonic decrease of the double distribution
$f(x,\alpha)$ in the $x$-direction and a universal profile 
in the $\alpha$-direction, one may expect that 
 $H_a (\tilde x,\xi;t)$ is larger than  $f(x_+)$ but smaller than  $f(x_-)$.
Inequalities between forward and nonforward distributions
were discussed in Refs.~31, 19, 32. 
They are based on the application of the Cauchy-Schwartz
inequality 
\begin{eqnarray}&&
| \langle N[(1-\xi)P]; (\tilde x - \xi)P ; S | 
\, N[(1+\xi)P]; (\tilde x + \xi)P ; S \rangle|^2  
\nonumber \\ &&  \leq  
\sum_S \langle N[(1+\xi)P]; (\tilde x + \xi)P ;  S | 
\, N[(1+\xi)P]; (\tilde x + \xi)P ;  S \rangle \nonumber \\ 
&& \times 
\sum_{S'} \langle N[(1-\xi)P]; (\tilde x - \xi)P ; S' | 
\, N[(1-\xi)P ]; (\tilde x - \xi)P ;  S' \rangle  
\end{eqnarray}
to  the skewed parton 
distributions written generically
 as 
\begin{equation} 
H(\tilde x, \xi;t) =
\sum_S \langle N[(1-\xi)P]; (\tilde x - \xi)P ;  S | 
\, N[(1+\xi)P]; (\tilde x + \xi)P ;  S \rangle  \  , 
\end{equation}
where  $|  N[(1\pm \xi)P]; (\tilde x \pm \xi)P ;  S  \rangle$ 
describes 
 the probability amplitude that the
nucleon with momentum $(1\pm \xi)P$  converts
  into a parton  with momentum 
$(\tilde x \pm \xi)P $ and spectators $S$. 
The forward matrix elements give  the usual
parton densities  
\begin{equation}
\sum_S 
\langle  N[(1\pm \xi)P]; (\tilde x \pm \xi)P ;  S   \rangle \, = 
\frac1{1 \pm \xi} \,  f(x_\pm).
\end{equation}
As a  result, one obtains\,\cite{jirev,pisoter,ddee} for the quark 
distributions 
\begin{equation}
 H_a(\tilde x, \xi;t) \leq \sqrt{\frac{f_a(x_+)f_a(x_-) }{1-\xi^2}}  \, .
\end{equation} 
For the gluon distribution, one has 
\begin{equation}
 H_g(\tilde x, \xi;t) \leq \sqrt{x_+x_-{f_g(x_+)f_g(x_-) }}  \, .
\end{equation}

It is clear  that the whole consideration 
makes sense only if $\tilde x  >\xi$.
If $\tilde x  < \xi$,  the integration line 
  $x=\tilde x -\xi \alpha$ intersects the line 
  $x=0$, where the usual
parton densities
are infinite.
Furthermore,
when negative $x$ are involved,
the behavior of DDs along the line cannot be monotonic.  
 Another deficiency of the Cauchy-Schwartz-type inequalities is that
they do not give the lower bound for nonforward 
distributions though our  graphical interpretation
suggests that   $ H(\tilde x, \xi;t=0)$ is larger than $f( x_+)$
if the $x$-dependence of the double distribution 
$f(x,\alpha)$ along the lines $\alpha =k (1-|x|) $ is monotonic.

  \subsection{SPDs at small skewedness} 
 
 To study  the deviation of skewed distributions  
 from their forward counterparts for small $\xi$ (or $\zeta$), 
let us consider 
 the 
$x \geq \xi$ part of $H(x,\xi)$
(see   Eq.~(\ref{710}))  and use its expansion  in powers of $\xi$
\begin{eqnarray}
H(\tilde x;\xi)|_{\tilde x \geq \xi} 
 &=& f(\tilde x) +  \xi^2 \left [ 
\frac12 \int_{- (1- \tilde x)}^{(1- \tilde x)}
\frac{ \partial^{2}f (\tilde x,  \alpha  )}{\partial \tilde x^{2}}
 \,\alpha  ^{2} \, d  \alpha   \right.  \nonumber \\  
  &+& \left. (1- \tilde x)^2
\left. \left (\frac{ \partial f (\tilde x,  \alpha  )}
 {\partial \alpha } 
 -2 \frac{ \partial f (\tilde x,  \alpha  )}
 {\partial \tilde x  } \right ) \right |_{\alpha = 1 - \tilde x}
 \right ] +  \ldots \, .
\label{exxi2} 
\end{eqnarray} 
where  $f(\tilde x)$ is  the forward distribution.
For small $\xi$, the corrections 
are formally $O(\xi^2)$.
However,   if $f(x,\alpha)$  
has a singular 
behavior like $x^{-a}$, then 
$$\frac{ \partial^{2} f(\tilde x,  \alpha  )}{\partial \tilde x^{2}}
\sim \frac{a (1+a)}{\tilde x^2} f(\tilde x,  \alpha  )\ , $$
and the relative suppression of the first correction 
is $O(\xi^2/\tilde x^2)$.   Though the corrections 
are tiny for  $\tilde x \gg \xi $, in  the region  $\tilde x \sim
\xi$  they have    no parameter  smallness.
 It is easy to write explicitly all  the   terms 
 which are not suppressed in the $\tilde x \sim \xi \to 0$ limit
  \begin{equation}
H(\tilde x;\xi) = \sum_{k=0} 
\frac{\xi^{2k}}{(2k)!}   
 \int_{- 1}^{1}
\frac{ \partial^{2k} f (\tilde x,  \alpha  )}{\partial \tilde x^{2k}}
 \,\alpha  ^{2k} \, d  \alpha   \, + \ldots = 
 \int_{- 1}^{1} 
\tilde  f (\tilde x -\xi   \alpha, \alpha  )  \, d \alpha
  + \ldots \, 
  \, ,
 \end{equation}
 where the ellipses denote the terms vanishing in this  
 limit. This result can be directly obtained 
 from Eq. (\ref{710}) by noting that 
 for small $x$, we can neglect the $x$-dependence in  the 
 limits $\pm (1-|x|)$ of the $\alpha$-integration. 
 Furthermore, for small $x$ one  can also 
 neglect the $x$-dependence of the  profile
 function $h(x,\alpha)$ in Eq. (\ref{65n})
 and  take the model $\tilde f(x,\alpha) = \tilde f(x) \rho (\alpha)$
 with $\rho (\alpha)$ being a symmetric normalized weight 
 function on $-1 \leq \alpha \leq 1$. Hence, in the region where both
  $\tilde x$ and $\xi$ are small, we can approximate Eq. (\ref{710}) by 
 \begin{equation}
 H(\tilde x;\xi) = {\rm ``P"}    
 \int_{- 1}^{1} 
\tilde  f (\tilde x -\xi   \alpha  ) \rho (\alpha) \, d \alpha
  + \ldots \, , \label{smallxi} 
 \end{equation}
 i.e., the OFPD $H(\tilde x;\xi)$ is obtained 
 in this case by averaging  the  usual (forward) 
 parton density $f(x)$ over the region 
 $\tilde x - \xi \leq x \leq \tilde x+\xi$ 
 with the weight $\rho (\alpha)$. 
 The  principal value prescription 
 ``P'' is only necessary in the case of singular quark 
 singlet distributions which are odd in $x$.   
 In terms of NFPDs, the relation is 
  \begin{equation}
\tilde  {\cal F}_{\zeta} (X )  = {\rm ``P"}    
 \int_{- 1}^{1} 
\tilde   f (X- \zeta (1+   \alpha)/2  ) \rho (\alpha)
 \, d \alpha
 + \ldots \, , \label{smallzeta}
 \end{equation}
 i.e., the average is taken over the region 
 $X -\zeta \leq x \leq X$.
 
 In fact,  for small values of the skewedness parameters
 $\xi, \zeta$, one can use  Eqs. (\ref{smallxi}), (\ref{smallzeta}) 
 for all values of $\tilde x$ and $X$:
 if $\tilde x \gg \xi$, Eq. (\ref{smallxi}) gives the
 correct result 
 $\tilde H(\tilde x;\xi) = \tilde  f (\tilde x  ) + O(\xi^2)$.
 Hence, to get SPDs at small skewedness,
 one only needs to know the shape 
 of the normalized profile function $\rho (\alpha)$.

 The imaginary part of  hard exclusive 
 meson electroproduction amplitude is determined by
 the skewed distributions at 
 the border point $\tilde x = \xi$ (or $X=\zeta$). 
 For this reason, the magnitude 
 of  ${\cal F}_{\zeta} (\zeta)$ [or $H(\xi, \xi)$]
   and its relation to the  forward densities
 $f(x)$   has a practical interest. 
 This example also gives a  possibility to study  
 the sensitivity of the results to the choice of the
 profile function.  
 Assuming   the infinitely narrow weight 
 $\rho(\alpha) = \delta (\alpha)$,
 we have ${\cal F}_{\zeta} (X ) = f(X-\zeta/2) + \ldots $ 
 and $H(x,\xi) = f(x)$.
 Hence, both ${\cal F}_{\zeta} (\zeta)$ and  $H(\xi, \xi)$
 are given by $f(x_{Bj}/2)$ because  $\zeta = x_{Bj}$
 and $\xi =x_{Bj}/2 +\ldots$.  Since the argument
 of $f(x)$ is twice smaller than in deep inelastic scattering,
this results in an enhancement factor. In particular, if 
  $f(x) \sim x^{-a}$ for small  $x$, the ratio
  ${\cal R} (\zeta ) \equiv {\cal F}_{\zeta} (\zeta ) /f(\zeta )$ is 
  $2^a$. 
 The use of a wider profile  function $\rho (\alpha)$ produces further
 enhancement. For example, taking the normalized profile 
 function 
 \begin{equation}
 \rho_b (\alpha)\equiv  \frac{\Gamma (b+3/2) }{ \Gamma (1/2) 
 \Gamma (b+1) } (1- \alpha^2)^b = 
  \frac{\Gamma (2b+2)}{2^{2b+1} \Gamma^2 (b+1)}
 (1-\alpha^2)^b \label{rhon}
 \end{equation} 
 and
$f(x) \sim x^{-a}$ we get 
 \begin{equation}{\cal R}^{(b)} (\zeta) \equiv 
 \frac{{\cal F}_{\zeta}^{(b)}   (\zeta ) }{f(\zeta )}
  = \frac{\Gamma (2b+2)\Gamma (b-a +1)}  
  {\Gamma (2b-a+2)\Gamma (b +1)}
 \end{equation}
 which is larger than $2^a$ for any finite $b$ and $0< a <2$.
 The $2^a$ enhancement appears as the $n \to
  \infty$ limit of Eq.~(\ref{rhon}).
 For small integer $n$, Eq.~(\ref{rhon})  reduces
 to simple formulas obtained in Refs.~21,~22.  
 For $n=1$, we have 
 \begin{equation}
 \frac{{\cal F}_{\zeta}^{(b=1)}   (\zeta ) }{f(\zeta )}
  = \frac1{(1-a/2)(1-a/3)}  \, ,
  \label{rho1} 
 \end{equation}
 which gives the factor of 3 for the  enhancement if $a=1$.
 For $b=2$, the ratio (\ref{rhon}) becomes
 \begin{equation}
 \frac{{\cal F}_{\zeta}^{(b=2)}   (\zeta ) }{f(\zeta )}
  = \frac1{(1-a/3)(1-a/4)(1-a/5)}  \, ,
  \label{rho2} 
 \end{equation}
  producing a smaller enhancement factor $5/2$ for $a=1$. 
 Calculating the enhancement factors,
one should remember that the gluon SPD ${\cal F}_{\zeta}(X)$ 
reduces to $Xf_g(X)$ in the $\zeta =0$ limit. 
Hence, to get the enhancement factor corresponding to  the 
$ f_g(x) \sim x^{-\lambda}$  small-$x$ behavior of the forward 
gluon density, one should take $a= \lambda -1$ in Eq.~(\ref{rhon}). 
As a result, the $1/x$ behavior of the 
singlet quark distribution gives the factor of 3
for the ${\cal R}^{(1)} (\zeta)$ ratio, but the same shape of 
the gluon distribution 
results in no enhancement.

Due to evolution, the effective parameter $a$ 
characterizing the small-$x$ behavior 
of the forward distribution 
is an increasing function of $Q^2$.
Hence, for fixed $b$, the ${{\cal R}^{(b)} 
  (\zeta ) }$  
ratio  increases with $Q^2$.  
In general, the  profile of $ f (\tilde x,  \alpha  )$
 in the $\alpha$-direction
is also  affected by the PQCD  evolution. 
In particular, in Ref.~21 
 it was shown that
  if one takes the  ansatz  corresponding 
  to an extremely  asymmetric profile function $\rho (\alpha) 
  \sim  \delta (1+\alpha)$,  the shift 
  of the profile function to a more symmetric 
  shape  is clearly visible in the evolution 
  of the relevant SPD. 
In the next sections, we  will 
discuss the evolution of GPDs and study
 the interplay between 
 evolution of $x$ and  $\alpha$  profiles 
 of DDs. 

\section{Evolution equations} 

\subsection{Evolution kernels for double distributions}

The QCD perturbative expansion 
for  the matrix element  in Eq. (\ref{17})
generates at one loop level the terms 
proportional to $\ln z^2$.   In other words, the 
 limit $z^2 \to 0$ is singular and 
the distributions  $F(x,y;t)$, $f(x,\alpha;t)$, etc.,
 contain 
logarithmic ultraviolet divergences 
which require an additional 
$R$-operation characterized by some subtraction scale
$\mu$:  $F(x,y;t)\to F(x,y;t\,  |     \,  \mu)$.
The $\mu$-dependence of  $ F(x,y;t\,  |     \,  \mu)$ 
 is governed by the
evolution equation
\begin{equation} 
 \mu \frac{d}{d\mu}  \, F_a(x,y;t\,  |     \,  \mu) =
\int_0^1 \int_0^1 \, \sum_b \,  R^{ab}(x,y;\xi,\eta) \, 
  F_b(\xi,\eta;t\,  |     \,  \mu)\, \theta(\xi +\eta \leq 1) \,
d \xi \, d \eta \,  , 
 \label{41} \end{equation}
where $a,b = g,q$.
A similar set of equations, with the kernels  denoted by 
$\Delta R^{ab}(x,y;\xi,\eta)$  governs  the evolution of 
 the parton helicity sensitive 
distributions  $G^a(x,y;t\,  |     \,  \mu)$. 
Since the evolution kernels do not depend on $t$,
from now on we will drop the   $t$-variable from the arguments
of $F(x,y;t\,  |     \, \mu)$ in all cases when this dependence 
is inessential (likewise, the $\mu$-variable  
will be ignored in our notation when it is not important).

Since  integration over $y$ converts $F_a(x,y;t=0\,  |     \, \mu)$ 
into the parton distribution function $f_a(x\,  |     \, \mu)$,
whose evolution is described by the Dokshitzer-Gribov-Lipatov-Altarelli-Parisi
(DGLAP)$^{33-35}$ 
equations  
\begin{equation}
\mu \frac{d}{d \mu}  f_a(x\,  |     \,  \mu) =
\int_x^1 \frac{d\xi}{\xi} P_{ab}(x / \xi) 
f_b( \xi\,  |     \,  \mu) \, d \xi  \,  , 
\label{42} \end{equation}
the kernels $R^{ab}(x,y; \xi, \eta)$ must
satisfy the reduction relation
\begin{equation}
\int_0^ {1-x}  R^{ab}(x,y; \xi, \eta) \,  d y =
\frac{1}{\xi}\,   P^{ab}(x/\xi)\, . 
\label{43} \end{equation}
Alternatively, integration over $x$ converts  $F_a(x,y;t=0\,  |     \, \mu)$ 
into an object similar to a meson distribution amplitude (DA), so 
one may expect that
the result of integration of 
 $R^{ab}(x,y; \xi, \eta)$
over $x$ should be related 
to  the    kernels  governing the  evolution 
of distribution amplitudes  
[Efremov-Radyushkin-Brodsky-Lepage (ERBL)$^{5,6,37}$ 
evolution],  
e.g., in case of the $qq$ kernel 
\begin{equation}
\int_0^{1-y}   R^{qq}(x,y; \xi, \eta) d x = V^{qq}(y,\eta) \,  .  
\label{44} \end{equation}
These  reduction properties  of 
the $R^{qq}(x,y;\xi, \eta)$ kernel 
can be illustrated using its explicit form,\cite{compton} 
\begin{eqnarray}
&& 
R^{qq}(x,y;\xi, \eta;g) = 
\frac{\alpha_s}{\pi} C_F \frac1{\xi}
\biggl \{  \theta  (0 \leq x/\xi \leq 
{\rm min} \{ y/\eta, \bar y / \bar \eta \} ) \nonumber
 \\
 && 
  +  
\frac{\theta (0 \leq x/\xi \leq 1) x/\xi}{ (1-x/\xi)} 
\biggl  [ \frac1{\eta}\delta(x/\xi - y/\eta) + 
\frac1{\bar \eta} \delta(x/\xi - \bar y/ \bar \eta) \biggr ]
\nonumber \\   &&  
-\delta(1-x/\xi) \delta(y-\eta)
\biggl [ \frac12 + 2 
\int_0^1 \frac{z}{1-z} \, dz \biggr ] \biggr \}. \label{46}
\end{eqnarray} 
Here the last (formally divergent) term, as usual,
provides the regularization for the $1/(x-\xi)$  singularities 
present in  the kernel. This singularity 
can be also written as $1/(\eta -y)$  for the term containing 
$\delta(x/\xi - y/\eta)$ 
and as $1/(\bar \eta - \bar y)$ for the term with 
$\delta(x/\xi - \bar y/ \bar \eta)$.
Depending on the chosen form
of the singularity,  incorporating  
 the $1/(1-z)$ term into a plus-type  distribution,
one should treat $z$ as $x/\xi$, $y/\eta$ or $\bar y /\bar \eta$. 
 One can check that  integrating  $R^{qq}(x,y;\xi, \eta)$ 
over $y$ or $x$ gives 
the DGLAP splitting function    
\begin{eqnarray}
 P^{qq}(z;g) = \frac{\alpha_s}{\pi} 
C_F \left (\frac{1+z^2}{1-z} \right )_+ , 
\label{47} 
\end{eqnarray}
and  the 
Brodsky-Lepage   evolution kernel\,\cite{bl} 
\begin{eqnarray}
 V^{qq}(y, \eta; g)  &=& \frac{\alpha_s}{\pi} C_F 
\biggl \{  \left (\frac{y}{\eta} \right )
\left [1+ \frac1{ \eta -y} \right ]
\theta(y \leq \eta) 
\nonumber \\ &+& 
\left (\frac{\bar y}{\bar \eta} \right )
\left [1+ \frac1{y- \eta} \right ]
\theta(y \geq \eta) 
  \biggr \}_+  \,  . 
\label{48} \end{eqnarray}
Here, ``+'' denotes the standard ``plus''
regularization.\cite{ap} 

\subsection{Light-ray evolution kernels}

The interrelation between different 
types of evolution kernels 
follows from the fact 
that, in  the leading logarithm 
approximation, the  evolution equations
can be written for  the light-cone operators 
themselves,$^{37-40}$ 
without any reference
to particular matrix elements 
\begin{equation}
 \mu \frac{d}{d \mu} \, 
{\cal O}_a(0,z)    =
\int_0^1  \int_0^{1}  
\sum_{b} B^{ab}(u,v ) {\cal O}_b( uz, \bar vz) \,
\theta (u+v \leq 1) \, du \, d v  \,  , 
\label{49} \end{equation}
where $\bar v \equiv 1-v$ and $a,b =q,g$. 
For valence distributions, there is no mixing,
and their  
 evolution 
is governed by the $qq$-kernel  alone.
The kernels $B^{ab}(u,v )$ have the following  symmetry 
\beq
 B^{ab}(u,v ) = B^{ab}(v,u ) \ . 
  \eeq
  
For the  parton helicity averaged case, 
the kernels  $B^{ab}(v,u )$ 
were originally obtained in Refs.~37,~38. 
We present them in the form
given in Ref.~16, 
\begin{eqnarray}
B^{qq}(u,v ) &=& \frac{\alpha_s}{\pi} C_F \left 
(1 + \delta( u) [\bar v/v]_+  + 
\delta(v) [\bar u/ u]_+ - \frac1{2} \delta( u)\delta(v) \right ) \, ,
\\
   B^{GQ}(u,v ) &=& \frac{\alpha_s}{\pi} C_F \biggl 
(2 + \delta( u)\delta(v) \biggr  ) \, ,
\\ 
 B^{QG}(u,v ) &=& \frac{\alpha_s}{\pi} N_f \left 
(1 + 4uv -u -v \right ) \, , \\ 
  B^{gg}(u,v ) &=& \frac{\alpha_s}{\pi} N_c \biggl (
4(1 + 3uv -u -v) + \frac{\beta_0}{2 N_c} \,
\delta(u)\delta(v) \nonumber \\ 
&+& \left  \{ \delta(u) \biggl[ \frac{\bar v^2}{v} - \delta (v) \int_0^1
\frac{d \tilde v}{\tilde v} \biggr ] 
+ \{ u \leftrightarrow v \} \right \} 
 \biggr  )  \, .
\label{414} \end{eqnarray}
Here,   $\beta_0 = 11- \frac2{3} N_f$ 
is the lowest coefficient of the 
QCD $\beta$-function.
Evolution kernels for the parton helicity-sensitive
case are given in  Refs.~39,~40,  
\begin{eqnarray} 
&&
\Delta B^{qq}(u,v ) = B^{qq}(u,v )
\\
&&  \Delta B^{GQ}(u,v ) = \frac{\alpha_s}{\pi} \, C_F \biggl 
( \delta( u)\delta(v) - 2 \biggr  ) \, ,
\\ 
&&   \Delta B^{QG}(u,v ) = \frac{\alpha_s}{\pi} \, N_f \left 
( 1- u - v  \right ) \, , \\ 
&&  \Delta B^{gg}(u,v ) = B^{gg}(u,v ) - 
12 \,  \frac{\alpha_s}{\pi} \, N_c \, uv .
 \label{418} \end{eqnarray}

Inserting   the 
operator evolution equation (\ref{49})  between 
particular   hadronic states
and parametrizing the matrix elements by appropriate
distributions, 
one can get the relevant evolution   
kernels. In particular,  parametrizing nonforward matrix element 
 in terms of DDs, 
one  expresses  
$R^{ab}(x,y;\xi,\eta;g)$  in terms of   $B^{ab}(u,v )$ kernels,
e.g., for the $qq$-kernel 
one has 
\begin{equation}
 R^{qq}(x,y; \xi, \eta) = \frac1{\xi} 
B^{qq} ( y - \eta x/\xi, \bar y - \bar \eta x/\xi) \ . 
\label{410} \end{equation}
In a similar way, one can get the expression 
for the DGLAP kernel 
\begin{equation}
 P^{qq}(z) =  \int_0^{1-z} 
B^{qq} (1-v-z, v)\, dv 
\label{410dgl} \end{equation}
and for the Brodsky-Lepage kernel
\begin{equation}
 V^{qq}(y, \eta) = \frac{\theta (y \leq \eta)}{\eta} \int_0^{y} 
B^{qq} ( \bar v -(y-v)/\eta) ) dv  + 
\{ y \to \bar y, \eta \to \bar \eta \}  \  . 
\label{410bl} \end{equation}

\subsection{Evolution kernels for 
SPDs} 

The nonforward matrix elements 
can be also parametrized in terms 
of SPDs.  
In the case of nonforward parton distributions,
the evolution  equations have the form
\begin{equation}
 \mu \frac{d}{d\mu}  {\cal F}_{\zeta}^a(X;\mu) =
\int_0^1  \, \sum_b \,  W_{\zeta}^{ab}(X,Z) \, 
{\cal F}_{\zeta}^b( Z;\mu) \, d Z \,  .
\label{76} 
 \end{equation}
 Again, the new kernels $W_{\zeta}^{ab}(X,Z)$ 
 can be expressed in terms of the $B$-kernels, e.g.,
 \begin{equation}
W^{qq}_{\zeta}(X,Z) = \int_0^1  \int_0^1 B^{qq}(u,v) \, 
\delta(X- \bar u Z + v (Z- \zeta)) \,
\theta(u+v \leq 1)
\, du \, dv    \  .
\label{77} \end{equation}
As we discussed earlier, NFPDs ${\cal F}_{\zeta}^a(X)$ 
have  two
components corresponding to regions 
$X>\zeta$ and $X<\zeta$. 
For this reason, one can imagine four  different 
possibilities for the
 kernels 
$W^{qq}_{\zeta}(X,Z)$: 
\begin{itemize} 
\item both $X$ and $Z$
are larger than $\zeta$; 
\item  both $X$ and $Z$
are smaller  than $\zeta$; 
\item the original   fraction 
$Z$ is larger than $\zeta$, but the evolved 
fraction $X$ is smaller  than $\zeta$; 
\item  $Z<\zeta$ but $X>\zeta$.
\end{itemize} 
The last possibility, in fact, is excluded
by the delta function in Eq. (\ref{77}).
Since $X= (1-u-v)Z +v\zeta$, we always have 
$X<\zeta$ when $Z<\zeta$. 
In other words, if the initial 
fraction $Z$ is smaller than $\zeta$,
the evolved fraction $X$ is also smaller than $\zeta$:
the parton is trapped in the $Z<\zeta$ region.

{\it DGLAP region: $ Z>\zeta$,  $X>\zeta$. } 
Recall, that when  $X > \zeta$,  
the initial parton   momentum $Xp^+$ 
is larger than the momentum transfer $r = \zeta p$,
and we can treat the nonforward  distribution
 function ${\cal F}_{\zeta}  (X)$ 
as  a generalization of the 
usual  distribution function $f(X)$ for a 
somewhat skewed kinematics. 
Hence, we  can expect that evolution in the region 
$\zeta < X \leq 1$   ,  
 $\zeta < Z \leq 1$ is similar
to that generated by the  DGLAP 
equation.
In particular,  it has the basic property that
  the evolved fraction
$X$ is always smaller than the original
fraction $Z$. The relevant $qq$ kernel  is  
 given by 
\begin{equation} 
\displaystyle  
W_{\zeta }^{qq}(X,Z) |_{\zeta \leq X \leq Z \leq 1}
= \frac1{Z} \int_0^{\frac{1-X/Z}{1 -\zeta/Z}}    \, 
B_{qq} \biggl  ( [1- X/Z -v(1-\zeta/Z)] \, ,v \biggr ) dv \, . 
\label{93} \end{equation}
Changing  the integration variable to 
$w \equiv v(1- \zeta/Z)/(1 -X/Z) = v /(1-X'/Z')$,
we obtain  the  expression in which the arguments
of the $B$-kernels are treated in a more symmetric way
\begin{equation} 
\displaystyle  
W_{\zeta }^{qq}(X,Z) |_{\zeta \leq X \leq Z \leq 1}
= \frac{Z-X}{ZZ'} \int_0^{1}    \, 
B_{ab} \left   ( \bar w \, (1- X/Z)  \, , \, 
w\, (1- X'/Z')  \right ) dw \, , 
\label{94} \end{equation}
where $X' \equiv X - \zeta$ and $Z' \equiv Z - \zeta$
are the ``returning'' partners of the 
fractions $X,Z$. Moreover, since  $Z-X = Z' -X'$,  the kernel
$W_{\zeta }^{ab}(X,Z)$ is given by a function symmetric 
with respect to the interchange  of $X,Z$ with  $X',Z'$.  
This is also true for the $gg,qg$ and $gg$ kernels. 
Introducing the notation 
$P^{ab}_{\zeta}(X,Z)\equiv W_{\zeta }^{ab}(X,Z) 
|_{\zeta \leq X \leq Z \leq 1}$ we present 
below the explicit expressions for   the 
$P$-kernels,\cite{npd} 
\begin{eqnarray} 
&& \hspace{-1cm}
P^{QQ}_{\zeta}(X,Z) = \frac{\alpha_s}{\pi} \, 
C_F \! \left \{ \!  \frac1{Z-X} \left [ 1 + \frac{XX'}{ZZ'} \right ]\,  
-\delta(X-Z) \int_0^1 \, \frac{1+z^2}{1-z} \ dz \right \}    , \label{96} 
\\ &&\hspace{-1cm}
P^{Qg}_{\zeta}(X,Z) = \frac{\alpha_s}{\pi} \, N_f \, \frac{1}{ZZ'}
 \left \{ \left (1 - \frac{X}{Z}\right ) \left (1 - \frac{X'}{Z'}\right )
 + \frac{XX'}{ZZ'} \right \} \, , \label{97} \\ &&\hspace{-1cm}
P^{gQ}_{\zeta}(X,Z) = \frac{\alpha_s}{\pi} \, C_F \, 
\left \{ \left (1 - \frac{X}{Z}\right ) \left (1 - \frac{X'}{Z'}\right )
 + 1 \right \} \  , \label{98} \\ &&\hspace{-1cm}
P^{gg}_{\zeta}(X,Z) = \frac{\alpha_s}{\pi} \,  N_c \, 
\left \{ 2\left [ 1 + \frac{XX'}{ZZ'} \right ] \frac{Z-X}{ZZ'} 
+\frac1{Z-X} \left [ \left (\frac{X}{Z} \right )^2 +
 \left ( \frac{X'}{Z'} \right )^2 \right ] \right. \nonumber \\ && \left. 
\hspace{2cm} + \ 
\delta(X-Z) \left [ \frac{\beta_0}{2N_c} -
\int_0^1  \frac{du}{u} - 
\int_0^1  \frac{dv}{v}
 \, \right ]
\right \} \, .
\label{99} \end{eqnarray}
The formally divergent integrals over $u$ and $v$ 
provide here the usual ``plus''-type
regularization of the $1/(Z-X)$ singularities.
The prescription following 
from Eq. (\ref{94}) is that  combining the $1/(Z-X)$
and $\delta(Z-X)$ terms into
$[{\cal F}_{\zeta}(Z)-{\cal F}_{\zeta}(X)]/(Z-X)$  in 
 the convolution of 
$ P_{\zeta}(X,Z)$ with ${\cal F}_{\zeta}(Z)$ 
one should change $u \to 1-X/Z$ and $v \to 1-X'/Z'$.

In the $\zeta \to 0$ limit, the 
$P^{ab}_{\zeta}(X,Z)$ kernels are   directly related
to  the DGLAP kernels,
\begin{eqnarray} 
P^{QQ}_{\zeta}(X,Z) \rightarrow 
\frac1{Z} P_{QQ}(X/Z) \ , \ 
P^{Qg}_{\zeta}(X,Z)\rightarrow 
\frac1{Z^2} P_{Qg}(X/Z)  \ , \nonumber \\  
P^{gg}_{\zeta}(X,Z) \rightarrow 
\frac{X}{Z} P_{gQ}(X/Z)   
 \ , \ 
P^{gg}_{\zeta}(X,Z) \rightarrow 
\frac{X}{Z^2} P_{gg}(X/Z) \  . 
\end{eqnarray}
Here  one should take into account that
 the  nonforward  gluon distribution function 
${\cal F}_{\zeta}^g(X)$ reduces in the limit $\zeta=0$ 
to $Xf_g(X)$ rather than to $f_g(X)$. 

In the region $Z > \zeta$, the evolution is 
one-sided: the evolved fraction 
$X$ is smaller than the original $Z$. 
Furthermore, since if $Z \leq \zeta$ then also $X\leq Z$,
the   distributions in the $X > \zeta$ region are not affected by 
the distributions in the $X < \zeta$ regions.
Hence, just like in the DGLAP case,  information  about 
the initial 
distribution in the $Z > \zeta$ region
is sufficient for  calculating  its evolution in this region.
As we will see below,
this situation may  be contrasted with the evolution of distributions
in the $Z < \zeta$ regions: in that case one should know the 
nonforward  distribution functions in the whole domain $0 <Z <1$.

Qualitatively, the evolution in the $X, Z > \zeta$ region
proceeds just like in  the DGLAP evolution: the distributions shift to 
smaller and smaller values of $X$.
In the DGLAP case, the distributions approach the 
$\delta(x)$ form condensing at a single point $x=0$.
In the nonforward case, 
the whole region $Z < \zeta$ works like a ``black hole''
for the partons:  after  they end up  there, they will never
come back to the $X > \zeta$ region.  

{\it ERBL region: $ Z<\zeta$,  $X<\zeta$. } 
When  $\zeta =1$,  the initial momentum 
coincides with the momentum transfer and  
${\cal F}_{\zeta}(X)$ reduces to a distribution amplitude
whose  evolution is governed by the 
BL-type  kernels,
\begin{equation}
W_{\zeta =1}^{ab}(X,Z)= V^{ab}(X,Z). 
\label{81} \end{equation}
In fact, the nonforward kernels $W_{\zeta }^{ab}(X,Z)$
in the $ Z<\zeta$,  $X<\zeta$ region 
can be directly expressed in terms of the 
BL-type kernels   
even in the general $\zeta \neq 1,0$ case.
As explained earlier, if   $X$  is  in the
region $X \leq \zeta$, 
then the  function ${\cal F}_{\zeta}(X)$ 
can  be treated as a distribution amplitude
$\Psi_{\zeta}(Y)$ with $Y= X/  \zeta$. 
For this reason, when both $X$ and $Z$ are smaller than $\zeta$,
we would expect that the kernels 
$W_{\zeta}^{ab}(X,Z)$  must simply  reduce 
to the rescaled  BL-type  evolution kernels $V^{ab}(X/\zeta,Z/\zeta)$.
Indeed, the relation (\ref{77}) can be written as  
\begin{equation}
W^{qq}_{\zeta}(X,Z) = \frac1{\zeta} \int_0^1  \int_0^1 B_{qq}(u,v) \, 
\delta \left (\frac{X}{\zeta}- \bar u \frac{Z}
{\zeta}  - v ( 1-\frac{Z}{\zeta}) \right ) \,
\theta(u+v \leq 1)
\, du \, dv  \, . 
\label{85} \end{equation}
Comparing this expression with the representation 
for the $V^{qq}(X,Z)$ kernels, we conclude   that,  
in the region where $X/\zeta \leq 1$ and $Z/\zeta \leq 1$,
the kernel $W^{qq}_{\zeta}(X,Z)$ is given by 
\begin{equation}
W_{\zeta }^{qq}(X,Z) |_{0 \leq \{X,Z \} \leq \zeta}  =
\frac1{\zeta} \,  V^{qq}\left ({X}/{\zeta}, 
{Z}/{\zeta} \right )\, .
\label{86} \end{equation}

{\it Transition from $Z>\zeta$ to $X<\zeta$.} 
The BL-type kernels  also 
govern the evolution  corresponding to transitions
from a fraction $Z$ which is larger
than $\zeta$ to a fraction $X$ which is smaller 
than $\zeta$. Indeed,  
using   the $\delta$-function to calculate the integral
over $u$ in (\ref{77}), we get
\begin{equation}
W_{\zeta }^{qq}(X,Z) |_{X \leq \zeta \leq Z }
= \frac1{Z} \int_0^{X/\zeta}   \, 
B_{qq} \biggl  ( [1- X/Z -v(1-\zeta/Z)] \, ,v \biggr ) dv \, , 
\label{92} \end{equation}
which has the same analytic form (\ref{85}) 
as the expression for $W_{\zeta }^{qq}(X,Z) $
in the region $X \leq Z \leq \zeta$.

As already noted,    the evolution jump through  the critical
fraction $\zeta$ is irreversible: when the parton momentum 
degrades in the evolution 
process to values smaller than the momentum transfer
$\zeta p^+ \equiv r^+$,
further  evolution is like that for a distribution
amplitude: the momentum can decrease or increase
up to the $r^+$-value but cannot exceed this value.
 Inside the $Z < \zeta$
region, the ERBL evolution 
transforms the $\Psi_{\zeta}(Y)$ distribution amplitudes into 
their asymptotic forms like $Y \bar Y,  \,\, Y \bar Y (Y - \bar Y)$  
for the quarks
and $(Y \bar Y)^2, \,\, (Y \bar Y)^2(Y - \bar Y)$ for the gluons; 
a particular form is dictated by the symmetry properties
of the relevant operators.

\subsection{Asymptotic solutions of evolution 
equations} 
 
 At the leading logarithm (or one loop) 
 level,  the solution for  QCD evolution
equations  is known in the operator form.\cite{bb} Choosing specific
matrix elements one can convert the universal 
solution into four (at least)    evolution patterns: 
for usual parton densities ($\langle p | \ldots |p \rangle$
case), distribution amplitudes ($\langle 0 | \ldots |p \rangle$
case),   skewed and double  parton distributions 
($\langle p-r | \ldots |p \rangle$ case). 
In  the  simplest 
case of 
 flavor-nonsinglet (valence) functions, the
 multiplicatively renormalizable operators 
 were originally found in Ref. 5,  
\begin{equation}
{\cal O}_n^{NS}   = (z\partial_+)^n \, \bar \psi 
\lambda^a \hat z C_n^{3/2} 
( z\stackrel{\leftrightarrow}{D}/z\partial_+) \psi \, , 
 \label{100} \end{equation}
 where $C_n^{3/2}(\alpha)$ are   the Gegenbauer polynomials
and we use the symbolic  notation 
$ (z\stackrel{\leftrightarrow}{D}/z\partial_+)$
introduced in Ref.~5, 
$\stackrel{\leftrightarrow}{D}
\equiv \stackrel{\rightarrow}{D} -
\stackrel{\leftarrow}{D}$ , $\partial_+ \equiv 
 \stackrel{\rightarrow}{D} +
\stackrel{\leftarrow}{D}
= \stackrel{\rightarrow}{\partial} +
\stackrel{\leftarrow}{\partial}$.  
In contrast, the usual  operators $\bar \psi 
\lambda^a \hat z  
( z\stackrel{\leftrightarrow}{D})^n  \psi  $
mix under renormalization with the lower spin operators 
$ (z\partial_+)^{n-k} \bar \psi 
\lambda^a \hat z  
( z\stackrel{\leftrightarrow}{D})^k  \psi  $.
In Ref.  5, 
it was also noted 
that these  operators coincide with the  
free-field conformal tensors.

 Inside the nonforward matrix 
 element, one can substitute 
 $(z\stackrel{\leftrightarrow}{D}) \to (zk) = x(zP)$
 and $(z\partial_+) \to (zr) = \xi (zP)$.
 Thus, the multiplicative 
renormalizability
of ${\cal O}_n^{NS}$ operators means that 
 the   Gegenbauer moments 
\begin{equation}
{\cal C}_{n}^{NS}(\xi | \mu) = \xi^n \int_{-1}^1 C_n^{3/2} (z / \xi) \, 
H^{NS}( z, \xi| \mu) \, d z
\label{101} \end{equation}
of  the skewed parton distribution  
$H^{NS}(z, \xi;\mu)$ have a simple evolution,\cite{npd}
\begin{equation}
{\cal C}_{n}^{NS}(\xi| \mu)  = {\cal C}_{n}^{NS}(\xi, \mu_0)
\left [ \frac{ \ln \mu_0/\Lambda}{ \ln \mu/\Lambda} 
\right ]^{\gamma_n/ \beta_0} \,  ,
\label{102} \end{equation}
where  $\beta_0 = 11 -\frac23N_f$ is the 
lowest coefficient of the QCD $\beta$-function
and $\gamma_n$'s  are the nonsinglet anomalous dimensions,\cite{gw,gp}
\begin{equation}
\gamma_n= C_F \left [ \frac1{2} - 
\frac1{(n+1)(n+2)} +2 \sum_{j=2}^{n+1} \frac1{j}
\right ].
\label{103} \end{equation}
For $n=0$,  the Gegenbauer  moment  coincides with the 
ordinary one and, 
since  $\gamma_0 =0$, the area under the curve remains constant.
Other Gegenbauer moments decrease as  $\mu$ increases. 

Switching from SPDs to DDs, writing the SPD variable  $\tilde x$ 
in terms of DD variables $ \tilde x = x + \alpha \xi $ and using 
\begin{equation}
C_n^{3/2} (x/\xi + \alpha) = \sum_{l=0}^{n} 
\frac{\Gamma(n-l+3/2)}{ \Gamma (3/2) (n-l)!}
 \, (2 x/\xi)^{n-l} \, 
C_l^{3/2+n-l} (\alpha) \,  ,
\end{equation}
one can express the Gegenbauer moments ${\cal C}_{n}(\xi, \mu)$ 
in terms of the combined 
[$x$-ordinary $\otimes $ $\alpha$-Gegenbauer]  moments 
of the relevant DDs:
\begin{eqnarray} 
{\cal C}_{n}^{NS}(\xi | \mu) 
&=& \sum_{k=0}^{[n/2]}  \xi^{2k} \,  
\int_{-1}^1 d x  \int_{-1+|x|}^{1-|x|}
2^{n-2k} \, \frac{\Gamma(n-2k+3/2)}{ \Gamma (3/2) (n-2k)!}
\nonumber \\ 
 &\times&  x^{n-2k}
C_{2k}^{3/2+n-2k} (\alpha)\,  \,  f^{NS} (x, \alpha | \mu) 
\, d \alpha \, . 
\label{gedd}
\end{eqnarray}
Hence,  each $x^m C_l^{3/2+m}(\alpha)$ 
moment of $ f^{NS} (x, \alpha;\mu)$ is multiplicatively
renormalizable and its evolution
is governed by the anomalous dimension
$\gamma_{l+m}$.\cite{npd,ddee}
In  Eq. (\ref{gedd}), we took into account that
the DDs $ f  (x,
\alpha)$ are  always even in $\alpha$, which  gives an expansion
of the Gegenbauer moments in powers of $\xi^2$. 
In the nonsinglet case, the Gegenbauer
moments ${\cal C}_{n}(\xi, \mu)$ are nonzero  
for  even $n$  only. A similar representation can be written
for the Gegenbauer
moments of the 
singlet quark 
distributions. In the latter case,  the DD $\tilde f^{S}  (x, \alpha)$ 
is  odd in $x$, and only odd Gegenbauer
moments ${\cal C}_{n}^{S} (\xi, \mu)$ do not vanish.

Another simple case is the evolution 
of the gluon distributions in
pure gluodynamics. 
Then the  multiplicatively renormalizable operators with the same 
Lorentz spin 
$n+1$ as in Eq. (\ref{100}) 
 are 
\begin{equation}
{\cal O}^g_n = z^{\mu} z^{\nu} (z\partial_+)^{n-1} G_{\mu
\alpha} C_{n-1}^{5/2} 
( z\stackrel{\leftrightarrow}{D}/z\partial_+) \, G_{\alpha \nu} \, . 
\end{equation}
Due to the symmetry properties of  gluon DDs, only   Gegenbauer moments 
\begin{equation}
{\cal C}_{n}^{G}(\xi | \mu) = 
\xi^{n-1} \int_{-1}^1 C_{n-1}^{5/2} (z / \xi) \, 
 H^{G}( z, \xi| \mu) \, d z
\label{1010} \end{equation}
with odd $n$ do not vanish.
 The  Gegenbauer moment  can also be 
written in terms of DD:
\begin{eqnarray}
{\cal C}_{n}^G(\xi, \mu) 
&=& \sum_{k=0}^{[(n-1)/2]}  \xi^{2k} \,  
\int_{-1}^1 d x  \int_{-1+|x|}^{1-|x|}
2^{n-2k-1} \, \frac{\Gamma(n-2k+3/2)}{ \Gamma (5/2) (n-2k-1)!}
 \nonumber \\ &\times& x^{n-2k}
C_{2k}^{3/2+n-2k} (\alpha)\,  \, f^g (x, \alpha | \mu) 
\, d \alpha \, . 
\label{geddglu}
\end{eqnarray}
 Note, that two shifts, $n \to n-1$ and $3/2 \to 5/2$,  
 compensate each other. Again, each   
 combined $x^m C_l^{3/2+m}(\alpha)$ moment 
 of $ \tilde f^G (x, \alpha)$ is multiplicatively
renormalizable and its evolution
is governed by the anomalous dimension
$\gamma_{l+m}^{GG}$.\cite{npd,ddee}

Since the Gegenbauer polynomials 
$C_l^{3/2+m}(\alpha)$ are orthogonal with the 
weight  $(1- \alpha^2)^{m+1}$, evolution of 
the $x^m$-moments  of DDs 
in both cases is given\,\cite{ddee}  by the  formula
\begin{eqnarray} 
f_m (\alpha\,  |     \,  \mu)\!
&\equiv &\! \! \int_{-1}^1 x^m  f (x, \alpha\,  |     \,  \mu) \, dx 
\nonumber \\
\! &=& \!\!
(1- \alpha^2)^{m+1} \sum_{k=0}^{\infty} A_{ml}
C^{m+3/2}_l(\alpha ) \left [\log (\mu /\Lambda) \right]^
{- \gamma_{m+l}/\beta_0} \,  , 
\label{eq:fnnons}
 \end{eqnarray}
where the coefficients $A_{ml}$  
are proportional to  $x^m C_l^{3/2+m}(\alpha)$
moments of DDs.
A similar representation 
holds in the singlet case, with 
 $\left [\log (\mu /\Lambda) \right]^
{- \gamma_{m+l}/\beta_0}$ substituted by
a linear combination of  terms involving 
 $\left [\log (\mu /\Lambda) \right]^
{- \gamma_{m+l}^+/\beta_0}$ and  
$\left [\log (\mu /\Lambda) \right]^
{- \gamma_{m+l}^-/\beta_0}$ with  singlet 
anomalous dimensions $\gamma_{m+l}^{\pm}$ 
obtained by diagonalizing the coupled 
quark-gluon evolution equations.\cite{ddee} 

Let us consider first two simplified situations.
In the quark nonsinglet case, the evolution is governed 
 by 
$\gamma_{n+k}^{QQ}$ alone
\begin{equation} 
 f_n^{NS}(\alpha  |     \,  \mu) = 
(1-\alpha^2)^{n+1} \sum_{k=0}^{\infty} A^{nk}
C^{n+3/2}_k(\alpha) \left [\log (\mu /\Lambda) \right]^
{2\gamma_{n+k}^{QQ}/\beta_0} \ .
\label{eq:fnnon}
 \end{equation}
Since $\gamma_0^{QQ} =0$ while all the  anomalous dimensions
$\gamma_{N}^{QQ}$ with $N \geq 1$ are negative,
only  $f_0^{NS}(\alpha\,  |     \,  \mu)$ survives in the 
asymptotic limit $\mu \to \infty$
while all the moments  $f_n^{NS}(\alpha\,  |     \,  \mu)$
with $n \geq 1$ evolve to zero values.
Hence, in the formal $\mu \to \infty$
limit, we have  $$f^{NS}(x,\alpha \,  |     
\, \mu \to \infty)\sim \delta(x)(1-\alpha^2)\, , $$ $i.e.,$ 
in each of its variables the limiting function 
 $f^{NS}(x,\alpha\,  |     \,  \mu \to \infty)$  
acquires the characteristic asymptotic form dictated by
the nature of the variable:
$\delta(x)$ is specific for the distribution functions,\cite{gw,gp}
while  the $(1-\alpha^2)$-form  is  
the asymptotic shape   for the lowest-twist two-body 
distribution amplitudes.\cite{tmf,bl}
For the off-forward distribution of a valence quark $q$ this gives
$${H}^{val \, ;  q} (\tilde x,\xi \, | \, \mu \to \infty) =\frac34  
  (1-\tilde x^2/\xi^2) /\xi \, . $$

Another example   is  the evolution of the gluon distribution 
in pure gluodynamics which is governed by
$\gamma_{n+k}^{gg}$ with $\beta_0 = 11 N_c/3$.
Note that the lowest local operator in this case corresponds
to $n=1$. Furthermore, in pure gluodynamics,
$\gamma_{1}^{gg}$ vanishes while $\gamma_{N}^{gg}<0$
if $N \geq 1$.
This means that in the $\mu \to \infty$ limit we have
 $$x f^{g}(x,\alpha \,  |     
\, \mu \to \infty) =\frac{15}{16} \, \delta(x) (1-\alpha^2)^2$$ 
for the double distribution which results in 
$${H}^{g} (\tilde x, \xi\, | \,  \mu \to \infty) = \frac{15}{16} \, 
 (1-\tilde x^2/\xi^2)^2 /\xi$$
for the off-forward distribution.
In the formulas above, the total momentum carried 
by the gluons (in pure gluodynamics!) was normalized to unity.

In QCD,  we should take into account the effects
due to quark gluon mixing.
The diagonalization procedure gives 
two multiplicatively renormalizable
combinations
 \begin{equation} 
P_{nk}^{\pm} = f_{nk}^q + c_{nk}^{\pm}  f_{nk}^g 
\end{equation}
where  (omitting the $nk$ indices) 
\begin{equation} 
 c^{\pm} = \frac{ \gamma^{gg} - \gamma^{qq} \pm
\sqrt{(\gamma^{gg} - \gamma^{qq})^2 + 
4 \gamma^{gq} \gamma^{qg}} }{2 \gamma^{gq}} \, .
\end{equation}
Their evolution is governed by the anomalous dimensions
\begin{equation} 
 \gamma_{nk}^{\pm} = \frac12 \left ( \gamma^{gg} + \gamma^{qq} \pm
\sqrt{(\gamma^{gg} - \gamma^{qq})^2 + 
4 \gamma^{gq} \gamma^{qg}} \right ) \, .
\end{equation}
In particular, $\gamma^+_{10} =0$ and $\alpha^+_{10} =1$
which means that the sum $f_{10}^q + f_{10}^g$ does not evolve:
the total momentum carried by the partons is conserved.
Another multiplicatively renormalizable combination 
involving  $f_{10}^q$ and $ f_{10}^g$ is 
$$f_{10}^q - \frac{C_F}{4N_f} f_{10}^g.$$
It vanishes in the $\mu \to \infty$ limit, and we have
 \begin{equation}
f_{10}^q (\mu \to \infty) \to \frac{N_f}{4 C_F + N_f}   \  \ ;  \  \ 
f_{10}^g (\mu \to \infty) \to \frac{4C_F}{4 C_F + N_f} 
 \, .
\end{equation}
Since  all  the combinations $P_{nk}^{\pm}$ with $n+k \geq 2$
vanish in the $\mu \to \infty$ limit, we have
\begin{equation} 
 xf^g(x,\alpha \, | \, \mu \to \infty) \to  
\frac{15}{16} \, \frac{4C_F}{4 C_F + N_f} \, \delta(x)
(1-\alpha^2)^2\,,
 \end{equation}
 \begin{equation} 
 xf^q(x,\alpha \, | \, \mu \to \infty) \to  
 \frac{15}{16} \, \frac{N_f}{4 C_F + N_f} \, \delta(x)
 (1-\alpha^2)^2 \, , 
\end{equation}
or
\begin{equation} 
f^q(x,\alpha  \, | \, \mu \to \infty) \to  
- \frac{15}{16} \, \frac{N_f}{4 C_F + N_f} \, \delta'(x)
(1-\alpha^2)^2 \, .
 \end{equation}
In terms of off-forward distributions this is equivalent to
\begin{equation} 
 {H}^{g} (\tilde x, \xi \, | \, \mu \to \infty) \to 
\frac{15}{16}\, \frac{4C_F}{4 C_F + N_f} \,   
 (1-\tilde x^2/\xi^2)^2 /\xi\, ,
\end{equation}
\begin{equation} 
  {H}^{q} (\tilde x, \xi \, | \, \mu \to \infty) \to 
\frac{15}{4}\, \frac{N_f}{4 C_F + N_f} \, 
\tilde x (1-\tilde x^2/\xi^2)^2 /\xi^2\,  . 
\end{equation}
Note, that  in the $\mu \to \infty$
limit both the functions
${H}^{q,g} (\tilde x, \xi \, | \, \mu)$
and their derivatives $(\partial/\partial \tilde x) 
{H}^{q,g} (\tilde x, \xi \, | \, \mu)$  vanish at the 
border-point $\tilde x = \xi$.

 \subsection{Reconstructing   SPDs  from usual parton 
 densities}

The anomalous dimensions $\gamma_{n}$ increase with raising $n$,
and, hence,   the $m$th $x$-moment of 
$ \tilde f (x, \alpha;\mu )$ is  asymptotically 
dominated by the $\alpha$-profile 
$(1-\alpha^2)^{m+1}$. 
Such a correlation between $x$-  and 
$\alpha$-dependences
of $ \tilde f (x, \alpha;\mu )$  is 
not something exotic. 
 Take a DD which is  constant in its  support region.
 Then  its   $x^m$-moment behaves like 
 $(1- |\alpha|)^{m+1}$, i.e., the  width of the 
 $\alpha$ profile   decreases  with 
 increasing $n$.
 This result is easy to understand:
 due to the spectral condition $ |\alpha| \leq 1 -|x|$,
 the $x^m$ moments with larger $m$ are 
 dominated by regions which are narrower
 in the $\alpha$-direction.

 These   observations
suggests to try  a model in which the moments 
 $ \tilde f_m   (\alpha; \mu)$ have the asymptotic $(1-\alpha^2)^{m+1}$
 profile
even at non-asymptotic  $\mu$. This is equivalent
to assuming that all the combined moments
$x^m C_l^{3/2+m}(\alpha)$ with $l>0$ vanish. 
Note that this assumption 
is stable with respect to  PQCD evolution. 
Since integrating $\tilde f_m (\alpha\,  ;     \,  \mu) $
over $\alpha$  one should get the moments $ \tilde f_m (\mu) $ of the forward 
density $f(x ; \mu)$, the DD moments  $ \tilde f_m   (\alpha; \mu )$ 
in this model are given by 
   \begin{equation}
  \tilde f_m   (\alpha; \mu )   
   = \rho_{m+1} (\alpha)
   \tilde f_m (\mu)  
   \end{equation}
 where  $\rho_{m+1} (\alpha) $ is  the normalized profile 
function   (cf. Eq.(\ref{rhon})). In the explicit form: 
 \begin{equation}
  \int_{-1}^{1} x^m  
\tilde f(x,\alpha\,  ;     \,  \mu) \, dx \, 
 =
 \frac{\Gamma(m+5/2)}{\Gamma (1/2) \, (m+1)!}
  (1-\alpha^2)^{m+1}  
  \int_{-1}^{1} \tilde f (z;\mu) z^m dz \  .
\label{10810} 
\end{equation}
 In this relation,  all the dependence on  $\alpha$ 
 can be trivially shifted 
to the  left-hand side  of this equation, and we immediately see
that  
$\tilde f(x,\alpha\,  ;     \,  \mu)$  in this model 
is a function of $x/(1-\alpha^2)$,
  \begin{equation}
 \tilde f(x,\alpha\,  ;     \,  \mu)  = F(x/(1-\alpha^2);\, \mu ) \, 
 \theta (0 < x/(1-\alpha^2 ) < 1) \, .
  \end{equation} 
A direct   relation  
between $\tilde f(z,\mu)$ and $F(u;\mu)$ 
can be easily obtained using the basic fact that
integrating $\tilde f(x,\alpha\,  ;     \,  \mu)$ 
 over $\alpha$ one should get  
the forward density $\tilde f(z,\mu)$;
 e.g., for positive $z$ we have 
 \begin{equation}
 f(z) = z  \int_z^1 \frac{F(u)}{u^{3/2} \sqrt{u-z}} \, du  \, . 
 \end{equation} 
This relation  has the structure of the Abel equation.
Solving it for $F(u)$ we get 
\begin{equation}
F(u) = - \frac{u^{3/2}}{\pi} \int_u^1 
\frac{ \left [ f(z)/z \right ]^{\prime}}{\sqrt{z-u}}  \, dz \,  .
\label{inabel}
\end{equation}
Thus, in this model, knowing the forward  density $f(z)$ 
one can calculate the double distribution function 
$\tilde f(x,\alpha)  = F (x/(1-\alpha^2))$.

Note, however, that the model derived above  
violates  the DD support condition $|x|+|\alpha| \leq 1$:
the restriction $|x| \leq 1- \alpha^2$  
defines a larger area. Hence, the model
is only applicable in a situation
when the difference between two 
spectral conditions can be neglected. 
A practically important case is the shape of 
$ \tilde H(\tilde x, \xi )$  for small $\xi$. 
Indeed, calculating  $\tilde H(\tilde x, \xi ) $
 for small $\xi$ one integrates  the relevant
 DDs $\tilde f(\tilde x)$ over practically 
 vertical lines. 
 If $\tilde x$ is also small,  both the correct 
$|\alpha| \leq 1 - |x|$ and model $\alpha^2 \leq 1- |x|$ 
conditions 
can be substituted by $|\alpha| \leq 1$. 
Now,  if $ \tilde x \gg \xi$, a   slight deviation
of the integration line from  the vertical 
direction can be neglected and $\tilde H(\tilde x, \xi ) $
can be approximated by the forward limit $\tilde f(\tilde x)$.

 Specifying  the  ansatz for $f(z)$, one 
  can  get an explicit expression for the model DD by 
 calculating $F(u)$ from Eq.~(\ref{inabel}).  
However, in the  simplest  case when $f(x) =  A x^{-a}$ 
for small 
$x$, the result is evident without any  calculation:
the DD $f(x,\alpha)$ which is a function of the
ratio $x/(1-\alpha^2)$ and reduces to $A x^{-a}$
after an integration over $\alpha$ must be given by 
$$f(x,\alpha)  = \rho_{a} (\alpha) f(x)\, , $$   
where $\rho_{a} (\alpha)$ is  the normalized profile
function of Eq.~(\ref{rhon}),
\begin{equation}
f(x,\alpha)  =   
A \, \frac{\Gamma(a+5/2)}{\Gamma (1/2) \, \Gamma (a+2)}
  (1-\alpha^2)^{a}  \, x^{-a} \, .
  \end{equation} 
 This DD   is 
 a particular  case of the general
 factorized ansatz  $f(x,\alpha)  = \rho_{n} (\alpha) f(x)$
 considered in the previous section. 
 Its most nontrivial feature is the correlation
 $n=a$ between the   profile function parameter $n$ 
 and the power $a$ characterizing the small-$x$
 behavior of the forward distribution.  
 
    Knowing the DDs, the relevant 
    SPDs $\tilde H_a(\tilde x, \xi )$  can be obtained   
in the standard way from   $\tilde f_a(x,\alpha)$
for  quarks and from  $x \tilde f^q (x,\alpha)$ 
in the case of   gluons. 
In particular,  the SPD enhancement  factor ${\cal R} (\zeta)$ 
for small $\zeta$ 
in this  model is given by 
\begin{equation}
 \frac{{\cal F}_{\zeta}^{q}   (\zeta ) }{f^q(\zeta )}
  = \frac{\Gamma (2a+2)}  
  {\Gamma (a+2)\Gamma (a +1)}
 \end{equation}
for quarks and  by 
\begin{equation}
 \frac{{\cal F}_{\zeta}^{g}   (\zeta ) }{\zeta f^g (\zeta )}
  = \frac{\Gamma (2a+2)}  
  {\Gamma (a+3)\Gamma (a +1)}
 \end{equation}
for gluons.

The use of the asymptotic profiles 
for DD  moments $\tilde f_n (\alpha)$ 
is the basic assumption of the  model 
described above.  However, if one is interested 
in SPDs  
for small $\xi$, the impact of 
deviations  of  $\tilde f_n (\alpha)$ from 
the asymptotic profile is suppressed. 
Even if the higher harmonics are  present 
in $\tilde f_n (\alpha)$, i.e.,  if  
the $x^{n-2k} C_{2k}^{3/2+n-2k} (\alpha)$
 moments  of $\tilde f(x,\alpha)$ 
 are nonzero for $k \geq 1$ values, their contribution
 into the Gegenbauer moments  ${\cal C}_n (\xi, \mu) $ 
  is strongly suppressed by $\xi ^{2k}$ factors (see Eq.~(\ref{gedd})).
  Hence, for small $\xi$, 
  the shape of $\tilde H(\tilde x, \xi )$  
  for a wide variety of model $\alpha$-profiles 
   is very close to that based  on the asymptotic profile model. 
   
   Absence of   higher harmonics in $\tilde f_n (\alpha)$
   is equivalent to absence of the $\xi$-dependence 
   in the Gegenbauer moments ${\cal C}_n (\xi, \mu) $.
   The assumption that  the moments  ${\cal C}_n (\xi, \mu) $  
   do not depend on $\xi$ was the starting point 
   for the model of SPDs $\tilde H(\tilde x, \xi )$ 
   constructed in Ref.~43.  
   Though the formalism of DDs was not 
   used there,  
   both approaches lead to identical  results:
   the final result of Ref.~43 
   has the form 
   of a DD representation for $\tilde H(\tilde x, \xi )$.

\section{DVCS amplitude at leading twist and beyond}

\subsection{Twist--2 DVCS amplitude for the nucleon}

Using the parametrization for the matrix elements
of the quark operator, we can easily write  the leading twist 
 contribution\,\cite{ji2,npd} to the DVCS amplitude 
\begin{eqnarray} 
&&T^{\mu \nu}_{tw-2} (P,r,q') =    
\sum_a 
e_a^2  \int_{-1}^{1} \frac{d\tilde x}{\tilde x -\xi +i0} \biggl [
\biggl (-g^{\mu \nu} + \frac1{(P  q') } 
(P^{\mu}q'^{\nu} +P^{\nu}q'^{\mu}) \biggr ) \nonumber \\ 
 &&\times \biggl  \{  \bar u(p') \hat q'  u(p) H_a (\tilde x,\xi;t) 
  +  \frac1{4M} \bar u(p') 
(\hat q' \hat r - \hat r \hat q' )u(p)  E_a (\tilde x,\xi;t) \biggr \} 
\nonumber \\ && \hspace{2cm} +
 i \epsilon^{\mu \nu \alpha \beta} \frac{P_{\alpha} q'_{\beta}}{(Pq') }
\,  
\biggl \{ \bar u(p') 
\hat q' \gamma_5  u(p) 
\,  \tilde H_a (\tilde x,\xi;t) \nonumber \\ && \hspace{3cm}
+  \frac{(q'r)}{2M} \bar u(p') \gamma_5 
  u(p)  \tilde E_a (\tilde x,\xi;t)    \biggr \}  \biggr ] \ .
\label{122} \end{eqnarray} 
Note, that the functions $H_a (\tilde x,\xi;t), E_a (\tilde x,\xi;t)$
parametrizing the 
matrix element
of the ${\cal O}_{\sigma}$
operator are odd in $\tilde x$ while the
distributions $\tilde H_a (\tilde x,\xi;t)$ and  
$\tilde E_a (\tilde x,\xi;t)$ 
related to ${\cal O}_{5\sigma}$ term are even in $\tilde x$. 
Alternatively, one can use the combinations
$$\frac12 [1/(\tilde x -\xi +i0) \pm 1/(\tilde x +\xi +i0)]$$ 
in which the contributions of the $s$- and $u$-channel
handbag diagrams are 
explicitly separated. 

Thus, the skewed parton distributions
appear in the DVCS amplitude in an  integrated form.
Note that 
the relevant integrals  
\beq h_a(\xi;t) = \int_{-1}^{1} H_a (\tilde x,\xi;t)\frac{d\tilde x}
{\tilde x -\xi +i0} \eeq
have both real and imaginary parts.
The latter are given by the values of the relevant SPDs
at the border point $\tilde x = \xi$ 
\beq{\rm Im } \, h_a(\xi;t) = -\pi H_a (\xi, \xi;t) \ . \eeq
For a fixed $t$, 
the ``effective'' SPD $H_a (\xi, \xi;t)$ 
is  a function of the Bjorken 
ratio $x_{Bj} = 2\xi/(1+\xi)$,
just like DIS structure functions.
A linear combination  of the effective (or border-point) 
 SPDs $H_a (\xi, \xi;t)$
is directly accessible  through the measurement
of the single-spin asymmetry.\cite{ji2} 
Another function of $x_{Bj}$ 
corresponds to the  real part of  $h_a(\xi;t)$.
It is given by the principle value integral 
 \beq {\rm Re }\, h_a(\xi;t) = {\rm P}
  \int_{-1}^{1} H_a (\tilde x,\xi;t)\, \frac{d\tilde x}{\tilde x -\xi} \ . \eeq
The real part of the DVCS amplitude
can be accessed through the measurement
of the lepton charge asymmetry.\cite{ji2}

 In Eq. (\ref{122}), the final photon momentum 
$q'$ is used  as a natural light-cone 4--vector
specifying  the ``minus'' direction. 
In this form, the amplitude  $T^{\mu \nu} (P,r,q')$
exactly satisfies the transversality condition
$T^{\mu \nu} (P,r,q')q'_{\nu} =0 $ with respect
to the final photon momentum. 
However, the convolution $T^{\mu \nu} (P,r,q')q_{\mu} $
is proportional to $r^\mu - P^\mu (rq')/(Pq') \equiv \Delta^\mu_{\perp}$,
the transverse component of the momentum transfer $\Delta \equiv r_{\perp}$. 
Hence, the accuracy of the twist--2 approximation
is not sufficient to satisfy 
the transversality condition $T^{\mu \nu} (P,r,q')q_{\mu} =0 $.
Guichon and Vanderhaeghen~\cite{Guichon} (GV) proposed 
to add a non-leading   $O(\Delta)$ term  
producing  the   expression
\beq
 T^{\mu \nu}_{GV} = T^{\mu \nu}_{tw-2} - 
\frac{\Delta^{\mu}}{ (\Delta q)} (q_\lambda T^{\lambda \nu}_{tw-2})
\eeq 
which satisfies both $ q_{\mu} T^{\mu \nu}=0$
and $ q'_{\nu} T^{\mu \nu}=0$.

It is important to note that the use of 
the GV prescription changes the $\{\mu \leftrightarrow \nu \}$ 
symmetry structure of the DVCS amplitude.
In particular, the GV expression
constructed from the  $\{\mu \leftrightarrow \nu \}$ 
symmetric part of $T^{\mu \nu}_{tw-2}$ 
satisfies the transversality conditions
but it is not symmetric in $\mu \nu$ anymore.
It is easy to see that this is a common feature. Indeed, 
the transversality conditions written 
 in    symmetric  variables
$Q= (q+q')/2$, $r=q'-q$ and $P=(p+p')/2$  convert 
into two  relations 
\begin{eqnarray} 
             Q^{\mu}T_{\{\mu\nu\}}= 
            {r^{\mu}\over 2} T_{[\mu\nu]} \   \   ,  \   \
               Q^{\mu}T_{[\mu\nu]}= 
            {r^{\mu}\over 2} T_{\{\mu\nu\}}
            \label{trans} 
\end{eqnarray}
connecting the 
symmetric $T_{\{\mu\nu\}}\equiv (T_{\mu\nu}+ T_{\nu\mu})/2$
and  antisymmetric 
$T_{[\mu\nu]} \equiv (T_{\mu\nu}- T_{\nu\mu})/2$
parts of $T^{\mu\nu}$.
In the $r=0$ forward limit,   the two relations
decouple  to give  the DIS 
transversality conditions $q^{\mu} T_{\{\mu\nu\}}=0$,
$q^{\mu} T_{[\mu\nu]}=0$. 

The GV  prescription  was  supported  
by several groups$^{46-48,24}$  
 who derived this term
 in a regular way  as a   kinematical twist-3 contribution. 
Note, that   the twist--3 quark--gluon 
 operators $\bar q G q$ are
dynamically independent from the 
twist--2  $\bar q q$ ones. Hence,  to 
get  a gauge invariant extension of the
twist--2 contribution,  it is sufficient to 
retain only the part of the 
twist--3 SPD's induced  by 
the twist--2 distributions, i.e., the  Wandzura--Wilczek (WW)
type terms.

A very convenient way  to analyze the DVCS amplitude 
beyond   the leading--twist level is provided by   
the  approach of Balitsky 
and Braun~\cite{bb} (see, however, 
Refs. 48-51 where other versions
of the light cone analysis are used). 
We combine it with  
the formalism of 
 double distributions\,\cite{ddee,sssdd}
 which provides 
 a simple way of deriving relations
 between SPD's describing the 
 kinematical twist--3 effects and the
 basic twist--2 distributions.

 \subsection{Twist decomposition}
 
The nonlocal  operators ${\cal O}_\sigma$,
${\cal O}_{5 \sigma}$ in Eq. (\ref{handbag}) do not have a 
definite twist.   
The   twist--2 part of these operators  is defined 
by formally Taylor--expanding the nonlocal  operators 
in the relative coordinate
 $z$ and retaining only the totally symmetric traceless 
parts of the coefficients in the expansion:
\begin{eqnarray}&&
\left[ \bar\psi (X - z/2) \gamma_\sigma \psi (X + z/2) \right]^
{\rm twist-2}
\equiv 
\sum_{n = 0}^\infty \frac{1}{ n!} \;
z_{\mu_1} \ldots z_{\mu_n} \nonumber \\ && \hspace{2cm}
\bar\psi (X) \left[
\gamma_{\left\{ \sigma \right. } \stackrel{\leftrightarrow}{D}_{\mu_1}
\ldots \stackrel{\leftrightarrow}{D}_{\left. \mu_n \right\}} 
- \mbox{traces} \right] \psi (X) \ , 
\label{gamma_LT}
\end{eqnarray}
and similarly for the operator with  
$\gamma_\sigma \gamma_5$ (cf., e.g., Ref.~41). 
 The symmetrization  
can be carried out directly at the level of non-local 
operators.\cite{bb}
Indeed, the part of the nonlocal operator corresponding to totally symmetric
local tensor operators is projected out by 
\beq
\left[ \bar\psi (X - z/2) \gamma_\sigma \psi (X + z/2) \right]^{\rm sym}
\;\; = \;\; 
\frac{\partial}{\partial z_\sigma} \int_0^1 dt \; 
\bar\psi (X - t z/2) \hat{z} \psi (X + t z/2) \  . 
\label{string_sym}
\eeq
The subtraction of traces in  the local operators implies that 
the twist-2 string operator contracted 
with $z_\sigma$ should satisfy the d'Alembert 
equation with respect to $z$,
\beq
\Box_z  \left[ \bar\psi (X - t z/2) \hat{z} \psi (X + t z/2) 
\right]^{\rm twist-2}
= 0.
\label{harmonic_O} \eeq

In the center-of-mass  $X$ and relative $z$
coordinates, the transversality conditions (\ref{trans})
 are 
\begin{eqnarray}
\frac{\partial}{\partial z_\mu} \Pi_{\{\mu\nu\}} (z|X) 
= \frac{1}{2} \frac{\partial}{\partial X_\mu} \Pi_{[\mu\nu ]} (z|X)
 \  \  ,  \\ 
\frac{\partial}{\partial z_\mu} \Pi_{[\mu\nu ]} (z|X)
= \frac{1}{2} \frac{\partial}{\partial X_\mu} 
\Pi_{\{\mu\nu\}} (z|X) \  . 
\label{transversality_antisym}
\end{eqnarray} 
  Consider the part of the current product  given  by 
Eq. (\ref{handbag})  with the nonlocal  operators replaced by 
their twist-2 parts. 
From  Eq. (\ref{harmonic_O})  and 
$$\frac{\partial}{\partial z_\rho} \frac{ z_\rho}{2\pi^2 z^4} 
 = -i \delta^{(4)} (z)$$  
it  follows  that  
\beq\frac{\partial}{\partial z_\mu} \, \Pi_{\{\mu\nu\}}^{\rm twist-2}=  0 \ 
\  \  , \ \ \    
\frac{\partial}{\partial z_\mu} \,\Pi_{[\mu\nu ]}^{\rm twist-2} 
 = 0 \ .\eeq 
 Since forward matrix elements  are zero 
for  all total derivative operators,  
 this  guarantees the 
transversality of the  twist--2 contribution
in  the case of deep inelastic scattering.     
In the  non-forward case,  we have 
$$({\partial}/{\partial X_\mu}) \Pi_{\{\mu\nu\}}^{\rm twist-2}
 \neq  0, \ 
({\partial}/{\partial X_\mu}  \Pi_{[\mu\nu ]}^{\rm twist-2}
 \neq  0\ , 
$$ and (\ref{transversality_antisym})  is violated. 
 The non-transverse terms 
in the twist--2 contribution can   only be 
compensated by contributions from 
operators of higher twist.
In fact, the necessary 
operators 
are contained in the   part of the 
string operator which was dropped in taking the twist--2 part.  
Incorporating   QCD equations of motion, it is possible 
 to show\,\cite{bb} that the twist$>2$  part  
 involves  the total derivatives  
 of nonlocal operators 
\begin{eqnarray} &&
\bar\psi (-z/2) \gamma_\alpha \psi (z/2)
- \left[ \bar\psi (-z/2) \gamma_\alpha \psi (z/2) \right]^{\rm sym} 
\nonumber \\ &&
\hspace{1cm}  = \frac{i}{2} \epsilon_{\alpha\xi\rho\kappa} z_\xi
\frac{\partial}{\partial X_\rho} \int_0^1 dt \, t \;
\bar\psi (-tz/2) \gamma_\kappa \gamma_5  \psi (tz/2) +\ldots  \   \   .
\label{rest_total_derivative} 
\end{eqnarray}
The ellipses stand for 
quark--gluon operators  (we do not write them
explicitly since  they  are not needed to  restore transversality of the 
twist--2 contribution, but, in principle they can be kept).  
The   relation  for the operator containing 
$\gamma_\alpha\gamma_5$ is obtained by changing  
$\gamma_\alpha \to \gamma_\alpha \gamma_5$,
$\gamma_\kappa \gamma_5 \to \gamma_\kappa $.  

Note, that the operators 
appearing under the total derivative 
on the right hand side  of Eq. (\ref{rest_total_derivative}) 
and its $\gamma_\alpha\gamma_5$ analog are  still the  
 full string operators  with 
  no definite twist. Hence, one  can   decompose them 
into a symmetric ({\it i.e.}, twist--2) part and total derivatives,
and so on; thus expressing  the original string operator 
as the sum of its symmetric part and an infinite series of  
arbitrary order total derivatives  
of symmetric operators. 
This series can be summed up in a closed 
form.\cite{Belitsky:2000vx,Kivel:2000rb,rw}  
 Up to operators whose matrix elements give $O(t)$ contributions 
  to the Compton amplitude, the result is~\cite{rw} 
\begin{eqnarray} &&
\bar\psi (-z/2) \gamma_\sigma \psi (z/2) 
= \int_0^1 dv \; 
\left\{  \cos \left[ \frac{i \bar v}{2} 
\; \left (z  \frac{\partial}{\partial X} \right) \right ] 
\frac{\partial}{\partial z_\sigma} \right. \nonumber \\  
 && \left. \hspace{0.5cm} +  \frac{i v}{2}
\sin \left[ \frac{i \bar v}{2} 
\; \left (z  \frac{\partial}{\partial X} \right) \right ]
\frac{\partial}{\partial X_\sigma}
\right\} \bar\psi (-vz/2) \hat z \psi (vz/2) 
\nonumber \\
&& \hspace{0.5cm} + \; 
\frac{i}{2} \epsilon_{\sigma\alpha\beta\gamma} z_\alpha 
\frac{\partial}{\partial X_\beta} \frac{\partial}{\partial z_\gamma}
\int_0^1 dv  \nonumber \\ &&  \hspace{0.5cm} \times
 \int_v^1 du 
\cos \left[ \frac{i \bar u}{2} 
\; \left (z  \frac{\partial}{\partial X} \right) \right ] 
\; \bar\psi (-vz/2) \hat z \gamma_5 \psi (vz/2) 
 + \ldots \  
\label{string_deconstructed_scalar}
\end{eqnarray} 
(see also Refs.~48, 49). 
An analogous   formula applies to the operators with
$\gamma_\sigma \rightarrow \gamma_\sigma\gamma_5$;  one 
should just replace $\hat z \rightarrow \hat z \gamma_5 , \; \hat z \gamma_5 
\rightarrow \hat z$. 

\subsection{Parametrization  of nonforward matrix elements}
\label{subsec_spectral}

 {\it Double distributions.} To get 
the amplitude for deeply virtual Compton scattering off a hadronic 
target we need  parametrizations of the hadronic 
matrix elements of the uncontracted 
twist--2 string operators ${\cal O}_{\sigma},{\cal O}_{5\sigma}$
appearing 
in  Eq. (\ref{handbag}). 
We will derive them from Eq. (\ref{string_deconstructed_scalar}).   
For simplicity, we  consider  here one quark flavor and  the  pion target,
which has zero spin and practically vanishing mass. 
In this  case, the matrix element of 
the contracted {\it axial}  operator 
$z^\sigma {\cal O}_{5\sigma}(z\,|\,0)$ 
(parametrized in the forward limit by the polarized parton density) 
is identically 
zero.  Thus we need only the 
parametrization for the matrix element 
$\langle P - r/2 \,|\, {\cal O}(z\,|\,0)\, |\, P + r/2 \rangle$
of the contracted {\it vector}
   operator ${\cal O}(z\,|\,0) 
   \equiv z^\sigma {\cal O}_{\sigma}(z\,|\,0)$. 
With respect to $z$,  it can be regarded as a 
function of three invariants $(Pz), (rz)$ and $z^2$. 
For dimensional reasons, 
the dependence on $z^2$ is through 
the combinations $t z^2$ and $P^2 z^2$ only. 
We are going to  drop $O(t)$ and $O(P^2)$ terms 
 in the Compton amplitude, so we may 
  ignore the dependence on $z^2$ and 
 treat this matrix element as a  function
of just two variables $(Pz)$ and $(rz)$.
Incorporating the spectral properties
of nonforward matrix elements,\cite{sssdd}
we  write   the plane wave expansion in the form 
\begin{eqnarray} 
\langle P - r/2 \,|\, {\cal O}(z\,|\,0)\, | \, P + r/2 \rangle 
 \! &=&\!   2(Pz)  \int_{-1}^1 d\tilde x  
\int_{-1 + |\tilde x|}^{1-|\tilde x|} e^{-i (kz)} 
 f(\tilde x,\alpha) \, d \alpha  \nonumber \\
\! &+&\!  (rz) \, \int_{-1}^1 e^{-i \alpha (rz)/2}\, 
D (\alpha) \,  d \alpha    \  , 
\label{para1}  
\end{eqnarray} 
where 
$k =  \tilde x P +  \alpha r /2$, 
$f(\tilde x,\alpha)$ is the  double distribution (DD) and  
 $D (\alpha)$ is the Polyakov-Weiss ({\rm PW}) distribution 
 amplitude\,\cite{PW}  
absorbing the $(Pz)$-independent terms.    
From this parametrization,   we can obtain  matrix elements
of   original uncontracted   operators, 
 (\ref{string_deconstructed_scalar}),    including  the kinematical 
twist--3 contributions. 
We consider first the part coming from the double distribution 
term in Eq. (\ref{para1}); the contributions from the PW--term will 
be included separately. 
In matrix elements, the total derivative turns into the momentum
transfer, $
i{\partial}/{\partial X_{\sigma}} \rightarrow 
r_{\sigma}=  2 \xi P_{\sigma} + \Delta_{\sigma}
$.
Similarly, we  write $k=(\tilde x+\xi \alpha)P + {\alpha} \Delta / 2$.
This gives   
\begin{eqnarray} 
 \frac{1}{2} \, 
\langle P - r/2\, |\, {\cal O}_\sigma (z\,|\,0)\, | \, P +r/2  \rangle 
=  \int_{-1}^1 d \tilde x \int_{-1 + |\tilde x|}^{1-|\tilde x|} 
d \alpha \, f(\tilde x,\alpha)  \nonumber \\  \times \biggl \{P_{\sigma}
e^{-i (\tilde x+ \xi \alpha )(Pz) -i\alpha (\Delta z)/2 }  
+ \frac12\, \bigl [\Delta_{\sigma} (Pz) - P_{\sigma}(\Delta z)\bigr ]
\nonumber \\  \times 
\int_0^1 dv \, v\, e^{-i v (\tilde x + \xi \alpha )(Pz)-iv\alpha (\Delta z)/2 } 
 \, 
  \bigl [  \sin  (\bar v (rz)/2) 
- i  \alpha  \cos (\bar v   (rz)/2)  \bigr ] \biggr \} \ . 
\label{para5} 
\end{eqnarray}
\par

 {\it Skewed distributions.} Expanding $\exp[-i\alpha (\Delta z)/2]=
 1-i\alpha (\Delta z)/2 +\ldots$ and keeping only terms
 up to  those linear  in  the transverse 
 momentum
  $\Delta$ we get an 
 expression\footnote{Because of  this truncation, the  
  $\Delta_\mu \Delta_\nu$ terms in the expression
  for the amplitude $T_{\mu\nu}$ will be  lost.
  If needed, they can be kept;  see 
  the discussion after Eq. (\ref{xider}).}  in which 
the spectral parameter $\tilde x$ appears in the exponential factors 
 only in the combination 
$
x  \equiv  \tilde x + \xi \alpha $. 
Thus,  we can introduce two skewed parton distributions,
\begin{equation}
\left.
\begin{array}{r}
H(x, \xi) \\[1.5ex]
A (x, \xi)
\end{array}
\right\}
\;\; \equiv \;\; \int_{-1}^1 d\tilde x  
\int_{-1 + |\tilde x|}^{1-|\tilde x|}  d \alpha   \, 
\delta (x  - \tilde x  -  \xi \alpha ) \, f(\tilde x, \alpha) 
\; \left\{
\begin{array}{r}
1 \\[1.5ex]
\alpha
\end{array}
\right.\, .
\label{reduction}
\end{equation}
Note, that  in  our 
case  the DD  $f(\tilde x, \alpha)$ is 
even in $\alpha$ and odd in $\tilde x$. 
As a result, the  functions $H$ and $A$ 
satisfy the symmetry relations
\begin{eqnarray}  
H(x , \xi) = -H(-x , \xi) \  \  ,  \  \  H(x , \xi) =   H(x , -\xi)  \    ,  
\nonumber \\ 
A(x , \xi) =  A(-x , \xi)  \  \  ,  \  \   A(x , \xi) = -A(x , -\xi) \ .
\label{symm}
\end{eqnarray} 
Furthermore, because of the antisymmetry of the combination 
$ \alpha f (\tilde x, \alpha)$  
with respect both to
$x$ and $\alpha$ we have
\begin{equation} 
\int_{0}^1 dx \, A (x, \xi) \;\; = \;\; 0 .
\end{equation}
Hence, the distribution $A (x, \xi)$ 
cannot be a  positive-definite function  
on $0 \leq x \leq 1$.

Uniting the cosine and sine functions
with the overall exponential factor $e^{-i vx (Pz)}$
one gets $vx \pm \bar v \xi$ combinations.
Using  (\ref{symm}), 
one can   arrange that only $vx + \bar v \xi$
would appear,
\begin{eqnarray}  && \hspace{-0.8cm}
\frac{1}{2}  \, 
\langle P-r/2 \, | \, {\cal O}_{\sigma} (z\, |\, 0) \,| \,P+r/2  \rangle =
P_{\sigma} \int_{-1}^1 d x  \, 
   e^{-i x(pz) } \biggl [ H(x, \xi) 
-  \frac{i (\Delta z)}{2} A(x, \xi) \biggr ]
\nonumber \\ 
&& \hspace{1.5cm} + \frac{i}{2} \, \bigl [\Delta_{\sigma} (Pz)  - 
P_{\sigma}(\Delta z)\bigr ]
\int_{-1}^1 d x \, \bigl [
H (x, \xi)- 
 A(x, \xi)  
 \bigr ]\nonumber \\ && \hspace{2cm} \times \int_0^1 dv \, v \, 
 \cos [(v x +\bar v \xi) (Pz)]  \  .
\label{para7} 
\end{eqnarray}
In a similar fashion, we get parametrization
for  the matrix element of the axial  
string operator (\ref{string_deconstructed_scalar}),
\begin{eqnarray} && \hspace{-0.8cm}
 \frac{1}{2}\,
\langle P-r/2 \, | \, {\cal O}_{5\sigma } (z\,|\,0) \,  |\,  P+r/2 \rangle =  
\frac{i}{2}  \,  \epsilon_{\sigma\alpha\beta\gamma} \, z_\alpha 
 \Delta_{\beta}    p_{\gamma}
\int_{-1}^1 d x \, \bigl [ 
H (x, \xi)- 
 A(x, \xi)  
 \bigr ] \nonumber \\&& \hspace{3cm} \times 
    \int_0^1 dv \, v \, 
  \sin[(v x +\bar v \xi) (Pz)] 
\ . 
\label{parax5} 
\end{eqnarray}
 Note that it is expressed in terms of the
same skewed distributions $H(x, \xi)$ and $A(x, \xi)$ which,  
in turn, are determined by the
original double distribution  
$f(\tilde x, \alpha) $, see Eq. (\ref{reduction}). 

\subsection{DVCS amplitude for pion target} 

 {\it DD-generated contribution.} 
Substituting the parametrizations (\ref{para7}) and (\ref{parax5}) 
into Eq. (\ref{handbag})  
and performing the Fourier integral over the separation $z$ one 
obtains the  Compton amplitude 
\begin{eqnarray} 
T_{\mu\nu}   &=& \;  - \frac{2}{(PQ)} 
 \; 
 \left[ P_\mu Q_\nu + Q_\mu P_\nu - g_{\mu\nu} (PQ) + 2 \xi P_\mu P_\nu
+  \frac{\Delta_\mu}{2} P_\nu- P_\mu\frac{\Delta_\nu}{2}    
\right] \nonumber \\ & & \hspace{-1cm} \times
  \int_{-1}^1 dx \frac{H(x, \xi)}{x - \xi + i0}    \, 
-  \frac1{(PQ)} \, \int_{-1}^1 dx \,  R(x, \xi)  
\int_0^1 dv  \,  
\frac{ (Q_\mu + 3 \xi P_\mu)\Delta_\nu }{\xi + v x + \bar v \xi - i0 }
 \nonumber \\ &-& \frac1{(PQ)} \, \int_{-1}^1 dx \,  R(x, \xi)  
\int_0^1 dv \; \frac{\Delta_\mu (Q_\nu + \xi P_\nu) }
{-\xi + v x + \bar v \xi + i0} 
 \  ,
\label{Compton_Anikin}
\end{eqnarray}
where $R (x, \xi)$ is  a new  SPD describing the kinematical
twist-3 contributions,
\begin{eqnarray}
 R (x, \xi) \equiv \frac{{\partial H(x, \xi)}}{\partial x}
 - \frac{{\partial A(x, \xi)}}{\partial x} \  . 
 \label{R}
\end{eqnarray} 
All three terms in 
Eq. (\ref{Compton_Anikin})  
are individually transverse up to terms of
order $t,P^2$. 

{\it Singularities.} 
 The 
first term is the  twist--2 part 
with the  tensor structure corrected exactly as suggested by 
Guichon and Vanderhaeghen.\cite{Guichon}
The integral over $x$ exists if  $H(x,\xi)$ is 
continuous at $x = \xi$, which is the case for SPD's derived from 
the DD's that are less singular than $1/\tilde x^2$ for $\tilde x =0$ 
and are continuous otherwise (see Ref. 23). 
 In particular, continuous SPD's were  obtained in
model calculations of  SPD's at a low scale in the instanton 
vacuum.\cite{bochum} 
The second term contributes only to the helicity amplitude
for a longitudinally polarized initial photon.
 The parameter integral over $v$ gives the function  
\mbox{$ [ \ln ( x + \xi - i0) - \ln (2 \xi-i0) ]/(x-\xi)$}  
which  is 
regular at $x = \xi$ and has a 
logarithmic singularity at $x = - \xi$.
The integral over $x$ exists if  
$
R(x, \xi) $ 
is bounded at $x=-\xi$, which again is the case
in the DD-based models described in  Ref. 23. 
The third term of Eq. (\ref{Compton_Anikin}) 
corresponds to the transverse polarization  of the initial photon. 
In this case,  one faces  the integrand   
$1/[v(x-\xi)+i0]$  which produces $dv/v$  
 divergence for the $v$-integral
at  the lower limit. One may  hope to  get a finite result 
only if the integral 
\beq
I(\xi) \equiv \int_{-1}^1 dx
  \frac{R(x,\xi)}{x - \xi + i0} \label{Ixi} 
\eeq
vanishes. 
From the definition of the skewed 
distributions $H(x, \xi)$ and  $ A (x, \xi)$ (\ref{reduction}) 
it  follows that  
 $$\frac{{\partial} A (x, \xi)}{\partial x} =
-\frac{ {\partial} H (x, \xi)}{\partial \xi} .$$
Hence, one can substitute 
  $R (x, \xi) $ 
by the combination  ${\partial} H (x, \xi)/{\partial x} + 
{\partial} H (x, \xi)/{\partial \xi}$  
(see   Refs.~48, 49, 24). 
    Integrating the
 $ {\partial} H (x, \xi)/{\partial x}$ term by 
 parts~\cite{Kivel:2000rb,rw}
  gives 
\beq
I(\xi) = \frac{d}{d \xi} 
\int_{-1}^1 dx \frac{H(x, \xi)}{x - \xi + i0} \ ,
\label{xider}  
\eeq
i.e., the $\xi$ derivative of the  twist-2 
contribution. In general, the latter has a  nontrivial 
$\xi$-dependent form  determined 
by the shape of SPDs (see, however, the discussion of the
PW contribution below).  
Hence,  the twist-3 part of the tensor 
amplitude $T_{\mu\nu}$ diverges in case of 
the transverse polarization of the initial photon.\cite{Kivel:2000rb,rw} 
 However, it is easy to see that the relevant 
 tensor structure $\Delta_\mu (Q_\nu +\xi P_\nu)$
 is just a truncated version of the exactly  gauge 
 invariant combination
 $\Delta_\mu {q'}_{\nu}$ which 
 has zero projection onto
 the polarization vector 
 $\epsilon'^\nu$ of the final real  photon: $(\epsilon' q')=0$.
  
 The structure $\Delta_\mu {q'}_{\nu}$ is obtained if one uses 
 the original full form of the DD parametrization
 (\ref{para5}).  
 It appears  from the term with   the   exponential
  factor of the argument 
  $  -i [v(\tilde x +\xi \alpha)(Pz)+
  \bar v \xi (Pz) + (v \alpha +\bar v) (\Delta z)/2] $
  which is 
  obtained by combining 
  the   sine/cosine functions  and the  
  exponential factor  in the second
  term of Eq. (\ref{para5}). In the Compton amplitude, 
  it gives rise to a  contribution in which the argument 
  of the quark propagator is
  $q+v(\tilde x +\xi \alpha)P  + \bar v \xi P + (v \alpha + \bar v)\Delta/2$.
   Since $(\Delta Q)$,   $(\Delta P$) and $\Delta^2 $ are negligible, 
     the denominator factors
  in Eq. (\ref{Compton_Anikin}) remain unchanged. 
 In  numerators,  representing $(v \alpha + \bar v)\Delta/2$
 as $[1-(1-\alpha)v]\Delta/2$,  we observe that 
   $\Delta_{\mu} (Q_\nu+\xi P_\nu)$  
      converts into the 
 $\Delta_{\mu} (Q_\nu+\xi P_\nu +\Delta_\nu/2 )= \Delta_\mu {q'}_{\nu}$ 
 term  
 plus a  $v \Delta_{\mu} \Delta_{\nu}$  type 
 contribution corresponding to a new SPD built  from
 the $(1-\alpha)^2 f(\tilde x, \alpha)$ DD (cf. (\ref{reduction})). 
 Due to the  extra $v$ factor, the $v$-integral for 
 the latter contribution is finite. 
   Hence, for the physical DVCS amplitude, 
 we find  no evidence for factorization breaking
 in the  kinematical twist--3 contributions, both in their
$1/\sqrt{-q^2}$ and $1/{q^2}$ terms.
It is quite possible that factorization breaks down 
at the $1/{q^2}$ level,  
but one needs to analyze $O(z^2)$  suppressed 
terms (i.e., twist--4 contributions) to see if it really happens.  

 {\it WW-type representation.}
These results  can be expressed in 
another  form\,\cite{Belitsky:2000vx,Kivel:2000rb,rw} 
by introducing new skewed distributions  related to 
$R(x,\xi)$  $via$  the integral  transformation 
similar to that used by Wandzura and Wilczek.\cite{WW}
Treating the combination $xv+\bar v \xi$ in 
(\ref{Compton_Anikin}) as a new 
variable  we  define 
\begin{eqnarray}  
R_W (x, \xi) \! &\equiv&\!    
\int_{-1}^1 R (y, \xi) \, dy \int_0^1   
\, \delta ( yv + \bar v \xi -  x)   \, dv   
\nonumber \\ \! &=&  \! \theta ( x > \xi) \int_{x}^1
\frac{R(y,\xi)}{y-\xi}\, dy  -  \, \theta ( x < \xi) \int_{-1}^x 
\frac{R(y,\xi)}{y-\xi}\, dy  \,  . 
 \label{rw} 
 \end{eqnarray} 
In terms of this transform, the matrix element of the 
vector operator (\ref{para7}) 
can be expressed as 
\begin{eqnarray}  && \hspace{-1.2cm}
\frac{1}{2} \langle \, P-r/2 \, |\,  
{\cal O}_{\sigma} (z\, |\, 0) \, | \, P+r/2 \rangle  
= 
  \int_{-1}^1 d x  \, e^{-i x(Pz) }
\biggl \{ 
P_{\sigma}   H(x, \xi)  \nonumber \\ && \hspace{-1cm} 
-  \frac{i}{2} \, P_{\sigma} \, 
 (\Delta z)\,  A(x, \xi)   +  
\frac{1}{4} \, \biggl (\Delta_{\sigma}   - 
P_{\sigma}\frac{(\Delta z)}{(Pz)}\biggr )
\biggl [ 
R_W (x, \xi)- 
 R_W (-x, \xi)  
 \biggr ] \biggr \}   \ .   
\label{para8} 
\end{eqnarray}
Note that only the odd part of $R_W (x, \xi)$ contributes 
here. In case of the axial operator (\ref{parax5})
\begin{eqnarray} 
&&\lefteqn{\frac12 \langle \, P-r/2 \, |\,  {\cal O}_{5 \sigma} 
(z\, |\, 0) \, | \, P+r/2 \rangle} \nonumber \\  &&
 =    \frac{i}{4} \,   \epsilon_{\sigma\alpha\beta\gamma} \,
\frac{ z_\alpha}{(Pz)} 
 \Delta_{\beta}    P_{\gamma}
\int_{-1}^1 d x \,e^{-i  x (Pz) } 
 \biggl [ R_W (x,\xi) + R_W (-x,\xi) \biggr ]  
\label{parax6} 
\end{eqnarray}
only the even   part of $R_W (x, \xi)$ appears. 
The  part  
 of the 
Compton amplitude (\ref{Compton_Anikin})  
containing  $R(x,\xi)$  
can   be written  in terms of this new
function as  
\begin{eqnarray} 
  - \frac1{(QP)}
\int_{-1}^1 
\, 
\biggl [ \frac{\Delta^{\mu} ( Q^{\nu} + \xi  P^{\nu})}
{x-\xi +i0} + \frac{( Q^{\mu} + 3 \xi  P^{\mu})\Delta^{\nu} }
{x+\xi -i0 }\biggr ]
R_W(x,\xi) \, dx \  . 
 \label{comfin} 
\end{eqnarray}

The integrals with $1/(x \pm \xi \mp i0)$  converge
only if the function $ R_W(x,\xi)$
is continuous for $x=\pm \xi$. According to Eq. (\ref{rw}),  
$ R_W(x,\xi)$  is  given by the integral
of $R(y,\xi)/(y-\xi)$ from $x$ to 1 if
$x> \xi$ and from  $x$ to $-1$ if $x < \xi$.
Evidently,  $x=-\xi$ is not a special 
point in the integral transformation (\ref{rw}), hence 
the function $ R_W(x,\xi)$ 
is continuous at $x=-\xi$. 
However, it  is extremely unlikely 
 that the limiting values approached by 
 $ R_W(x,\xi)$ for $x=\xi$  
from below and from above do coincide. 
Indeed, the difference of the two limits 
can be written as the principal value integral,\cite{Kivel:2000rb,rw} 
\begin{eqnarray}
 R_W(\xi+0, \, \xi) -  R_W(\xi-0,\, \xi )
 = {\rm P} \int_{-1}^1 \frac{R(y,\xi)}{y-\xi}\, dy \ , 
 \label{pv} 
 \end{eqnarray}
 which  can be converted  into the $\xi$-derivative
 of the real part of the twist--2 contribution. 
 This means that  the singularity,  
which was obtained  as a  straight divergence of the $dv/v$ 
integral, in the WW-type  approach 
appears due to an unavoidable  discontinuity  of the 
$R_W (x, \xi)$ transform   
at $x = \xi$.

{\it Contribution from the PW term.} 
The contribution of the PW term to the 
vector operator 
\begin{eqnarray} 
\frac{1}{2} \, 
\langle P - r/2 \, | \, {\cal O}_{\sigma}(0\,|\,z)  \,  
| \, P + r/2 \rangle_{\mbox{\scriptsize PW term}}
 =
\frac{r_{\sigma}}{2}  \int_{-1}^1 
 \, e^{-i \alpha (rz)/2} \, 
D (\alpha)\,  d \alpha 
\label{para_PW} 
\end{eqnarray}
 has a simple structure corresponding to a 
parton picture in which the partons carry the fractions $(1 \pm \alpha)/2$ 
of the momentum transfer $r$. Since only one momentum  $r$  is 
involved, this term can contribute only to the totally symmetric part
of the vector string operator: it  ``decouples'' in the 
reduction relations (\ref{rest_total_derivative}). In particular, the  
{\rm PW} term  does not contribute 
to the second contribution  in Eq. (\ref{string_deconstructed_scalar})
which is generated by decomposition
of the axial string operator:  both derivatives, with respect to  
$X$ and $z$, give rise to the momentum transfer $r$,  whence the 
contraction with the $\epsilon$--tensor in 
(\ref{string_deconstructed_scalar}) gives zero.
Thus, the PW-contribution  should be transverse by 
itself. Indeed, a straightforward calculation gives 
\begin{equation}
T_{\mu\nu}|_{\mbox{\scriptsize PW}}  \;\; = \;\; 
- \frac{2}{(rq)}
\biggl [ r_{\mu} q_{\nu} +  q_{\mu} r_{\nu}  -  g_{\mu \nu} (rq)
+ r_{\mu} r_{\nu} \biggr ] \; \int_{-1}^1 
\frac{D (\alpha)}{\alpha-1}  \,  d \alpha  \  , 
\label{PW} 
\end{equation} 
which evidently 
satisfies  
$
q_{\mu} T_{\mu \nu}|_{\mbox{\scriptsize PW}}  =  0, \ 
r_{\mu} T_{\mu \nu}|_{\mbox{\scriptsize PW}}  = 0.
$
Hence,  this term can be treated as a separate contribution. 

Alternatively, one may include it into the basic SPD $H(x,\xi)$ 
and all SPD's derived from $H(x,\xi)$.
Specifically, for $\xi >0$, the PW contribution 
to $H(x,\xi)$ is $D (x/\xi) \, \theta (|x| \leq \xi)$;\cite{PW} 
it contributes   $(\xi-x)D'(x/\xi)\, \theta (|x| \leq \xi)/\xi^2$
[where $D'(\alpha) \equiv (d/d\alpha) D(\alpha)] $ to 
$R(x,\xi)$; 
furthermore,  the  PW contribution to  $R_W(x,\xi)$ is 
$D (x/\xi)\, \theta (|x| \leq \xi)/\xi$. 
Inserting these functions into Eqs. (\ref{Compton_Anikin}) and 
(\ref{comfin})
one rederives Eq. (\ref{PW}). 
One can also observe that the PW term gives zero contribution
into $I(\xi)$, Eq. (\ref{Ixi}).

\section{Real Compton scattering} 

\subsection{Compton amplitudes and light--cone dominance}

The Compton scattering in its various versions 
provides a unique tool for studying 
 hadronic structure.
The  Compton amplitude probes the hadrons through a 
 coupling of two electromagnetic currents
 and in this aspect it  can be considered 
as a generalization of hadronic form factors.  
In QCD, the photons interact with the 
quarks of a hadron through 
a vertex which, in the lowest
 approximation,  has a   pointlike structure.
However, in the soft regime, 
strong interactions  produce large 
corrections uncalculable within the
perturbative QCD framework.
To take advantage of  the basic pointlike structure of the
photon-quark coupling and the asymptotic
freedom feature of QCD, one should choose 
a specific kinematics  in which the behavior
of the relevant amplitude is dominated by 
short (or, being more precise, lightlike) distances.
The general feature of all such types of kinematics is the 
presence of a large momentum transfer.
For  Compton amplitudes, there are several
situations when large momentum transfer induces
dominance of configurations 
involving lightlike distances:  \\ 
$i)$ both photons are far off-shell 
and have  equal spacelike virtuality:
 virtual forward Compton amplitude,
 its imaginary part determines structure
 functions of deep inelastic scattering (DIS); \\ 
 $ii)$ initial photon is highly virtual,
 the final one is real and the momentum transfer 
 to the hadron is small: deeply virtual 
 Compton scattering (DVCS) amplitude; \\ 
 $iii)$ both photons  are real but the 
 momentum transfer
 is large:  wide-angle Compton 
 scattering (WACS) amplitude.

 The first two cases were discussed
 in the previous sections. As argued in Ref. 25, 
 at accessible momentum transfers
 $|t| < 10$ GeV$^2$, the  WACS amplitude  is also dominated 
 by  handbag diagrams, just like  in  DIS and DVCS.
 In the most general case, the nonperturbative 
 part of the handbag contribution
 is described by  double distributions (DDs) 
 $f(x,\alpha;t),   g(x,\alpha;t)$, etc.,  which,
 as discussed earlier,  can be 
 related  to  usual parton 
 densities $f(x)$, $\Delta f(x)$ and nucleon form factors
 like $F_1(t),G_A(t)$. 
 Among the arguments of DDs,   $x$ is 
  the fraction of the initial hadron momentum carried 
 by the active parton and $y$ is the  fraction   
 of the momentum transfer $r$.
The description of the WACS amplitude simplifies 
when one can neglect the $\alpha$-dependence 
 of the hard part and integrate out
    the $\alpha$-dependence
 of the double distributions. In that case,
 the long-distance dynamics is described by nonforward
 parton densities~\cite{realco} (NDs)  
 ${\cal F}(x;t), {\cal G}(x;t),$ etc. 
 The latter 
 can be interpreted as  the usual parton densities $f(x)$ 
 supplemented by a form factor type $t$-dependence.  
    A simple model 
 for the relevant NDs was proposed in Ref. 25. 
It both satisfies the relation between  ${\cal F}(x;t)$ and
   the  usual parton densities $f(x)$ and  produces 
   a good description of 
  the  $F_1(t)$ form factor up to $t \sim  - 10$ GeV$^2$. 
 This model was used to calculate  the WACS amplitude in a   
  rather close agreement to existing data.\cite{shupe}  

\subsection{Modeling nonforward densities}

Let us apply the DD formalism 
to the   large-$t$ real Compton scattering.
 Since the initial   photon  is also     real,  
 $q^2=0$ (and hence $x_{Bj}=0$), it is natural to expect that 
 the nonperturbative functions which appear in WACS 
 correspond   to the $\xi = 0$ limit of 
 the  skewed  parton distributions. 
 In the $\xi=0$  limit, 
 the SPDs reduce to the  nonforward parton 
densities 
\begin{equation} 
{\cal F}^a(x;t) = \int_{-1+|x|}^{1-|x|}  
 f^a(x,\alpha;t)  \, d\alpha \, .\end{equation}
 Note that NDs depend on  
 ``only   two''  variables $x$ and $t$,
 with this dependence  
   constrained by reduction
 formulas
  \begin{equation}  {\cal  F}^a(x;t=0) 
= f_a(x)  \  \  ;  \  \   \sum_a e_a \int_0^1 
\left [ {\cal F}^a(x;t) -   
 {\cal F}^{\bar a} (x;t)  \right ] 
 \,  dx  
 =F_1(t) \, . \label{3b} \end{equation}

 Furthermore,    it is possible 
to  interpret the nonforward  densities  
in terms of the  light-cone wave functions.
Consider for simplicity 
a two-body bound state  whose 
lowest Fock component is described by a light cone   
wave function $\Psi(x,k_{\perp})$.
Choosing a frame where the momentum transfer $r$  is purely 
transverse  $r=  r_{\perp}$, we can write  
the two-body contribution into the form factor 
as\,\cite{bhl}
\begin{equation}
F^{(two-body)} (t) = \int_0^1 \,  dx \, \int \, 
\Psi^* ( x, k_{\perp}+\bar x r_{\perp}) \,  
\Psi (x, k_{\perp}) 
\, {{d^2 k_{\perp}}\over{16 \pi^3}}  \, , 
\end{equation}  

\begin{figure}[htb]
\vspace{1cm} \mbox{
   \epsfxsize=12cm
 \epsfysize=5cm
  \epsffile{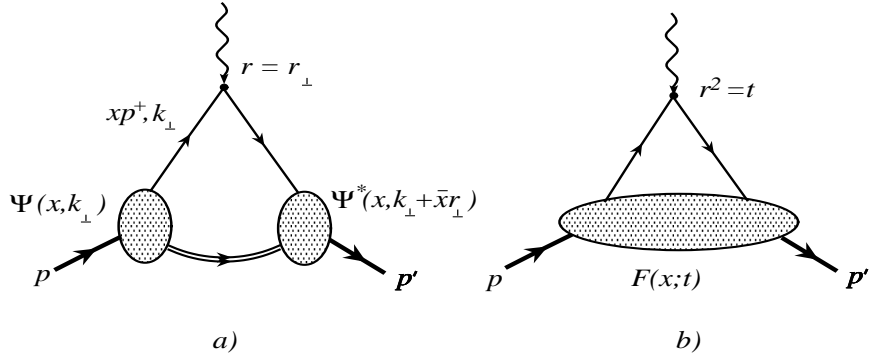}  }
{\caption{\label{fig:psifor} $a)$ Structure of the 
effective two-body contribution 
to form factor in the light cone formalism.
$b)$ Form factor as an $x$-integral  of  nonforward parton densities.
   }}
\end{figure}
\noindent where $\bar x \equiv 1-x$. 
  Comparing this expression with the reduction formula 
  (\ref{3b}), we conclude that 
 \begin{equation} 
 {\cal F}^{(two-body)} (x,t) = \int  \, 
 \Psi^* ( x, k_{\perp}+ \bar x r_{\perp}) \,
\Psi (x, k_{\perp}) \, 
{{d^2 k_{\perp}}\over{16 \pi^3}}
\end{equation}  
 is the  two-body contribution into the nonforward  parton density
 ${\cal F} (x,t)$.
Assuming  a Gaussian dependence on the transverse momentum $k_{\perp}$
(cf. Ref. 54) 
\begin{equation}\Psi (x,k_{\perp}) =  \Phi(x) 
 e^{-k^2_{\perp}/2x \bar x \lambda^2} \,  , \label{11} 
 \end{equation}
we get 
\begin{equation}
{\cal F}^{(two-body)} (x,t) = 
f^{(two-body)}(x) e^{\bar x t /4 x \lambda^2 } \, , \label{8}
\end{equation}
where 
\begin{equation}
f^{(two-body)}(x) = 
\frac{x \bar x  \lambda^2}{16 \pi^2} \, \Phi^2(x) 
= {\cal F}^{(tb)} (x,t=0)
\end{equation}
is the two-body part of the relevant parton density.
Within the light-cone approach, to get the total
result for either  usual $f(x)$
or 
nonforward parton densities   ${\cal F}(x,t)$,
one should 
add the contributions due to  higher Fock components.
These contributions
{\it are  not small}, e.g., the  valence $\bar d u$  contribution
into the normalization of the $\pi^+$  form factor 
at $t=0$ is less than 25\% (see Ref. 54). 
In the absence of a formalism providing  explicit expressions
for an infinite tower of light-cone wave functions 
one can  choose to treat  Eq. (\ref{8}) as a guide
for  fixing interplay between the $t$ and $x$ dependences
of NDs and propose to 
 model them by 
\begin{equation}
{\cal F}^a(x,t) = f_a(x) e^{\bar x t /4 x \lambda^2 }  = 
{{f_a(x)}\over{\pi x \bar x  \lambda^2}}\,
\int  \, e^{-(k^2_{\perp}+ (k_{\perp}+
\bar x r_{\perp})^2)/2x \bar x \lambda^2}
d^2 k_{\perp} \,  . \label{13}
\end{equation}
The functions  $f_a(x)$  here are the 
usual parton densities. They can   be 
taken from existing  parametrizations like GRV, MRS, CTEQ, etc.
In the $t=0$
limit (recall that $t$
is negative)  this model, by construction,  
satisfies the  first of  reduction formulas (\ref{3b}). 
 Within the Gaussian ansatz (\ref{13}), 
 the basic scale $\lambda$ specifies  the 
  average transverse momentum carried by the quarks.
   In particular, for valence quarks 
  \begin{equation} \langle  k^2_{\perp} \rangle ^a = 
  \frac{\lambda^2}{N_a}\int_0^1 
  x \bar x f_a^{val}(x)  \,  dx   \, , 
  \end{equation}
  where $N_u=2, N_d=1$ are the numbers of 
  the valence $a$-quarks in the
  proton.

  The magnitude of  $\lambda$ can be fixed 
  using the second reduction formula in 
  (\ref{3b}) relating nonforward densities ${\cal F}^a(x,t)$
  to the $F_1(t)$ form factor.
  The following simple expressions for
  the valence distributions
  \begin{equation} f_u^{val} (x) = 1.89 \,
   x^{-0.4} (1-x)^{3.5} (1+6x) \, , \end{equation}
  \begin{equation} f_d^{val} (x) = 0.54 \, 
   x^{-0.6} (1-x)^{4.2} (1+8x) \, . \end{equation}
   \begin{figure}[htb]
\vspace{2cm}\mbox{
   \epsfxsize=8cm
 \epsfysize=8cm
 \hspace{2cm}  
  \epsffile{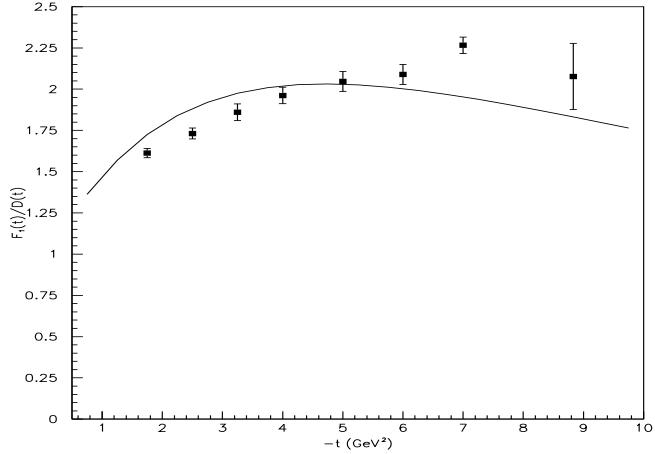}  }\vspace{-1cm}
{\caption{\label{fig:ff} Ratio $F_1^p(t)/D(t)$ 
of the $F_1^p(t)$ form factor to
the dipole fit $D(t) =1/(1-t/0.71\,{\rm GeV^2})^2$. Curve
is based on  Eq. (47) with $\lambda ^2 = 0.7 \,
{\rm GeV}^2$. Experimental
data are taken from Ref. 56.
   }} 
\end{figure} 
 closely reproduce the  relevant 
 curves given by the GRV parametrization\,\cite{grv} 
 at a low normalization point $Q^2 \sim 1$  GeV$^2$. 
 The best agreement between the   model 
 \begin{equation}
 F_1^{\rm soft}(t) =  \int_0^1 \left [ e_u\, f_u^{val} (x) +e_d
  \, f_d^{val} (x) \right ] e^{\bar x t / 4x \lambda^2} dx \label{14}
  \end{equation}
  and experimental data\,\cite{ff} in the
  moderately large $t$ region
   {1  GeV$^2$ $< |t|< 10$ GeV$^2$} is reached for 
   $\lambda^2 =0.7 \,$ GeV$^2$ (see Fig. \ref{fig:ff}). 
   This value gives a reasonable magnitude 
   \begin{equation}  \langle  k^2_{\perp} \rangle^u = (290 \,  {\rm MeV})^2 
   \  \   \  ,  \ \ \   \langle  k^2_{\perp} \rangle^d = (250 \, {\rm MeV})^2
   \end{equation}
   for the average transverse momentum of the valence $u$ and $d$ quarks
   in the proton.

 Similarly, building a model  for the 
parton helicity sensitive NDs ${\cal G}^a(x,t) \equiv \tilde H(x, \xi=0;t) $
one can take their  $t=0$ shape  from existing 
parametrizations for spin-dependent 
parton distributions $\Delta f_a(x)$
and  then fix the relevant $\lambda$ parameter by fitting
the $G_A(t)$ form factor. 
The case of hadron spin-flip distributions ${\cal K}^a(x,t)
\equiv E (x, \xi=0;t)$ 
and ${\cal P}^a(x;t)\equiv \tilde E(x, \xi=0;t)$ is more complicated  
since the distributions $e_a(x)$, $\tilde e_a(x)$ are unknown.

At $t=0$, the model by construction gives a  correct 
normalization $F_1^p(t=0)=1$ for the form factor.
Moreover, the curve  is within 5\% from the data  points\,\cite{ff} for
$1 \, {\rm GeV}^2 < -t < 6$ \, GeV$^2$ and does not deviate 
from them by more than 10\% up to 9 GeV$^2$.
Modeling the $t$-dependence by a more complicated formula
(e.g., assuming  a slower decrease at large $t$, and/or
choosing different $\lambda$'s for
$u$ and $d$ quarks and/or splitting NDs
into several components with different 
$\lambda$'s,  etc., (see Ref. 30 
for an example of such an attempt) or changing the shape 
of parton densities $f_a(x)$ 
one can improve the quality of the fit
and extend  agreement with the data to higher $t$. 
However, the very fact that  
a reasonable  description of the  
$F_1(t)$ data in a wide region 
\mbox{1  GeV$^2$ $< |t|< 10$ GeV$^2$ } 
was obtained by fixing just a single parameter
 $\lambda$ reflecting the proton size 
 is   an 
 evidence that the model
correctly catches the gross features of the 
underlying physics.

Since 
the    model implies  a Gaussian  
dependence on the transverse momentum,
it includes  only what is usually referred to as   
an overlap of soft wave functions.
It completely  neglects effects due to 
hard PQCD gluon exchanges  generating 
the power-law  $O( (\alpha_s / \pi)^2 /t^2)$
tail of the nonforward densities at large $t$.
Note also that  taking   the   nonforward 
densities  ${\cal F}^a(x,t)$ with  an exponential 
dependence on $t$, one gets 
a power-law asymptotics    $F_1^{\rm soft} (t) \sim (-4 
\lambda^2/t)^{n+1}$ for the
$F_1(t)$ form factor, with the power dictated  
 by the   $(1-x)^n$ behavior of 
the parton densities for $x$ close to 1.
This connection arises because  
the integral (\ref{14}) over $x$ 
is dominated at large $t$ by  the region 
$\bar x \sim  4 \lambda^2 /|t|$. 
In other words,  the large-$t$ behavior of $F_1 (t) $ in our model
is governed by the Feynman mechanism.\cite{feynman}
One should realize, however, that  the relevant scale
$4 \lambda^2 =2.8$ GeV$^2$ is rather large.
For this reason, when 
 $|t| < 10$ GeV$^2$,  
 it is premature to rely on asymptotic estimates 
 for the soft contribution. 
  Indeed, with $n=3.5$, the asymptotic estimate is  
 $F_1^{\rm soft} (t) \sim t^{-4.5}$,
in an apparent contradiction with 
the ability of our curve to follow 
the  dipole behavior.
The resolution of this paradox is very simple:
the maxima of nonforward densities  ${\cal F}^a(x,t)$ 
for $|t| < 10$ GeV$^2$ are 
at rather low $x$-values $x < 0.5$.
Hence, the $x$-integrals producing $F_1^{\rm soft} (t)$
are not dominated  by the $x \sim 1$ region yet and the 
asymptotic  estimates are not applicable:
the functional dependence of $F_1^{\rm soft} (t) $ in 
our model  is much more complicated than 
a simple power of $1/t$.    
 
  The fact that the  soft overlap    model  closely reproduces 
the experimentally 
observed 
dipole-like behavior of the proton form factor is a 
clear demonstration 
  that 
such a  behavior does not necessarily reflect the 
quark counting rules~\cite{brofar,mmt} $F_1^p(t) \sim 1/t^2$ 
valid  for the
asymptotic behavior of 
the  hard gluon exchange 
contributions.
 Our explanation of the observed 
 magnitude and the $t$-dependence
 of  $F_1  (t)$ by a purely soft 
 contribution  is in  strong contrast   
 with that of the hard PQCD  approach to this problem. 
 
\subsection{Wide-angle  Compton scattering}

With both photons real, it is not sufficient to have
large photon energy to ensure short-distance dominance:
large-$s$, small-$t$ region is strongly affected by
Regge contributions.
Hence, having large $|t| > 1 \, $GeV$^2$ is a  necessary
condition for revealing   short-distance  dynamics.

\begin{figure}[htb]
\mbox{
   \epsfxsize=12cm
 \epsfysize=3cm
  \epsffile{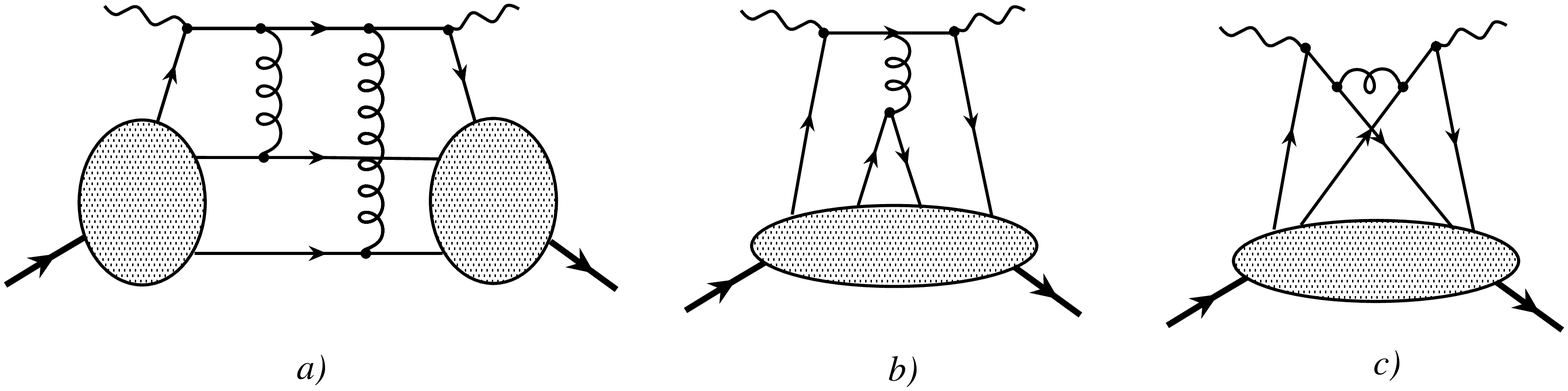}   }   
{\caption{\label{fig:cosubno} Configurations 
involving double and single gluon exchange.
   }}
\end{figure}

The simplest contributions for the WACS amplitude 
are given by the $s$- and $u$-channel
handbag diagrams Fig.~\ref{fig:cohan}b,c. 
The nonperturbative part in this case is given by the proton 
 DDs  
which determine the $t$-dependence of the total contribution.
Just like in the form factor case, 
the contribution dominating in the 
formal asymptotic limit $s,|t|, |u| \to \infty$,
is given
by diagrams corresponding to the pure SD regime,
see Fig.~\ref{fig:cosubno}a. The hard subgraph then 
involves two hard gluon exchanges which results in a  
suppression  factor $(\alpha_s/ \pi)^2 \sim 1/100$  
absent in  the handbag term. 
The total contribution  of all two-gluon exchange  
contributions  was calculated by Farrar 
and Zhang,\cite{far} recalculated by  
 Kronfeld and Ni\v{z}i\'{c},\cite{kroniz}
 by Vanderhaeghen~\cite{vanderwacs,vander} and by
 Brooks and  Dixon.\cite{brodix}   
 A sufficiently large 
 contribution is  only obtained if one uses  humpy DAs
 and   $1/k^2$ propagators with no finite-size effects included.
 Even with such propagators, 
 the WACS  amplitude 
calculated with the asymptotic DA is 
negligibly small\,\cite{vander} compared to existing  data. 
As argued in Ref. 25, 
 the    enhancements generated by 
the humpy DAs should not be taken at face  value both  
for form factors and  
wide-angle Compton scattering amplitudes. 
For these reasons, we ignore the  hard contributions 
to the WACS  amplitude as negligibly small.

Another type of configuration containing  
 hard gluon exchange is  shown in Fig. \ref{fig:cosubno}b. There are  
also the diagrams with 
 photons coupled to different 
quarks (``cat's ears'', Fig. \ref{fig:cosubno}c). 
Such contributions have   both higher order 
and   higher twist.
This brings in the 
$\alpha_s/\pi $ factor  and  
 an  extra  $1/s$ suppression.
The latter is  partially compensated by  a slower fall-off 
of the four-quark DDs with $t$ since only one valence quark
should change its momentum.

For simplicity, let us  neglect all the suppressed terms 
and deal only with the handbag contributions Fig. \ref{fig:cohan}b,c 
in which 
the highly virtual quark propagator connecting the photon vertices 
  is convoluted with  proton DDs
  parametrizing the 
overlap of soft wave functions.  
Since  the  basic scale $4 \lambda^2$ 
characterizing the $t$-dependence of 
DDs in our model  is 2.8 GeV$^2$, while existing data    are 
all  at  momentum transfers $t$ below   5  GeV$^2$,
we deal with the region where the asymptotic estimate 
(Feynman mechanism) for the overlap
contribution is  not working yet.

\begin{figure}[htb]
\hspace{5mm} \mbox{
   \epsfxsize=5cm
 \epsfysize=4cm
  \epsffile{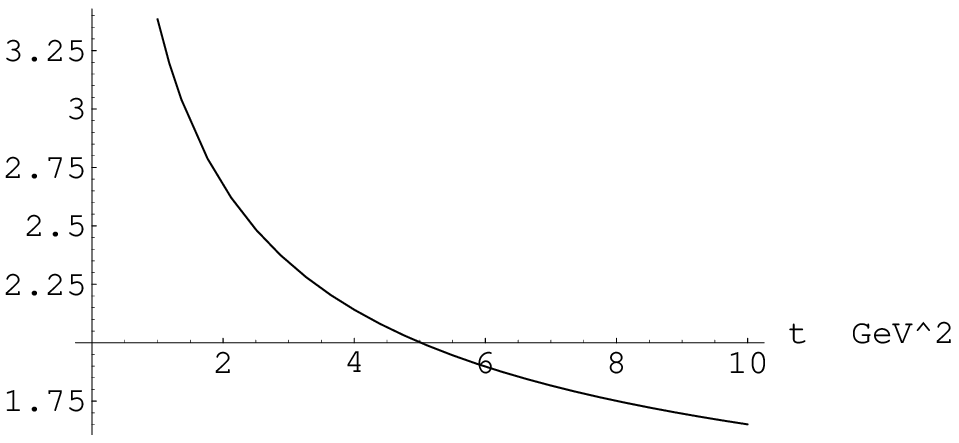} \hspace{1cm}    \epsfxsize=5cm
 \epsfysize=4cm
  \epsffile{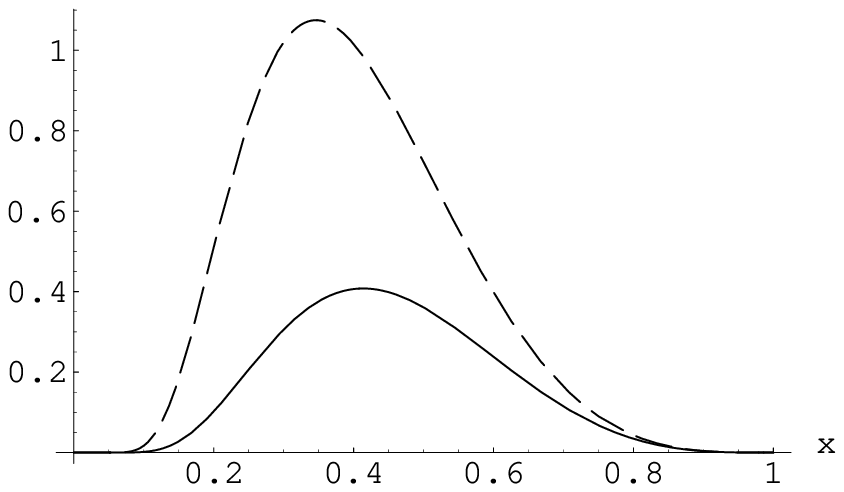} }
  \vspace{0.5cm}
{\caption{\label{fig:rcuux} $a)$ Ratio
$R_1^u(t)/F_1^u(t)$; $b)$ Functions  
${\cal F}^u (x;t)$ (solid line)
 and ${\cal F}^u (x;t)/x$ (dashed line) at  $t = - 2.5$ GeV$^2$. 
  }}
\end{figure}

  The hard quark propagators  
for the $s$ and $u$ channel handbag diagrams 
in this case are 
 \begin{equation}
\frac{x \hat P + \alpha \hat r/2 + \hat Q}{(xP+\alpha r/2 +Q)^2} = 
\frac{ x \hat P + \alpha \hat r/2 + \hat Q}{x \tilde s -
(\bar x ^2  -\alpha^2)t/4 +x^2 m_p^2} 
\end{equation}
and
\begin{equation}
\frac{x \hat P + \alpha \hat r/2  - \hat Q}{(xP+\alpha r/2-Q)^2} =
\frac{x  \hat P+  \alpha \hat r/2 -\hat Q}{x \tilde u -
(\bar x ^2  -\alpha^2)t/4 +x^2 m_p^2} \, , 
\end{equation}
respectively. We denote $\tilde s =2(pq)=s-m^2$ and  
 $\tilde u =-2(pq')=u-m^2$.
 Since DDs are even functions of $\alpha$, the $\alpha \hat r$ terms 
 in the numerators can be dropped. 
Note that it is legitimate to
keep $O(m_p^2)$ and $O(t)$ terms in the denominators: 
 the dependence of  hard propagators 
on target parameters $m_p^2$ and $t$ can be 
calculated exactly because of  the effect analogous to the 
$\xi$-scaling 
in DIS\,\cite{Georgi:1976ve} (see also Ref. 65). 
Note that the $t$-correction to hard propagators disappears 
in the large-$t$  limit dominated by the $x \sim 1$ integration.
The $t$-corrections are the largest for $\alpha=0$. 
At this value and   for $x=1/2$  and $t=u$, corresponding to 
 90$^{\circ}$ angle in the center of mass system (cms), 
the $t$-term in the denominator of the most important
second propagator is 
only 1/8 of the $u$ term. This ratio increases to 1/3 for 
$x=1/3$. However, at nonzero $\alpha$-values, the $t$-corrections 
are smaller. Hence, the $t$-corrections in the denominators
of hard propagators can produce $10\% -20 \%$ effects  
and should be included in a complete analysis.     
Here, we consider an approximation in which 
 these terms are neglected 
and  hard propagators are given 
by $\alpha$-independent 
expressions $(x \hat P + \hat Q)/x \tilde s$ 
and $(x \hat P + \hat Q)/x \tilde u$.
As a result, the  $\alpha$-integration acts only on the
  DDs ${f }(x,\alpha ;t)$ and 
converts them into  nonforward densities ${\cal F}(x,t)$. 
The latter    
appear  through two types of integrals
 \begin{equation}
\int_0^1 
{\cal F}^a(x,t) \, {dx} \equiv F_1^a(t) \ \ {\rm and} \ \ \int_0^1 
{\cal F}^a(x,t) \, \frac{dx}{x} \equiv   R_1^a(t),  
\end{equation}
 and similarly for ${\cal K,G,P}$.
 The functions $F_1^a(t)$ are the flavor components 
 of the usual $F_1(t)$ form factor while $R_1^a(t)$
 are  the flavor components of a new form factor
  specific to the wide-angle Compton scattering.
  In the formal asymptotic limit $|t| \to \infty$, the $x$ integrals 
  for $F_1^a(t)$ and $R_1^a(t)$ 
  are both dominated 
  by the $x \sim 1$ region: the large-$t$ 
  behavior  of these functions is governed 
  by the Feynman mechanism and their ratio tends to 1 as
  $|t|$  increases (see Fig. \ref{fig:rcuux}a). However, due to 
  large value of the effective scale $4 \lambda^2 =2.8$ GeV$^2$, 
  the  accessible momentum transfers
  $t < 5$  GeV$^2$ are very far from being asymptotic.

In Fig. \ref{fig:rcuux}b, the plotted functions 
are   
   ${\cal F}^u (x;t)$ and ${\cal F}^u (x;t)/x$ at  $t = - 2.5$ GeV$^2$. 
  It is clear that the relevant integrals  are dominated
  by rather small $x$ values $x < 0.4$ 
  which results in a strong
  enhancement of $R_1^u(t)$ 
   compared to $F_1^u(t)$ for $|t| < 5$  GeV$^2$.
   Note also that the 
   $\langle p' | \ldots x \hat P \ldots |p \rangle $ 
   matrix elements  can produce only  $t$ as a large variable 
   while $\langle p' | \ldots  \hat Q \ldots |p \rangle $ 
   gives $s$. As a result, the enhanced form factors 
   $R_1^a(t)$ are accompanied by extra $s/t$  factors
   compared to the $F_1^a(t)$ terms. In the cross section,
   these enhancements are squared, i.e.,  
   the contributions due to the non-enhanced form factors   $F_1^a(t)$ 
  are always accompanied by $t^2/s^2$ factors
  which are smaller than 1/4 
  for cms angles 
  below 90$^{\circ}$. Because of double suppression,
   the terms with  $F_1^a(t)$ 
 can be neglected  in a  simplified 
  approach.

 The  contribution due to the ${\cal K}$ functions  
  appears through  the flavor components $F_2^a(t)$ of the $F_2(t)$
  form factor and their enhanced analogues $R_2^a(t)$. 
  The major part of contributions due to  the ${\cal K}$-type NDs 
  appears in the  combination 
  \mbox{$R_1^2(t)-(t/4m_p^2)R_2^2(t)$.}
  Experimentally, $F_2(t)/F_1(t)\approx 1 \,{\rm GeV}^2/|t|$.
  Since $R_2/F_2 \sim R_1/F_1 \sim 1/ \langle x  \rangle $, 
  $R_2(t)$ is similarly suppressed compared 
  to $R_1(t)$,  and we  neglect contributions 
  due to the $R_2^a(t)$ form factors.  
  We also neglect here the terms with 
  another spin-flip distribution  ${\cal P}$  related
  to the pseudoscalar form factor $G_P(t)$ which is dominated 
  by the $t$-channel pion exchange.  Our  calculations
  show that the contribution due to  
  the parton helicity sensitive densities  ${\cal G}^a$ 
  is suppressed by the factor $t^2/2s^2$ compared to that due to the 
  ${\cal F}^a$ densities. This factor only reaches
  1/8  for the cm angle of  
  90$^{\circ}$,  and hence the ${\cal G}^a$ contributions are not 
  very significant  numerically. For simplicity,  we 
  approximate  ${\cal G}^a(x,t) $ by ${\cal F}^a(x,t)$. 
After these approximations,
the WACS  cross section is given by the product  
 \begin{equation}
 \frac{ d \sigma}{dt}  \approx \frac{2 \pi \alpha^2}{\tilde s^2} 
 \left [  \frac{(pq)}{(pq')} +  \frac{(pq')}{(pq)}
  \right ] \,  R_1^2(t) \,  , 
 \end{equation}
 of  the  Klein-Nishina  cross section 
(in which we dropped  $O(m^2)$ and $O(m^4)$ terms) and 
 the square of the $R_1(t)$ form factor 
  \begin{equation} 
 R_1(t) =  \sum_a e_a^2  \left [R_1^a (t) + R_1^{\bar a} (t) \right ] \, .
 \end{equation}
 In the model of Ref. 25, 
 the effective form factor
  $R_1(t)$ is given by 
 \begin{equation} 
R_1(t) =   \int_0^1 
\biggl [ e_u^2 \, f_u^{val} (x) +
  e_d^2  \, f_d^{val} (x) + 2( e_u^2+e_d^2+e_s^2) 
  \, f^{sea} (x) \biggr ] 
  e^{\bar x t / 4x \lambda^2} \frac{dx}{x} \, . 
\end{equation}
 The 
 sea   distributions are included  here assuming that they are all
  equal $f^{sea} (x)
=f^{sea}_{u,d,s} (x)= f_{\bar u, \bar d,\bar s} (x)  $ 
and given by   a  simplified parametrization 
\begin{equation}
f^{sea} (x) = 0.5 \,  x^{-0.75} (1-x)^7
\end{equation}
which  accurately reproduces  
 the  GRV formula for $Q^2 \sim 1$ GeV$^2$.
 Due to  suppression  of the small-$x$ region
 by the exponential $\exp [\bar x t / 4x \lambda^2]$,
 the sea quark contribution is rather 
 small ($\sim 10 \%$) even for $-t \sim  1$   GeV$^2$ and 
 is invisible  for  $-t > 3 $   GeV$^2$.

\begin{figure}[htb]
\mbox{
   \epsfxsize=8cm
 \epsfysize=11cm
 \hspace{2cm}  
  \epsffile{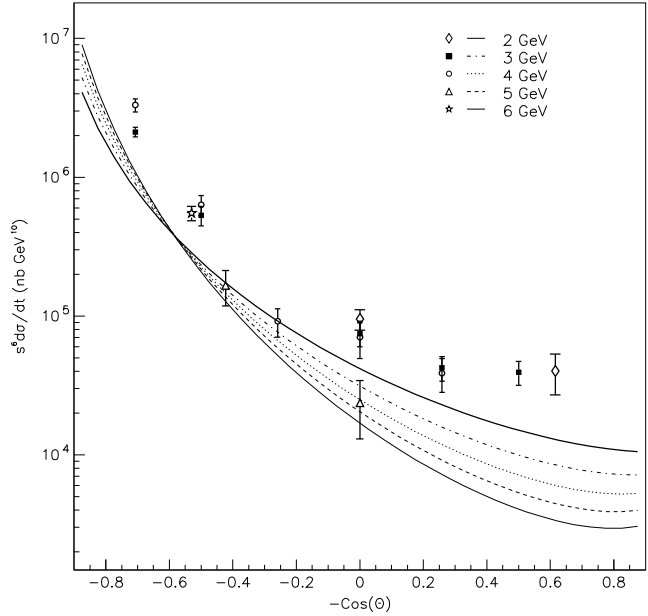}  }
  \vspace{-1.2cm}
{\caption{\label{fig:rctheta}  Angular dependence of 
the combination $s^6 (d \sigma /dt)$.
   }}
\end{figure}

Comparison with existing data\,\cite{shupe} 
is shown in Fig. \ref{fig:rctheta}.  
The curves  follow the data pattern
but are systematically lower  by a factor of  2,
with disagreement becoming more pronounced 
 as  the scattering angle  increases.
Since  several terms were neglected each capable 
of producing up to a $20 \%$ correction in  the amplitude, 
the agreement between  curves and the data 
may be treated as  encouraging.
 The most important corrections which should 
be included in a more detailed investigation
are the $t$-corrections in the denominators of 
hard propagators and contributions due to the ``non-leading''
${\cal K, G,P}$ nonforward densities.
The latter, as noted above, are usually accompanied 
by $t/s$ and $t/u$ factors, i.e., their contribution 
becomes  more significant at larger angles. 
The $t$-correction in the most important hard propagator term  
$1/[x \tilde u - (\bar x^2 - \alpha^2)t/4 +x^2 m_p^2]$
also enhances the amplitude at large angles.
   
Note that the curves  for the combination
$s^6 (d \sigma /dt)$ taken  for the initial photon energies 
2, 3, 4, 5 and 6 GeV
intersect each other at  $\theta_{\rm cm} \sim 60^{\circ}$. 
This  is in good agreement with experimental data
of Ref.~53
where the differential 
cross section at fixed cms angles was fitted by powers of $s$:
$d \sigma /dt \sim s^{-n (\theta)}$ with 
$n^{\rm exp}(60^{\circ}) = 5.9 \pm 0.3$. 
The  curves of Ref. 25 
correspond to 
$n^{\rm soft}(60^{\circ}) \approx 6.1$
and $n^{\rm soft}(90^{\circ}) \approx 6.7$ which also agrees 
with the experimental result $n^{\rm exp}(90^{\circ}) = 7.1 \pm 0.4$. 

This can be compared with the scaling behavior
of the asymptotic  hard contribution: 
modulo logarithms contained in the $\alpha_s$ factors,
they have    a universal angle-independent power
$n^{\rm hard}  (\theta) =6$.
For $\theta_{\rm cms}  = 105^{\circ}$, the experimental result
based on just two  data points is $n^{\rm exp}(105^{\circ}) = 6.2 \pm 1.4$,
while our model gives $n^{\rm soft}(105^{\circ}) \approx 7.0$.
Clearly, better data are needed to draw any conclusions here.
 
\section{Concluding remarks}
\label{sec_conclusions}

In this paper, I described the basic 
elements of the theory of generalized
parton distributions. 
For uniformity of presentation,
I heavily relied on  the approach developed 
in my papers, see Refs.~16--18 and 21--25.
In this concluding section,
I     briefly list  other   developments
in the theory of GPDs and  its applications,
not covered in the present review.  
Additional  references can be  found 
in the reviews and review sections of the original papers cited in 
Refs.~11, 14, 17, 19--21, 45, 66, 50, and 67.

{\it Introduction of GPDs and factorization.}  
In various ways (and under
different names: off-forward, nonforward, nondiagonal, off-diagonal, etc.)
GPDs were introduced~\cite{drm,ji,compton,gluon,cfs}
as nonperturbative functions describing the 
nonperturbative part of the  factorized representation
for the amplitudes of hard elastic electroproduction  processes.
The  PQCD factorization for DVCS and meson production processes was 
discussed in Refs.~17, 20, 68, 69. 
One loop corrections to the DVCS amplitude 
were calculated in Refs.~70--72. 
  
{\it Evolution of GPDs.} Evolution equations for GPDs
obtained originally
at   one loop level~\cite{ji2,npd,ffgs,bgr,gevol} 
were used for  numerical studies of evolution
using the orthogonal polynomials techniques~\cite{bello,lechlo}
and direct integration.\cite{gufre,golec,sinevol} 
Two-loop evolution was investigated in Refs.~78--80. 

{\it Applications of  GPDs to hard  electroproduction
processes.}  There is a growing
literature devoted to  practical aspects 
of  using  GPDs in the description 
of hard electroproduction processes.
In particular, deeply virtual Compton 
scattering  is discussed in Refs.~11, 44, 66, 62, 81--84.
The  PQCD approach 
to  meson electroproduction at large $Q^2$ formulated
in Refs.~20, 16, 17 
was applied  to particular channels: 
$\rho$-meson production,\cite{Guichon,Mankiewicz:1998uy,Mankiewicz:1999aa}, 
pion production,\cite{Guichon,Mankiewicz:1999kg,Frankfurt:1999fp}
and, finally,  
electroproduction processes accompanied by excitation
of decuplet baryons.\cite{Frankfurt:2000xe}


\section*{Acknowledgments}

I would like to express my gratitude to   
I.~I.~Balitsky,  I.~V.~Musatov  and 
\mbox{C.~Weiss} in    collaboration with whom 
I obtained  some of the results described 
in the present review.   
 For stimulating discussions and correspondence
 I thank
 A. Bakulev, A. Belitsky,
 J. Blumlein, V. Braun, S. Brodsky, J. Collins, 
 M. Diehl, 
  L. Frankfurt,
 A. Freund, B. Geyer, K. Goeke,
 K. Golec-Biernat, P. Guichon, M. Guidal, 
 X. Ji, N. Kivel,
  G. Korchemsky, P. Kroll, E. Levin,
  L. Mankiewicz,  A.D. Martin, D. Muller, 
 G. Piller, B. Pire,
M. Polyakov,   D. Robaschik, R. Ruskov, M. Ryskin,
A. Sch\"afer, 
M. Strikman,
O. Teryaev, and M. Vanderhaeghen.   
This work was supported in part by the 
U.S. Department of Energy 
 under Contract No. DE-AC05-84ER40150.

\end{document}